\documentstyle[onecolumn,epsf]{mn}

\newif\ifAMStwofonts

\def\simless{\mathbin{\lower 3pt\hbox
   {$\rlap{\raise 5pt\hbox{$\char'074$}}\mathchar"7218$}}} 
\def\simgreat{\mathbin{\lower 3pt\hbox
   {$\rlap{\raise 5pt\hbox{$\char'076$}}\mathchar"7218$}}} 
\newcommand{\ol}{\overline}  
\newcommand{\bl}{\left \langle}
\newcommand{\br}{\right \rangle}

\newcommand{\beq}{\begin{equation}}
\newcommand{\eeq}{\end{equation}} 
\newcommand{\beqa}{\begin{eqnarray}} 
\newcommand{\eeqa}{\end{eqnarray}}  

\newcommand{\btu}{\bigtriangleup}
\newcommand{\tr}{\rmn Tr}
\newcommand{\erf}{{\rmn erf}}

\newcommand{\kms}{\rmn km\ s^{-1}}
\newcommand{\pc}{\rmn pc} 
\newcommand{\kpc}{\rmn kpc}

\newcommand{\msun}{M_{\odot}}
\newcommand{\mbh}{M_{\rmn bh}}
\newcommand{\mbhcrit}{M_{\rmn bh,crit}}
\newcommand{\fhalo}{f_{\rmn halo}}
\newcommand{\vrel}{V_{\rmn rel}}
\newcommand{\vreli}{V_{{\rmn rel},i}}
\newcommand{\rhm}{r_{\rmn hm}}
\newcommand{\mhigh}{M_{\rmn high}}
\newcommand{\psic}{\tilde{\psi}_0}
\newcommand{\vbh}{V_{\rmn bh}}
\newcommand{\vcl}{V_{\rmn cl}} 

\newcommand{\odm}{{\cal O}(\btu M/M)}
\newcommand{\odmtwo}{{\cal O}(\btu M/M)^2} 
\newcommand{\rt}{r_{\rmn t}}
\newcommand{\rcore}{r_{\rmn core}}
\newcommand{\mcyl}{M_{\rmn cyl}}
\newcommand{\rmnd}{{\rmn d}}
\newcommand{\bd}{b_{\rmn d}}
\newcommand{\bdiff}{b_{\rmn diff}}
\newcommand{\nbh}{n_{\rmn bh}}
\newcommand{\rhobh}{\rho_{\rmn bh}}
\newcommand{\fdest}{f_{\rmn dest}}
\newcommand{\fking}{f_{\rmn king}}
\newcommand{\vesc}{v_{\rmn esc}}
\newcommand{\sigmabh}{\sigma_{\rmn bh}}
\newcommand{\ej}{({\rmn ej})}
\newcommand{\bound}{({\rmn b})} 
\newcommand{\vesci}{v_{{\rmn esc},i}}
\newcommand{\pej}{p_{\rmn ej}}
\newcommand{\Nej}{N_{\rmn ej}}

\newcommand{\be}{{\bmath e}}
\newcommand{\bv}{{\bmath v}}
\newcommand{\dbv}{\delta {\bmath v}}
\newcommand{\bvrel}{{\bmath V_{\rmn\! rel}}}
\newcommand{\bb}{{\bmath b}}
\newcommand{\bR}{{\bmath R}}
\newcommand{\brr}{{\bmath r}}
\newcommand{\bvbh}{{\bmath V_{\rmn\! bh}}}
\newcommand{\bvcl}{{\bmath V_{\rmn\! cl}}}
\newcommand{\bs}{{\bmath s}}
\newcommand{\bXi}{{\bmath \Xi}}
\newcommand{\bLambda}{{\bmath \Lambda}}

\ifoldfss
  \newcommand{\rmn}[1] {{\rm #1}}

  \ifCUPmtlplainloaded \else
    \NewTextAlphabet{textbfit} {cmbxti10} {}
    \NewTextAlphabet{textbfss} {cmssbx10} {}
    \NewMathAlphabet{mathbfit} {cmbxti10} {} 
    \NewMathAlphabet{mathbfss} {cmssbx10} {} 
  \fi
  \ifAMStwofonts
    \ifCUPmtlplainloaded \else
      \NewSymbolFont{upmath} {eurm10}
      \NewSymbolFont{AMSa} {msam10}
      \NewMathSymbol{\upi}     {0}{upmath}{19}
      \NewMathSymbol{\umu}     {0}{upmath}{16}
      \NewMathSymbol{\upartial}{0}{upmath}{40}
      \NewMathSymbol{\leqslant}{3}{AMSa}{36}
      \NewMathSymbol{\geqslant}{3}{AMSa}{3E}

      \let\leq=\leqslant 
      \let\geq=\geqslant 
    \fi
  \fi  
\fi 
 
\ifnfssone
  \newmathalphabet{\mathit}
  \addtoversion{normal}{\mathit}{cmr}{m}{it}
  \addtoversion{bold}{\mathit}{cmr}{bx}{it}
  \newcommand{\rmn}[1] {\mathrm{#1}}

  \newmathalphabet{\mathbfit} 
  \addtoversion{normal}{\mathbfit}{cmr}{bx}{it}
  \addtoversion{bold}{\mathbfit}{cmr}{bx}{it}
  \newmathalphabet{\mathbfss} 
  \addtoversion{normal}{\mathbfss}{cmss}{bx}{n}
  \addtoversion{bold}{\mathbfss}{cmss}{bx}{n}
  \ifAMStwofonts
    \ifCUPmtlplainloaded \else
      \UseAMStwoboldmath
      \makeatletter
      \new@mathgroup\upmath@group
      \define@mathgroup\mv@normal\upmath@group{eur}{m}{n}
      \define@mathgroup\mv@bold\upmath@group{eur}{b}{n}
      \edef\UPM{\hexnumber\upmath@group}
      \new@mathgroup\amsa@group
      \define@mathgroup\mv@normal\amsa@group{msa}{m}{n}
      \define@mathgroup\mv@bold\amsa@group{msa}{m}{n}
      \edef\AMSa{\hexnumber\amsa@group}
      \makeatother
      \mathchardef\upi="0\UPM19
      \mathchardef\umu="0\UPM16
      \mathchardef\upartial="0\UPM40
      \mathchardef\leqslant="3\AMSa36
      \mathchardef\geqslant="3\AMSa3E

      \let\leq=\leqslant 
      \let\geq=\geqslant 
    \fi
  \fi
\fi 
 
\ifnfsstwo
  \newcommand{\rmn}[1] {\mathrm{#1}}

  \DeclareMathAlphabet{\mathbfit}{OT1}{cmr}{bx}{it}
  \SetMathAlphabet\mathbfit{bold}{OT1}{cmr}{bx}{it}
  \DeclareMathAlphabet{\mathbfss}{OT1}{cmss}{bx}{n}
  \SetMathAlphabet\mathbfss{bold}{OT1}{cmss}{bx}{n}
  \ifAMStwofonts
    \ifCUPmtlplainloaded \else
      \DeclareSymbolFont{UPM}{U}{eur}{m}{n}
      \SetSymbolFont{UPM}{bold}{U}{eur}{b}{n}
      \DeclareSymbolFont{AMSa}{U}{msa}{m}{n}
      \DeclareMathSymbol{\upi}{0}{UPM}{"19}
      \DeclareMathSymbol{\umu}{0}{UPM}{"16}
      \DeclareMathSymbol{\upartial}{0}{UPM}{"40}
      \DeclareMathSymbol{\leqslant}{3}{AMSa}{"36}
      \DeclareMathSymbol{\geqslant}{3}{AMSa}{"3E}

      \let\leq=\leqslant 
      \let\geq=\geqslant 
    \fi
  \fi
\fi 
 
\ifCUPmtlplainloaded \else
  \ifAMStwofonts \else 
    \def\upi{\pi}
    \def\umu{\mu}
    \def\upartial{\partial}
  \fi
\fi

\title[Limits on Massive Halo Black Holes]
{Canaries in a Coal Mine: Using Globular Clusters to Place Limits
on Massive Black Holes in the Galactic Halo}

\author[P. Arras and I. Wasserman]
{Phil Arras and Ira Wasserman \\
Center for Radiophysics and Space Research, 
Cornell University, Ithaca, NY 14853}

\pagerange{\pageref{firstpage}--\pageref{lastpage}}
 
\begin{document}
 
\maketitle
 
\label{firstpage}
 
\begin{abstract}

	We explore the possibility that massive black holes 
comprise a significant fraction of the dark matter of our galaxy
by studying the dissolution of galactic globular clusters bombarded
by them. In our simulations, we evolve the clusters along a sequence
of King models determined by changes of state resulting from 
collisions with the black holes. We include mass loss in collisions
as well as the heating of the remaining bound stars, and determine
the role which a finite number of stars plays in the variance of the
energy input and mass loss. Several methods are used to determine the
range of black hole masses and abundances excluded by survival of 
galactic globular clusters: simple order of magnitide estimates; 
collision-by-collision simulations of the energy input and mass 
loss of a stellar cluster; and a `smoothed' Monte Carlo calculation of
the evolution of cluster energy and mass. The results divide naturally
into regimes of `small' and `large' black hole mass. `Small' black
holes do not destroy clusters in single collisions; their effect is
primarily cumulative, leading to a relation between $\mbh$ and
$\fhalo$, the fraction of the halo in black holes of mass $\mbh$, 
which is $\fhalo\mbh < $ constant (up to logarithmic corrections). 
For $\fhalo=1$, 
we find $\mbh \simless 10^{3} \msun$ by requiring survival of the
same clusters studied by Moore \shortcite{moore93}, who neglected cluster 
evolution, mass loss, and stochasticity of energy inputs in his estimates,
but reached a similar conclusion. `Large' black holes may not 
penetrate a cluster without disrupting it; their effect is mainly 
catastrophic (close collisions), but also partly cumulative (distant
collisions). In the large $\mbh$ limit,
$\fhalo$ (but not $\mbh$) can be constrained by computing
the probability that a cluster survives a combination of close, 
destructive encounters and distant, nondestructive encounters.
We find that it is unlikely that $\fhalo \simgreat 0.3$ by requiring
$50$ per cent survival probability for Moore's clusters over 
$10^{10}$ years.

\end{abstract}


\section {Introduction}

In this paper, we reexamine the idea that black holes compose a 
substantial fraction of the dark matter in the halo of our galaxy.
Ostriker \& Lacey \shortcite{ol85} originally proposed that heating by
a significant population of halo black holes with masses
$\mbh \sim 10^6 M_{\odot}$ could explain Wielen's (1977) inference that
the velocity dispersions $\sigma_\star$ of disk stars of age
$t_\star$ follow the trend $\sigma_\star\propto t_\star^{1/2}$.
Although subsequent work questioned the basis of this argument,
both because it is unclear that $\sigma_\star$ rises as fast
as $t_\star^{1/2}$ (e.g. Carlberg et al. 1985; Str\"omgren  1987;
Gomez et al. 1990),
and that black holes are needed to explain the observed
trend of $\sigma_\star$ with $t_\star$ (e.g. Lacey 1991), the
heating of disk stars remains a constraint on the masses and
abundance of any hypothetical population of massive halo
objects; in the notation of Wasserman \& Salpeter \shortcite{iraned94},
if the halo consists of a fraction $\fhalo$ in the form of massive
black holes, then the Ostriker \& Lacey \shortcite{ol85} argument implies
$\fhalo \mbh \simless 10^{6.6}M_{\odot}$. Limits on the masses and
abundances of massive black holes could have important
cosmological implications (e.g. Loeb 1993; Umemura, Loeb, \& Turner, 1993; 
Loeb \& Rasio 1994).

Carr \shortcite{carr94} has reviewed various limits on baryonic
or black hole dark matter in the halo of our Galaxy. (See also
Carr \& Sakellariadou 1998.) One of the most powerful constraints
on black hole properties yet proposed was put forth by Moore 
\shortcite{moore93}
who concluded that the survival of a set of relatively
tenuous, low mass ($M \simless 4\times 10^4 M_{\odot}$)
globular clusters over a timespan $\simgreat 7\times
10^9$ years would be possible only if $\mbh \simless 10^3 M_{\odot}$.
The argument employed by Moore\ \shortcite{moore93} was first applied to this
problem by Wielen \shortcite{wiel85}, who explored perturbations of more
massive clusters ($M \sim 10^6 M_{\odot}$) by heavier black holes,
in the range advocated by Ostriker \& Lacey \shortcite{ol85}. 
[Klesson \& Burkert (1995)
extended Moore's and Wielen's calculations to
a range of globular cluster properties.] The idea
is to compute the heating of a globular cluster by passing
black holes over a chosen timescale comparable to the cluster
age [as noted above, Moore \shortcite{moore93} used $7\times 10^9$ years,
roughly half the age inferred for typical clusters]. If the
total energy imparted by black hole perturbations is
sufficiently large, then the cluster is said to be disrupted.

In view of the importance of this problem, we
have begun a more comprehensive
study of the disruption of the same set of globular clusters
considered by Moore \shortcite{moore93} by
a hypothetical population of halo black holes. This investigation aims
to tighten up Moore's argument in several different ways. Some
of the improvements are technical [e.g. Moore used a simple
analytic form for the energy input to a cluster by a passing
black hole which is only very approximate; Klesson \& Burkert (1995)
improved on his formula], but others are
qualitative. Of particular importance are: (1)
Moore computed the energy
input for a `static' cluster, whose structure was held fixed.
Our calculations evolve the clusters along a King sequence.
(2) Qualitatively, one expects a cluster that
gains a large amount of energy compared to its initial binding
energy from encounters with passing black holes to lose a large
fraction of its mass, too. Our calculations include mass
loss by the clusters.
(3) Although one can get a
rough idea of the survival probability by considering the mean
heating of a cluster, the energy transfer process is actually
stochastic, and the variance in the energy input may play
an important role in final estimates of critical masses
for disruption to be likely.
(4) Moore's calculations
pertain to black holes of relatively low mass, which are incapable
of disrupting a cluster in a single perturbation. For black hole
masses $\mbh \simgreat M\vrel/\sigma_{\rmn cl}$, 
where $\sigma_{\rmn cl}$ is 
the characteristic
velocity dispersion of the cluster and $\vrel$ its characteristic
speed relative to the approaching black hole 
($\vrel \sim 330 \kms$ is
a fair estimate), a single encounter at the cluster's tidal radius
will likely destroy it. In this regime, one can obtain limits on
the halo mass density in massive black holes, but not on
$\mbh$ (e.g. Wielen 1988; Wasserman \& Salpeter 1994).

Our study combines analytic and Monte Carlo calculations.
Our most complete results come from Monte Carlo simulations in which we
simulate each encounter between a cluster and a black hole separately.
For simplicity, we shall consider clusters at fixed galactocentric
radius, as was done by Moore \shortcite{moore93}. We model clusters using
N point masses whose positions and
velocities are chosen from a King distribution (see, e.g., 
Binney \& Tremaine 1987). 
Velocity perturbations are
computed for a given black hole impact parameter and speed
relative to the cluster center of mass using the impulse approximation
(e.g. Binney \& Tremaine 1987). The change in velocity of the cluster
center of mass is subtracted from each individual
velocity perturbation to determine the change in cluster energy
and mass as a consequence of the collision. 
Determining which stars are ejected is tricky in any scheme 
that does not employ a direct
N body simulation of interparticle interactions, but as long as the
perturbations in individual encounters are not excessive, it should
be sufficient to designate for ejection those stars whose
post-collision velocities exceed their local pre-collision escape speed.
To find the new King model that describes the remaining cluster, we need
its tidal radius in addition to its mass and energy after the black
hole encounter; we get this by assuming that the tidal radius
is proportional to $M^{1/3}$, consistent with our assumption of
fixed galactocentric radius. Within these `rules of the game' we
simulate the evolution of the globular clusters studied by Moore 
\shortcite{moore93}
over a timespan of $10^{10}$ years, or until they are disrupted, whichever
comes first. Our simulations are more comprehensive than the
Monte Carlo calculations of Klesson \& Burkert \shortcite{kb95}, who
chose discrete black hole encounter times, relative velocities
$\vrel$ and impact parameters $b$ from the appropriate
probability distributions, but merely updated the cluster
velocity dispersion by a completely deterministic amount
$\btu \sigma(b,\vrel)$, without accounting for stochasticity
of heating, mass loss or the change in internal cluster
structure in individual encounters.

Several different criteria are employed to decide whether or not
a cluster has been destroyed. Some of these may be called `global'
in that they depend on integrated properties of the cluster, such as
total energy or total mass. 
We can regard a cluster as having been destroyed,
for example, when its mass or energy or energy per mass has changed by
a fractional amount in excess of some pre-set values (e.g. 0.5). Since
we evolve models along a sequence of King models which is limited, clusters
may also die when they reach the end of the sequence (see also 
Chernoff, Kochanek, \& Shapiro 1986).
The other criteria for cluster disruption
to be used are `local', in that they depend on changes in the properties
of a cluster as a consequence of a single collision. Thus, if the mass or
energy of a cluster changes by more than some pre-set fractional amounts
in an encounter with a black hole, we shall regard it as disrupted. This
should also allow us to control the inaccuracy of our criteria for 
determining the mass loss per encounter somewhat. We evaluate
survival probabilities for the various criteria separately.
 
The Monte Carlo calculations outlined above allow us to study both
the large and small $\mbh$ regimes. In the large $\mbh$ regime, where
destruction occurs only after relatively few encounters, the Monte Carlo
calculations ought to be reasonably fast computationally. However, for
small $\mbh$ where the encounters are more frequent
(the regime focussed on by Moore 1993), we expect the Monte
Carlo calculations to be more cumbersome, thus limiting the number of
stars we can use in realizing clusters. Ideally, one would like to
be able to represent the stars `one-by-one' since the variances in
energy input depend on the number of cluster particles. Fortunately,
the perturbations due to individual encounters are relatively gentle in
the small $\mbh$ regime, and the problem lends itself to a Fokker-Planck
treatment, which promises to be faster (at the price -- justified in our
view -- of losing the ability to resolve short timescale structure in
the cluster evolution). We present a two dimensional Fokker-Planck
scheme in which we follow cluster evolution in mass and energy.

The calculations reported here suffer from two or three principal
deficiencies. Most important is their reliance on the sequence of 
King models
and on the impulse approximation. In addition, one would like
to interweave perturbations by a hypothetical population of black holes
with well-established sources of heating, such as disk shocking (e.g.
Ostriker, Spitzer, \& Chevalier 1972; Spitzer \& Chevalier 1973;
Chernoff et al. 1986; 
Binney \& Tremaine 1987);
in the similar problem of
wide binary evolution, the interplay of perturbations by stars, molecular
clouds, and dark matter is known to be important (e.g. Retterer \& 
King 1982; Bahcall, Hut, \& Tremaine 1985;
Weinberg, Shapiro, \& Wasserman 1987; Wasserman \& Weinberg 1991). 
Moroever, globular clusters
evolve on their own as a consequence of internal relaxation, resulting
in evaporation and energy changes even 
without external perturbations; limits
on halo black hole properties may be altered when internally induced
changes in state are accounted for properly. These important effects
will be ignored here to concentrate solely on the cluster evolution
due to the collisions with black holes. We shall study some of these
issues in a subsequent paper \cite{murali98}.

\section{ Qualitative Overview of the Collision Process and 
Evolutionary Scenarios }

\subsection{ Approximations } 

Throughout this paper we employ the impulse approximation and 
ignore deflection of the perturbing black hole orbit from a straight
line. For nonpenetrating encounters at impact parameter $b$ and
relative velocity $\vrel$, these approximations are valid when
$\Omega b/\vrel \simless 1$. The characteristic frequency is roughly
$\Omega \sim (GM/\rhm^3)^{1/2}$, where $M$ is the cluster
mass and $\rhm$ is the cluster half mass radius, for the outer 
parts of the cluster where the tidal perturbation is strongest.
If we set $\vrel=\xi_1 V_c$, where
$V_c$ is the galactic circular speed, and let $\sigma_0=\xi_2
(GM/\rhm)^{1/2}$ be the cluster's central one-dimensional
velocity dispersion, then the
impulse and straight line approximations should hold for 
$b/\rhm \simless (\xi_1 \xi_2) V_c/\sigma_0$. For the clusters studied
in this paper, $V_c/\sigma_0 \simgreat 10-100$, 
and the approximations only
fail far outside $\rhm$.

For penetrating encounters, this assessment remains valid for
`typical' collisions, but may fail for perturbations of particles
deep in the cluster core. This is because the characteristic frequency
near the center of the cluster where the perturbations are largest
is $\Omega \simeq (4\pi G \rho_0/3)^{1/2}$, where $\rho_0$ is the
central mass density of the cluster. Adopting $b=\rhm$ as typical,
we find that the impulse and straight line approximations
hold for $\vrel \simgreat 0.1 \times \rhm(\pc) 
[\rho_0/\msun \pc^{-3}]^{1/2} \kms$. Although our approximations would
fail for the most concentrated clusters, they remain true for
the relatively tenous ones studied here (and indeed for many `normal' 
clusters). 

We also ignore all processes influencing cluster evolution except 
perturbations by black holes even though our limits are based on
survival probabilities over timescales long enough for tidal shocking
and internal relaxation to be important. In more realistic simulations,
including these effects could tighten limits
on properties of hypothetical halo black holes. 

\subsection{ Single Collisions }

Consider a single collision between a globular cluster and a black hole
which takes place in the halo of the Galaxy. Let the cluster
have tidal
radius $\rt$, $N$ stars, King model parameter
$W_0 \equiv \tilde{\psi}_0$, and total energy $E$. The 
black hole passes the cluster with a speed $\vrel$
and an impact parameter $b$. 
Define the `collision parameter', 
$\eta_c=(\mbh/M)(\sigma_0/\vrel)$.
In section $\S \ref{einputformulas}$, it will be shown that
in the impulsive limit
the energy input, mass loss and their variances are
\beqa
&&
\frac{ \ol{\btu E(b,\vrel)} }{|E|} = 
\eta_c^2 R_E(b,\eta_c,\tilde{\psi}_0)
\nonumber \\ &&
\frac{ \ol{\btu M(b,\vrel)} }{M} =-  
\eta_c^2 R_M(b,\eta_c,\tilde{\psi}_0)
\nonumber \\ && 
\frac{ \sigma^2_{\btu E}}{E^2} =   
\frac{\eta_c^2}{N} R_{EE}(b,\eta_c,\tilde{\psi}_0)
\nonumber \\ && 
\frac{ \sigma^2_{\btu M}  }{M^2} =   
\frac{\eta_c^2}{N} R_{MM}(b,\eta_c,\tilde{\psi}_0)
\label{deestimate}
\eeqa
where the $b$ dependent radial functions have the approximate limits
(see $\S \ref{tidallimit}$)
\beqa
&&
R_E \simeq C_{E}(\tilde{\psi}_0)\left( \frac{\rt}{b} \right)^4
\nonumber \\ &&
R_M \simeq C_{M}(\tilde{\psi}_0)\left( \frac{\rt}{b} \right)^4
\nonumber \\ &&
R_{EE} \simeq C_{EE}(\tilde{\psi}_0)\left( \frac{\rt}{b} \right)^4
\nonumber \\ &&
R_{MM} \simeq  C_{MM}(\tilde{\psi}_0)\left( \frac{\rt}{b} \right)^4
\label{deestimatetidal}
\eeqa 
when $b \simgreat \rt$. 
As much of the mass in
the cluster is contained within the core radius, one can actually use
the $b^{-4}$ dependence all the way into the core as a first 
approximation. Examples of the radial functions $R_E$, $R_M$, $R_{EE}$, 
and $R_{MM}$ for impacts inside the cluster 
are given in $\S \ref{einputformulas}$.

For large enough $\mbh$, a single collision will suffice to
disrupt the cluster. This will be referred to as the 
`high black hole mass limit'. Define `disruption' to occur 
for an energy input of size $f|E|$. Then the cluster is destroyed in a
single collision for
\beqa
\frac{b}{\rt} & \simless &\frac{\bd}{\rt} \equiv
 \left( \frac{C_E}{f} \right)^{1/4} \eta_c^{1/2}.
\eeqa
A safe {\it overestimate} of the mass, $\mhigh$, at which the 
cluster may be disrupted is given by the black hole mass at which 
$\bd=\rt$. Using $\vrel=\xi_1 V_c $, $\xi_1 \simeq 1.5$, and 
the Galactic circular speed $V_c=220\kms$, the result is 
\beqa
\mhigh & = & M \frac{V_c}{\sigma_0} 
\left( \frac{\xi_1^2f}{C_E} \right)^{1/2}
\eeqa
which is bigger than $M$ by a factor of $V_c/\sigma_0\gg 1$.

The probability distribution for energy input (and mass loss) can be
understood in the following qualitative terms. The ratio of the variance
to the mean energy input is
\beqa
&&
\frac{ \sigma^2_{\btu E}}{ \left( \ol{\btu E} \right)^2 }
= \frac{ R_{EE} }
{ N \eta_c^2 R_E^2  }
\simeq \frac{ C_{EE}}{NC_E^2} \frac{1}{\eta_c^2} \frac{b^4}{\rt^4}.
\eeqa
This becomes small for large $N$ and small $\mbh$, but is
also proportional to $b^4$, insuring that 
$\sigma_{\btu E} \geq \ol{\btu E}$ for
\beqa
&& \frac{b}{\rt} \simgreat \frac{\bdiff}{\rt}  =  
\left( \frac{C_E^2}{C_{EE}} \right)^{1/4}  N^{1/4} \eta_c^{1/2}.
\eeqa 
Note that $\bdiff>\rt$ for 
$\mbh/M \simgreat \vrel/(\sqrt{N}\sigma_0)$.
The distributions of mass loss and energy 
input are sharply defined about the mean for $b\simless \bdiff$, 
and become `fuzzy' for $b \simgreat \bdiff$.  
In particular, the energy input will {\it always} be sharply peaked
about the mean at the destructive radius $\bd$, because
$\bd/\bdiff = \left( C_{EE}/fC_EN\right)^{1/4} \ll 1$.
Hence the probability of getting an energy input which is {\it not}
destructive when $b<\bd$ is quite small. 

\subsection{ Evolution Over Many Collisions \label{themany} }

If $\mbh \ll \mhigh$ (`the small $\mbh$ limit'),
it will take many collisions to disrupt the cluster. Individual
collisions only `tickle' the cluster, and the evolution
can be approximated by averaging over the effects of
many collisions. In $\S \ref{diffcoeff}$, we find that the mean
rate of energy input, averaged over $b$ and $\vrel$, takes the form
\beqa
\frac{ \langle \dot E \rangle}{|E|} & =& 
d_E(\eta_{c0},\tilde{\psi}_0) \Gamma_0
\eeqa
where $\Gamma_0 \equiv \nbh\pi \rt^2 V_c$, 
$\nbh$ is the number density of black holes,
$\eta_{c0}=(\mbh/M)(\sigma_0/V_c)$ and 
$d_E(\eta_{c0},\tilde{\psi}_0) \simeq \kappa_E(\tilde{\psi}_0) 
\ln(1/\eta_{c0})\eta_{c0}^2$. The mean time for
disruption, which we define here to occur at $\btu E=f|E|$, is
\beqa
T_{\rmn disrupt} & = & \frac{ f |E|} { \langle \dot E \rangle}
= \frac{ f}{\Gamma_0 \kappa_E \ln(1/\eta_{c0}) \eta_{c0}^2 }
\propto \frac{1}{ \mbh}
\eeqa
so that for very small $\mbh$ it takes a long time to disrupt the 
cluster. Given a value for $T_{\rmn disrupt}$ there is a certain critical
value of $\mbh$, called $\mbhcrit$, 
above which the cluster is disrupted. 
The variance of the averaged energy input in a time $T$ takes 
on the form (see $\S \ref{diffcoeff}$)
\beqa
\frac{ \sigma^2_{\btu E} }{E^2} & = & \frac{d_{EE}}{N} \Gamma_0 T
\eeqa
where $d_{EE}(\eta_{c0},\tilde{\psi}_0) \simeq 
\kappa_{EE}(\tilde{\psi}_0)
\ln(1/\eta_{c0})\eta_{c0}^2$.

	Given an ensemble of clusters with initial energy $E_0$, 
a time $ T \gg 1/\Gamma_0$ later the cluster energies
will be roughly
$E(T) = E_0 + \bl \btu E(T) \br \pm \sigma_E(T)$. For small times the 
variance will dominate the mean corresponding to a very broad spread
in cluster states but for sufficiently long times the ensemble will
peak sharply about the mean.
We find that for times $T \simeq T_{\rmn disrupt}$
\beqa
&&\frac{ \sigma^2_{\btu E}(T_{\rmn disrupt}) }
{ (\bl \btu E(T_{\rmn disrupt}) \br)^2 }
\simeq \frac{\kappa_{EE}}{N\kappa_E} \frac{|E|}
{\langle \btu E(T_{\rmn disrupt}) \rangle}
= \frac{\kappa_{EE}}{Nf\kappa_E} \propto \frac{1}{N} \ll 1
\eeqa
so that the range of final states is always sharply defined about the 
mean for times long enough to disrupt the cluster when 
$\mbh \ll \mhigh$. The probability of survival, $P_s$, is given
(here) by the fraction of clusters with $\btu E < f|E_0|$. This fraction
will change from unity to a minimum value in a
`transition region' with size, $\delta \mbhcrit$,
dictated by the ratio
$\sigma_{\btu E}(T_{\rmn disrupt})/\bl \btu E(T_{\rmn disrupt}) \br $ 
so that $\delta \mbhcrit/\mbhcrit \sim N^{-1/2} \ll 1$.

When $\mbh \simgreat \mhigh$, any collision inside the destructive
radius $\bd$ will destroy the cluster. In the tidal limit, 
the expected number of destructive encounters, 
$N_d$, in the time $T$ is
\beqa
&& N_d  = 
\left(\frac{C_E}{f}\right)^{1/2}
\frac{\rhobh}{M} \pi \rt^2 \sigma_0 T
\eeqa
where $\rhobh=\mbh \nbh$ is the mass density in black holes.
For an ensemble of clusters with initial energy $E_0$, the destructive
$b<\bd$ collisions act as a `sink' for clusters while the nondestructive
$b>\bd$ encounters give rise to a slower, diffusive energy change.
The probability of survival, $P_s$, (which takes on its minimum value
in the $\mbh \simgreat \mhigh$ limit)
is the product of $\exp(-N_d)$, the probability that no single 
destructive encounters occur, and the probability that all the
$b>\bd$ collisions combined give $\btu E < f|E_0|$.
Since $N_d$ does not depend on $\mbh$, only $\rhobh$
(or equivalently $\fhalo$) can be constrained in the 
$\mbh \simgreat \mhigh$ limit, not the black hole mass. In addition, 
the cluster evolution is quite stochastic in this regime as it depends
on whether the destructive collisions do or do not occur.

The model for cluster evolution described in this section is 
summarized in the (purely
illustrative) diagram of Fig. $\ref{schematic}$. The fraction of clusters
destroyed in time $T$, called $f_{\rm dest}=1-P_s$, 
changes from $0$ to $\sim 0.8$
in a region of width $\sim 2000 M_{\odot}$ centered on $\mbhcrit \sim
5000 M_{\odot}$. Note that $f_{\rm dest}$ 
in this example does not asymptote
at one, but instead reaches $f_{\rm dest} \sim 0.8$ corresponding to 
$N_d \sim 1$.

\setcounter{figure}{0}
\begin{figure}

\begin{picture}(324,324)(0,0)

\put(81,0){\epsfxsize=4.5in
\epsffile{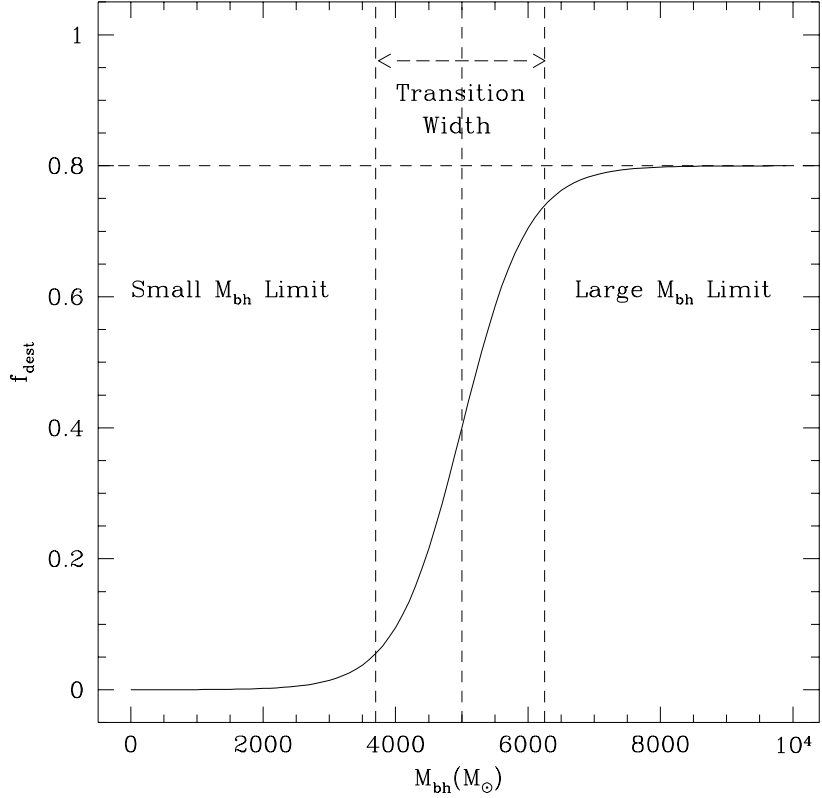}}
\end{picture}

\caption{ Schematic of fraction of cluster destroyed as a function
of $\mbh$}

\label{schematic}

\end{figure}

\section{The Cluster Model and Evolution of the Clusters} 
\label{clustermodel}

The nondimensional King models (e.g. Binney \& Tremaine 1987) 
are uniquely determined by the
normalized central potential $\tilde{\psi}_{0} 
\equiv \Psi(r=0)/\sigma_1^2 \equiv \Psi_0/\sigma_1^2$,
which is the parameter $W_0$ of King \shortcite{king66}. The distribution
function is given by
\beqa
\fking \left( \brr,\bv \right) {\rmnd}^3\brr {\rmnd}^3\bv & = &
{\rmnd}^3\brr {\rmnd}^3\bv \frac{\rho_1}{( 2\pi \sigma_1^2)^{3/2}}
\left( e^{(\vesc^2-v^2)/(2 \sigma_1^2)} - 1\right).
\label{kingpdf}
\eeqa
for stellar positions $\brr$, velocities $\bv$, and local
escape speed $\vesc(r)$. 
The dimensional King models can be specified 
by three independent quantities
such as $E$, $M$, and $\rt$.
Define $\nu(\psic)$
by $E=\nu(\psic) \left( GM^2/\rt \right)$. For $0.0<\tilde{\psi}_{0}<8.5$
, or $-0.60 > \nu > -2.13$, $\nu(\psic)$ is single-valued so
that the King model is known given $E$, $M$, and $\rt$.
A physical reason for excluding large $\tilde{\psi}_{0}$
is that simulations have shown  King models become
susceptible to gravothermal instability at $\tilde{\psi}_{0}\simgreat 7.40$
\cite{wiy85}. 
The clusters we study in this paper are not core collapsed.

        A further restriction on the clusters is that they be tidally 
limited by the galaxy. To include the time-dependent 
effect of tidal
stripping due to the galaxy would require detailed restricted three-body 
simulations for a range of both globular cluster orbits and orbits 
of stars in the clusters. In this paper we  consider a
first approximation in which the
the galactic tidal field
provides a relationship between $\rt$ and $M$, but does not contribute
a time-dependent perturbing force. For circular 
orbits, if we define $\rt$
as the distance from the cluster center to the Lagrange point of the 
cluster plus galaxy potential, we get 
\beqa
 \frac{M}{3 \rt^3}& \simeq &\frac{M_g}{R_g^3}
\label{hillseqn}
\eeqa
for a point mass galaxy with mass $M_g$ and galactocentric radius $R_g$,
and we adopt $M/\rt^3=$ constant for our clusters
as they evolve due to collisions with black holes.

A weakness of our paper is its dependence on the King model
sequence. As we shall see in $\S \ref{greensfunction}$, for 
some clusters black hole collisions may force $\psic \rightarrow 0$
after only modest energy input and mass loss, leading to very tight 
(but somewhat artificial) bounds on black hole properties. We shall
rectify this deficiency in a subsequent paper where cluster
structure is not restricted to the King sequence \cite{murali98}.

\setcounter{table}{0}
\begin{table*}
\begin{minipage}{5in}
\label{clustertable}
\caption{Parameters of the Globular Clusters}
\begin{tabular}{lrrrl} \hline \hline
Cluster Name & $M$ & $\rt$ & $\rcore$ & $R_g$ \\
  & $(M_{\odot})$ & (pc) & (pc) & (kpc)  \\ \hline
AM 4&           700&            11.7&   3.7&            30.0    \\
Arp 2&          18000&          75.0&   9.5&            20.4    \\
NGC 5053&       37700&          72.0&   10.9&           16.7    \\
NGC 7492&       10400&          45.4&   4.5&            18.7    \\
Pal 4&          24900&          92.3&   15.4&           96.0    \\
Pal 5&          13700&          76.4&   13.9&           16.5    \\
Pal 13&         3000&           30.0&   3.0&            25.7    \\
Pal 14&         10400&          117.5&  22.4&           69.9    \\
Pal 15&         15000&          41.9&   10.5&           30.0    \\
\hline
\end{tabular}
\end{minipage}
\end{table*}

We shall use two different sets of globular clusters to determine
a maximum allowed black hole mass, $\mbhcrit$. First we examine
the set of loosely bound 
globular clusters found in Moore \shortcite{moore93} and listed in Table 1,
but then we also investigate a larger cluster perhaps more
representative of the initial cluster population.

\section{The Black Hole Model} \label{bhmodel}

        Black holes of mass $\mbh$ are assumed to compose 
a spherical halo
with an isotropic velocity distribution. A fraction $\fhalo$ of the
total halo mass is presumed to be in the black holes. When no value
of $\fhalo$ is explicitly stated, $\fhalo=1$ is assumed. 
We model the black
hole population as a singular isothermal sphere with 
one-dimensional velocity
dispersion $\sigmabh$ and mass density
\begin{equation}
\rhobh(R_g) = \fhalo \frac{\sigmabh^2}{2\pi G R_g^2}
\label{rhoiso}
\end{equation}
(e.g. Binney \& Tremaine 1987, Ch.4) with
$\sigmabh=V_c/\sqrt{2}=220/\sqrt{2}\kms \simeq 156\kms$.
The number density of black holes at $R_g$ is
\beqa
\nbh(R_g) = \fhalo \frac{V_c^2}{4\pi G \mbh R_g^2}.
\eeqa

We ignore rotation of the black hole halo, so their velocity 
distribution is $f(\vbh)=
(\pi V_{c}^{2})^{-3/2}
\exp{ \left(- \vbh^{2}/V_{c}^{2} \right)} $.
For computing the impulsive mass loss and energy input to the cluster
due to the collision, we need the distribution of relative
speeds, $\bvrel$. Let $\bvcl$ be the cluster velocity, so
$\bvrel=\bvbh-\bvcl$. 
The distribution of relative speeds between the
halo of black holes and the cluster becomes
\begin{eqnarray}
f(\vrel)d\vrel & = & \frac{1}{\sqrt{\pi}} \frac {V_c}{\vcl}
\left\{\exp{\left(- \frac{\left(\vrel -\vcl\right)^2}{V^2_c} \right)} - 
\exp{\left(- \frac{\left(\vrel +\vcl\right)^2}{V^2_c} \right)}\right\}
\frac {\vrel d\vrel}{V_c^2}.
\label{fofvrel}
\end{eqnarray}
In this paper the clusters are on circular orbits with $\vcl=V_c$. 

	The mean relative speed given by this distribution is 
\begin{eqnarray}
\langle \vrel \rangle & = & V_c \left\{
\left( x + \frac{1}{2x}\right) \mbox{erf}(x) + 
\frac{1}{\sqrt{\pi}}\exp\left(-x^2\right)
\right\}
\label{vreltherm}
\end{eqnarray}
where $x\equiv \vcl/V_c$ and 
$\mbox{erf}(x)=
\left(2/{\sqrt \pi}\right)\int_0^x \exp(-t^2) \rmnd t $.
 As $x\rightarrow 0, 
\langle \vrel \rangle \rightarrow 2V_c/\sqrt{\pi}$ (the Gaussian result),
as $x \rightarrow \infty$,  $\langle \vrel \rangle \rightarrow \vcl$,
and for $\vcl=V_c$ we find $\langle \vrel(x=1) \rangle/V_c 
\equiv \xi_1 \simeq 1.47$. 
Note that the distribution in eq. ($\ref{fofvrel}$)
is different from that used by Moore \shortcite{moore93} 
and Klesson \& Burkert 
\shortcite{kb95}. In addition,
they chose to approximate the differential rate of collisions
$d\Gamma = \nbh 2\pi b db f(\vrel)\vrel d\vrel$ by
$\nbh 2\pi b db f(\vrel)V_c d\vrel$, which will lead to errors
in the number of collisions and the rate of energy input and mass
loss to the cluster.

	For a globular cluster at $R_g$, there will be a certain
value of $\mbh$ {\it below} which there will be more than one
black hole inside the cluster on average at any time (the `many-body' 
limit).
The number of black holes inside a cluster is
$N_{\rmn inside} =  \nbh 4\pi \rt^3/3  
= (V_c^2\rt^3)/(3G\mbh R_g^2)$.
Since $\rt \sim GM/\sigma_0^2$, eq. 
$\ref{hillseqn}$ and $\ref{rhoiso}$ give
$3(V_c^2/R_g^2)(\rt^2/\sigma_0^2) \sim 1$ and so
$N_{\rmn inside} \simeq M/(10\mbh)$.
This exceeds one for $\mbh<0.1M$ which we will see is close to
black hole mass limits for some clusters.
The mathematical description of the energy input is complicated in this
regime since the duration of a collision $\sim \rhm/\vrel$
is {\it longer} than the time between collisions
$ \sim (\nbh \pi \rhm^2 \vrel)^{-1}$.

\section{The Impulsive Energy Input for a Single Encounter}
\label{discrete}

In the impulse and straight line approximations, the velocity change
of a star due to the passage of a black hole is
\begin{eqnarray}
\btu \bv & = & - \frac{2G\mbh\bs}{\vrel s^2}
\label{delvimp}
\end{eqnarray}
where $\bs \perp \bvrel$ is the projected vector 
from the black hole to the star at
closest approach. Since the stellar velocity $v \ll \vcl$ and
$v \ll \vbh$, we may neglect $v$ in the relative velocity so that
$\bvrel \simeq \bvbh - \bvcl$. 
In this approximation, all stars recieve a
velocity kick in the same plane perpendicular to $\bvrel$. The
density of cluster stars can be projected onto this plane;
then for a star at projected position $\bR$
relative to the cluster center, $\bs=\bR-\bb$, where
$\bb$ is the impact parameter of the black hole relative to the
cluster center. 

The cluster is destroyed for impacts with $\btu v \simgreat 
\sigma_0$. An order of magnitude estimate of the ratio of these
two speeds is
\beqa
\frac{\btu v}{\sigma_0} & \sim &  \frac{\mbh}{M} \frac{\sigma_0}{\vrel}
\frac{\rt^4}{b^4}
\equiv \eta_c \frac{\rt^4}{b^4} 
\label{etac}
\eeqa
For a penetrating impact with
$b \simless \rt$, the cluster will be destroyed if $\eta_c \simgreat 1$. 
As the velocity kick, energy input, etc. in the impulsive limit must
scale as $\mbh/\vrel$, $\eta_c$ is a convenient dimensionless measure
of the destructiveness of the collision.

Next let star {\it i}
have a mass $m_i$ and a velocity $\bv_i$ with respect to the center
of mass before the collision. The energy of the cluster before the 
collision is then
\beqa
E_{\rmn before } & = & \sum_{i=1}^{N} m_i
\left( \frac{1}{2}v_i^2 + \frac{1}{2} \phi_i \right)
\eeqa
where $\phi_i=-GM/\rt-\psi_i$ is the gravitational potential, and
$\vesci=\sqrt{2\psi_i}$ is the escape speed for star $i$.
The velocity kick relative to the center of mass is
\beqa
\dbv_i & =&\btu \bv_i - \ol{\btu \bv} \equiv  \btu \bv_i - \frac{1}{N}
\sum_{i=1}^{N} \btu \bv_i 
\eeqa 
where $\ol{\btu \bv}$ is computed for a continuous cluster in Appendix A.
Star {\it i} is ejected if  
$ \left( \bv_i + \dbv_i \right)^2  \geq \vesci^2 $.
The total amount of mass ejected from the cluster can then be formally 
written down as
\beqa
\btu M & = & -\sum_{i=1}^{N} m_i \theta_i \ej
\eeqa
where 
$\theta_i \ej \equiv 
\theta \left[ \left( \bv_i + \dbv_i \right)^2 - 
\vesci^2 \right]$ 
is a useful shorthand notation; similarly \newline
$\theta_i \bound \equiv 
\theta \left[ \vesci^2 - 
\left( \bv_i + \dbv_i \right)^2 \right]$ 
for the bound stars. We neglect the possibility that stars unbound by 
our criterion remain near the cluster for a long time, possibly
to become bound once again in a subsequent encounter with a black hole
[Spitzer \shortcite{spit87} discusses orbits of this type].

To find the change in cluster energy we need the kinetic and potential 
energy changes, $\btu T$ and $\btu V$, respectively. 
The change in kinetic energy arises both from the kicks to the bound 
stars and the loss of the energy of the ejected stars:
\beqa
\btu T & = & - \frac{1}{2}  \sum_{i=1}^{N} m_i v_i^2 \theta_i\ej +
\sum_{i=1}^{N} m_i \left( \bv_i \cdot \dbv_i +
\frac{1}{2} \delta v_i^2 \right) \theta_i\bound.
\label{deltatdiscrete}
\eeqa
The change in the potential energy is $\btu V=V_{\rmn after}-V_{\rmn before}$,
where the potential energy before the collision is
\beqa
V_{\rmn before} & = & 
- \frac{G}{2} \sum_{i=1}^N \sum_{j \neq i= 1}^N
\frac{m_im_j}{r_{ij}} 
\left[ \theta_i\bound \theta_j\bound + 2  \theta_i\ej \theta_j\bound
+ \theta_i\ej \theta_j\ej  \right].
\eeqa
and potential energy of the bound stars remaining afterwards is
\beqa
V_{\rmn after} & =& - \frac{G}{2} \sum_{i=1}^N \sum_{j \neq i = 1}^N 
\theta_i\bound \theta_j\bound \frac{m_im_j}{r_{ij}} 
\eeqa
for separation vectors $\brr_{ij}$. Consequently,
\beqa
\btu V & = & 
  \sum_{i=1}^N \sum_{j \neq i = 1}^N \theta_i\ej 
\frac{Gm_im_j}{r_{ij}}
- \frac{1}{2} \sum_{i=1}^N \sum_{j \neq i = 1}^N
\theta_i\ej \theta_j\ej \frac{Gm_im_j}{r_{ij}} \\
& = & - \sum_{i=1}^N \theta_i\ej m_i \phi_i
- \frac{1}{2} \sum_{i=1}^N \sum_{j \neq i = 1}^N
\theta_i\ej \theta_j\ej \frac{Gm_im_j}{r_{ij}}
\label{deltavdiscrete}
\eeqa
where $\phi_i$ still denotes the potential of star {\it i} from 
{\it before} the collision.
The first term in eq. ($\ref{deltavdiscrete}$) is just the pre-collision
potential energy of the ejected stars, and is $\odm$, the fractional
mass loss. The second term is $\odmtwo$, and will be smaller than the 
first as long as $\btu M/M$ is small. In our calculations we neglect the 
smaller (and difficult to compute) second term, and evaluate the first
using the King model potential at the positions of the ejected stars.
To protect against gross inaccuracies, we terminate our simulations
if $|\btu M/M|> 0.2$ in a single collision.

	With these approximations, the energy change is
\beqa
\btu E & = & -  \sum_{i=1}^{N} m_i
\left( \frac{1}{2} v_i^2 + \phi_i \right)\theta_i\ej +
\sum_{i=1}^{N} m_i \left( \bv_i \cdot \dbv_i +
\frac{1}{2} \delta v_i^2 \right)\theta_i\bound.
\eeqa
It should be noted that the energy change due to
ejected stars is always positive, and hence mass loss always heats
the cluster. 
In the sum over bound stars, the $\bv_i \cdot \dbv_i$
term can have either sign and hence can heat or cool. The mean of this
term is nonzero but small since it depends on the mass loss to 
make the final distribution slightly anisotropic; it also contributes
substantial variance.  
The $\frac{1}{2} \delta v_i^2$ term gives rise to the familiar mean
heating (King 1966; Binney \& Tremaine 1987).

\subsection{ The Model for Energy Input and Mass Loss \label{newcks} }

	For the analytic estimates and Fokker-Planck calculations
of cluster survival, we shall need formulae for the mean energy and 
mass changes, $\ol{\btu E}$ and $\ol{\btu M}$, their variances, 
and the cross-term $\ol{\btu E \btu M}$. To get them we use a
model for energy input and mass loss which
is due to Chernoff et al. \shortcite{cher86} 
(hereafter called C.K.S.). In this model, 
the portion of phase space from which mass escapes is identified.
The energy 
and mass of the remaining cluster are then found as integrals over the 
distribution function of the remaining stars.  
This model relies only on the information given by the distribution
function from {\it before} the collision; no detailed solution of the
Vlasov equation is attempted. The higher order effects which arise from 
the detailed alteration of the distribution function by the perturbing
black holes are estimated and define the error in our method.

We shall also extend the the C.K.S. model
slightly by considering the fluctuations in the number of stars lost
in any collision resulting from a finite number $N$ of stars.
Before the collision, the phase space density
of the cluster is given by the King
model in eq. ($\ref{kingpdf}$).
After the collision, we do not have a full expression for the phase
space density
including the effect of the change in velocity to all the stars,
but we do know that all parts of phase space for which
$\theta\ej=1$ will have their mass ejected.
Define $f_{<}=\fking \theta\bound $ and $f_{>}=\fking \theta\ej$
which are nonzero only for bound and unbound stars, respectively.
The average of a quantity $x$ over the
bound, or ejected, stars will be written
\beq
\langle x \rangle_{<} = \frac{ \int {\rmnd}^6 \Gamma f_{<} x }
{\int {\rmnd}^6 \Gamma f_{<}} ,\ \ \
 \langle x \rangle_{>} = \frac{ \int {\rmnd}^6 \Gamma f_{>} x }
{\int {\rmnd}^6 \Gamma f_{>}}
\eeq
respectively. If a symbol $<$ or $>$ is not specified, it means that the
quantity is averaged over the entire phase space.

	The probability that a star is in the portion of phase
space from which mass is ejected is $\pej=\langle \theta\ej \rangle$,
which is an integral over the cluster model
and is independent of $N$. 
We then suppose that the distribution for 
losing $\Nej$ stars out of the total $N$ is given by
\beqa
P(\Nej) & = & \frac{ (\pej N)^{\Nej} }{\Nej!} \exp( - \pej N ).
\eeqa
In averaging the expressions of $\S \ref{discrete}$, we will only
need the moments $ \langle \Nej \rangle = N \pej$ and 
$ \langle \Nej^2 \rangle = (N \pej)^2 + N \pej$.

\subsection{ Moments of $\btu E$ and $\btu M$ for Individual 
Collisions \label{einputformulas} }

In this section,we integrate the first and second moments of $\btu E$
and $\btu M$ from $\S \ref{discrete}$ over the King cluster model
and the distribution for the number of ejected stars $P(\Nej)$. These
averages will be denoted by an overbar to distinguish them from the
averages over bound and ejected stars defined in the previous section.

	The mean mass loss and energy input are
\beqa
&& \frac{ \ol{ \btu M(b,\eta_c,\tilde{\psi}_0)}}{M} = 
- \langle \theta\ej \rangle
\label{dmofb}
\eeqa
and
\beqa
&& \frac{ \ol{ \btu E(b,\eta_c,\tilde{\psi}_0)}}{|E|} =
\frac{M}{|E|} 
\left\{
- \pej
\bl \frac{1}{2} v^2 + \phi \br_{>}
+ (1- \pej)
\bl \bv \cdot \dbv +
\frac{1}{2} \delta v^2 \br_{<}
\right \}
\nonumber \\ && =
\frac{M}{|E|} \left\{ - \pej \bl \frac{1}{2} v^2 + \phi \br_{>}
+ \bl \bv \cdot \dbv + 
\frac{1}{2} \delta v^2 \br_{<}
\right\} + {\cal O}(\eta_c^4).
\label{deofb} 
\eeqa
Consistent with our neglect of terms $\propto (\btu M/M)^2$ (and 
smaller), we only retain contributions at ${\cal O}(\eta_c^2)$
and ${\cal O}(\eta_c^3)$; note that $\pej \propto \eta_c^2$, 
$|\delta v| \propto \eta_c$, and consequently $\langle
 \bv \cdot \dbv \rangle_{<}$
-- which would vanish in absence of mass loss -- is 
$\propto \eta_c^3$. 
To the same accuracy, the variance of $\btu M$ is
\beqa
\frac{ \sigma^2_{\btu M}(b,\eta_c,\tilde{\psi}_0,N) } { M^2} =
\frac{\pej}{N}
\label{varmofb} 
\eeqa
and the variance of $\btu E$ is
\beqa
&& \frac{ \sigma^2_{\btu E}(b,\eta_c,\tilde{\psi}_0,N) } { E^2} =
\frac{M^2}{N E^2}
\left[
(1- \pej) \bl \left( \bv \cdot \dbv +
\frac{1}{2} \delta v^2 \right)^2 \br_{<}
+ \pej \bl \left( \frac{1}{2} v^2 + \phi \right)^2 \br_{>}
\nonumber  \right.\\ & &  \left. -
\left(-\pej \bl \frac{1}{2} v^2 + \phi \br_{>}
+  (1- \pej) \bl \bv \cdot \dbv
+ \frac{1}{2} \delta v^2 \br_{<}  \right)^2
\right]
\nonumber \\ && \simeq
\frac{M^2}{N E^2}
\left[
\bl \left( \bv \cdot \dbv 
\right)^2 \br_{<} 
+ \pej \bl \left( \frac{1}{2} v^2 + \phi \right)^2 \br_{>} 
\right]. 
\label{vareofb} 
\eeqa
Finally, the mixed moment is
\beqa
&& \frac{ \sigma^2_{\btu E \btu M}(b,\eta_c,\tilde{\psi}_0,N)}
 {M |E|} \equiv
\frac{ \ol{ (\btu M \btu E)(\eta_c,\tilde{\psi}_0,N)}}{M|E|}
- \frac{ \ol{ \btu M (\eta_c,\tilde{\psi}_0)}}{M}
\frac{ \ol{ \btu E(\eta_c,\tilde{\psi}_0)}}{|E|} 
\nonumber \\ && =
\frac{M}{N|E|}
\left[
+ \pej \bl \frac{1}{2} v^2 + \phi \br_{>}
\right],
\label{varemofb} 
\eeqa
keeping only terms ${\cal O}(\eta_c^3)$ and larger.

\setcounter{figure}{1}
\begin{figure}
 
\begin{picture}(468,468)(0,0)
\put(0,234){\epsfxsize=3.25in
\epsffile{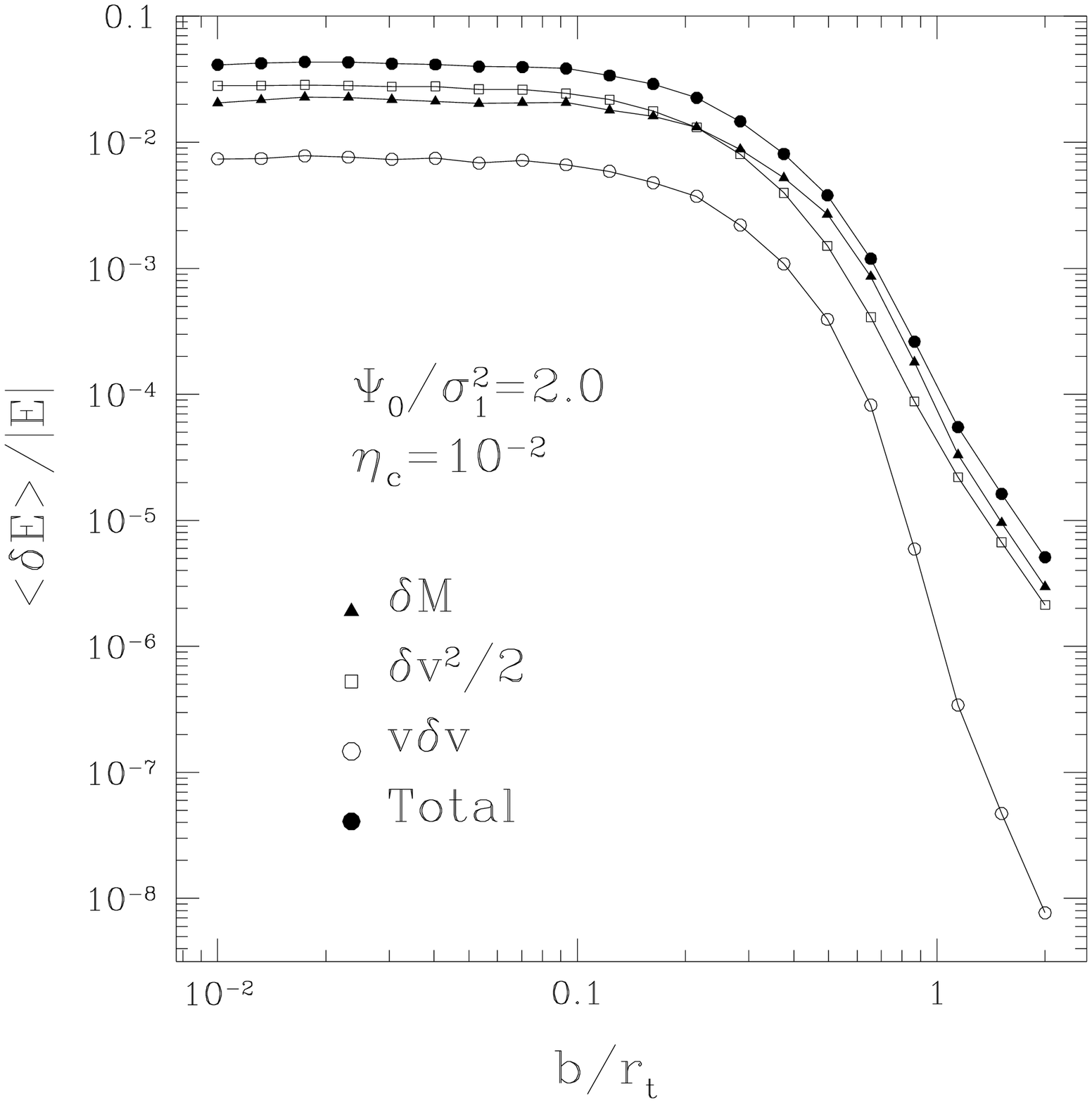}}
\put(234,234){\epsfxsize=3.25in
\epsffile{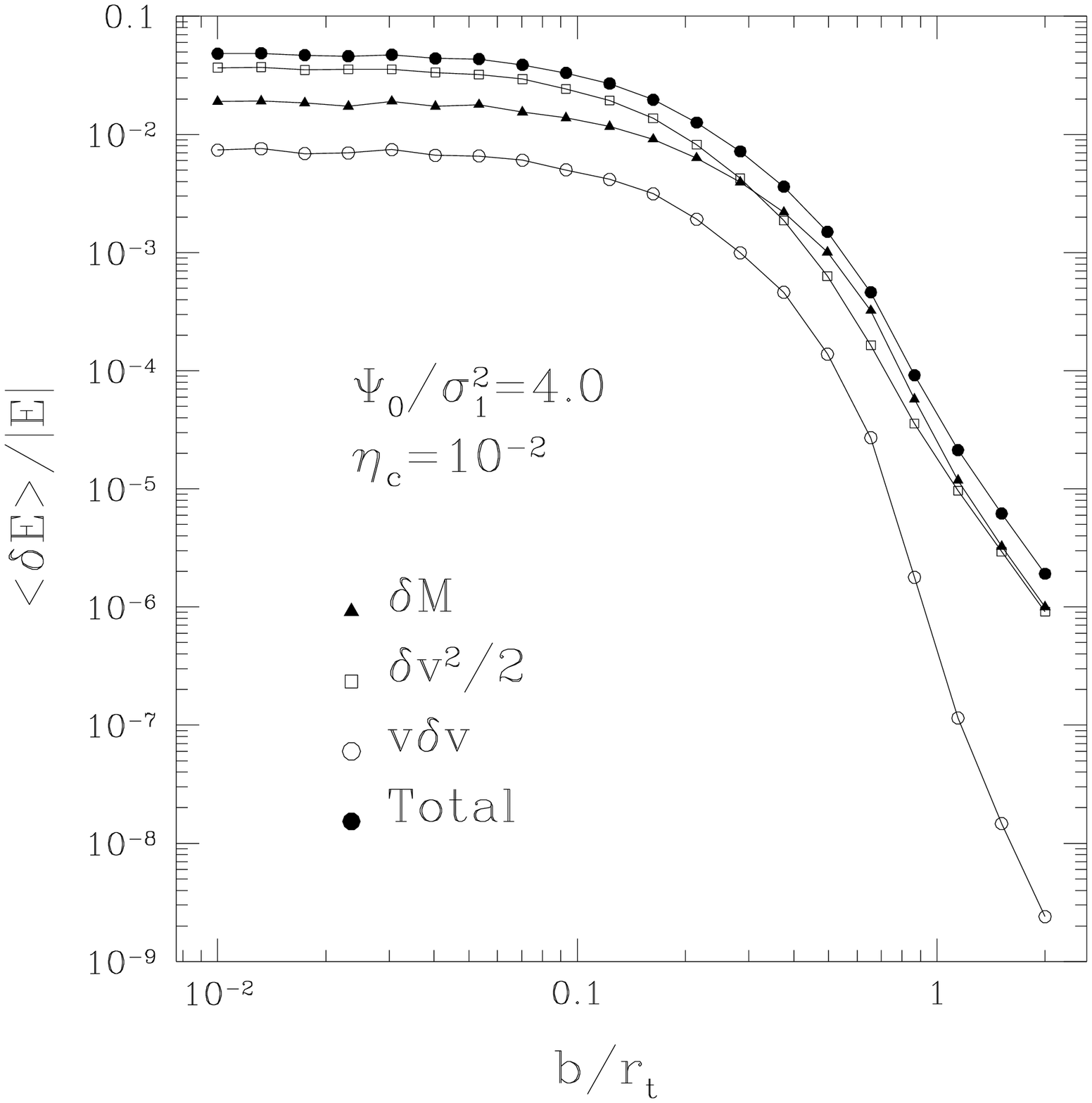}}
\put(117,0){\epsfxsize=3.25in
\epsffile{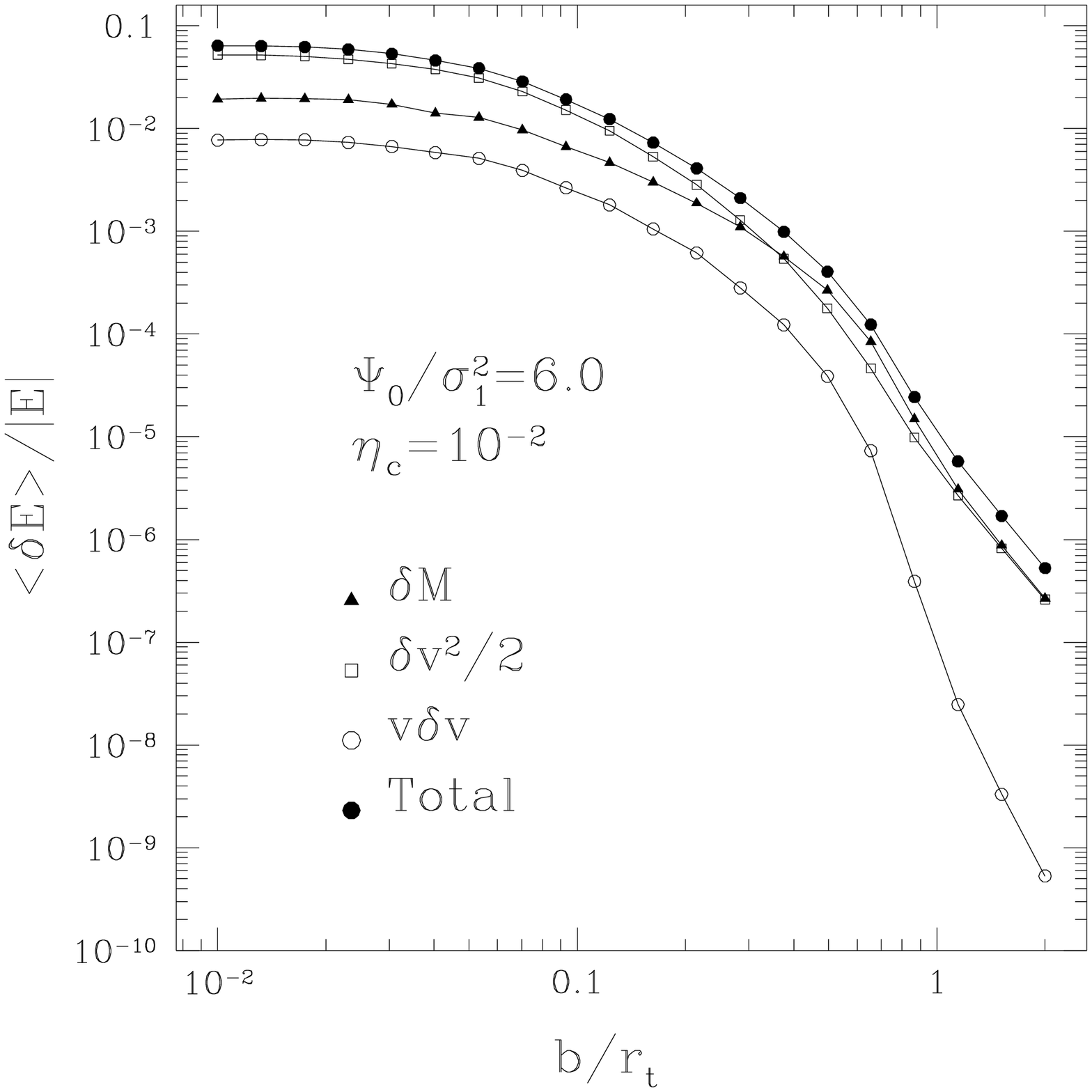}}
\end{picture}
 
\caption{ $ \langle \btu E(b,\eta_c) \rangle/|E|$. }
 
\label{deofbplot}
\end{figure}

We have calculated the integrals needed to evaluate these moments 
using Monte Carlo methods. The velocity kick is evaluated using eq. 
($\ref{delvimp}$) and the center of mass velocity kick is given in
eq. ($\ref{cmkick}$). A sample of the results
for $\psic=2.0,4.0,6.0$ and $\eta_c=10^{-2}$ are plotted in
Figs. $\ref{deofbplot}$, $\ref{dmofbplot}$, and $\ref{sigdeofbplot}$. 
For $\psic=2.0$, $4.0$, and $6.0$, the core radii are at 
$\rcore/\rt=0.2$, $0.1$, and $0.05$, respectively, while the half
mass radii are at $\rhm/\rt=0.3$, $0.2$, and $0.15$, respectively. 
The Monte-Carlo integrals were computed to an estimated fractional
error of $5$ per cent. 

\setcounter{figure}{2}
\begin{figure}

\begin{picture}(468,468)(0,0)
\put(0,234){\epsfxsize=3.25in
\epsffile{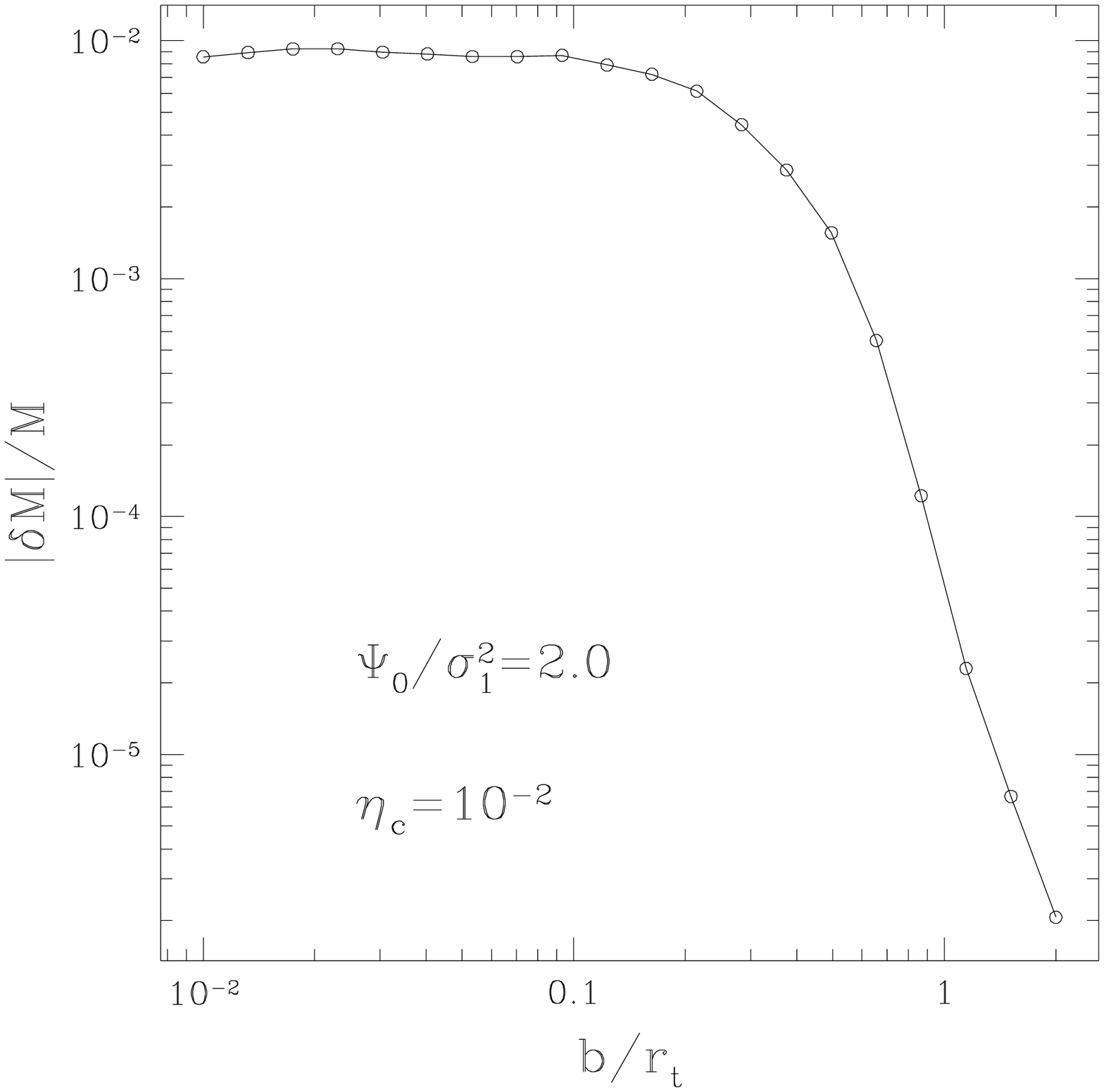}}
\put(234,234){\epsfxsize=3.25in
\epsffile{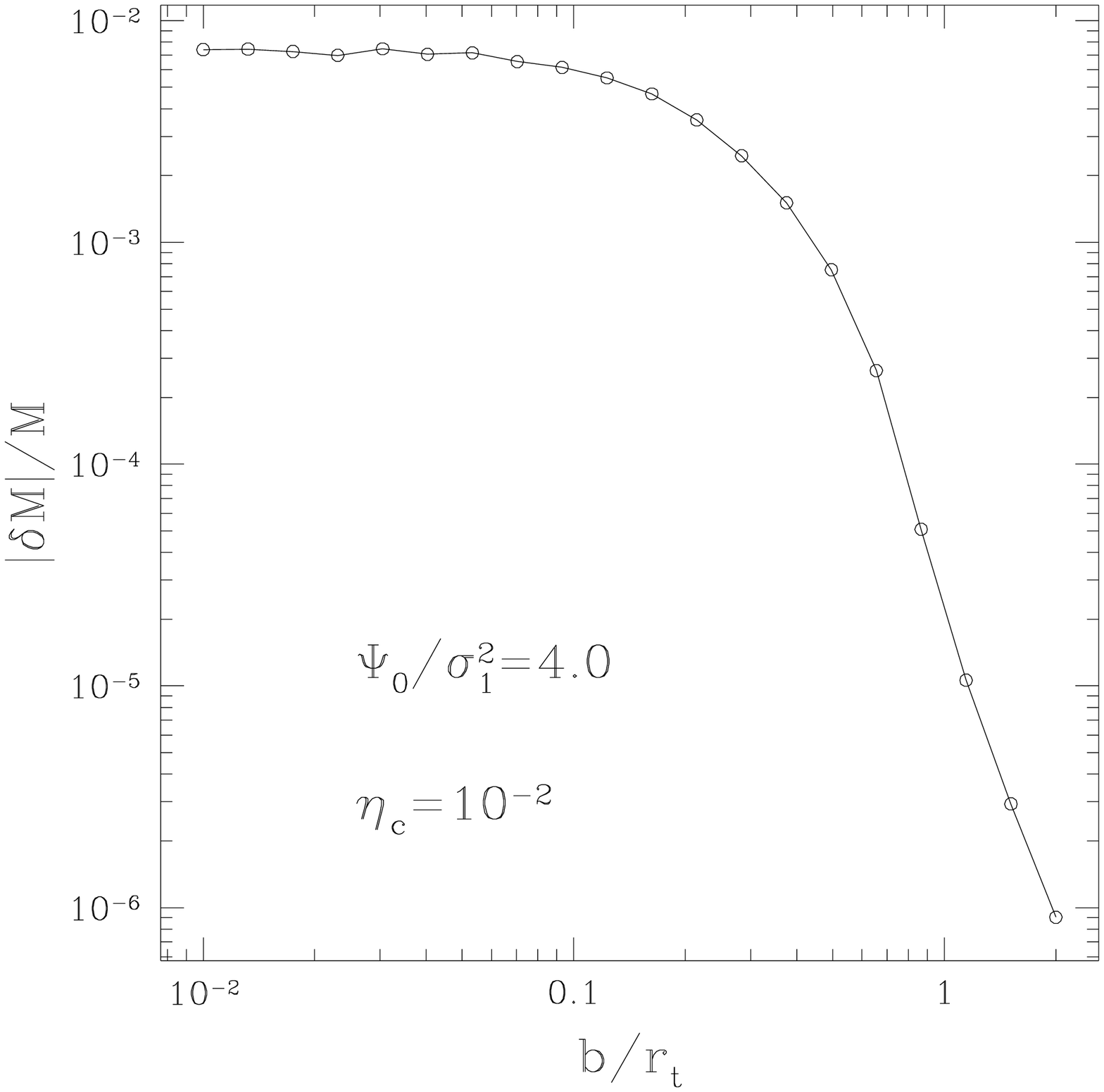}}
\put(117,0){\epsfxsize=3.25in
\epsffile{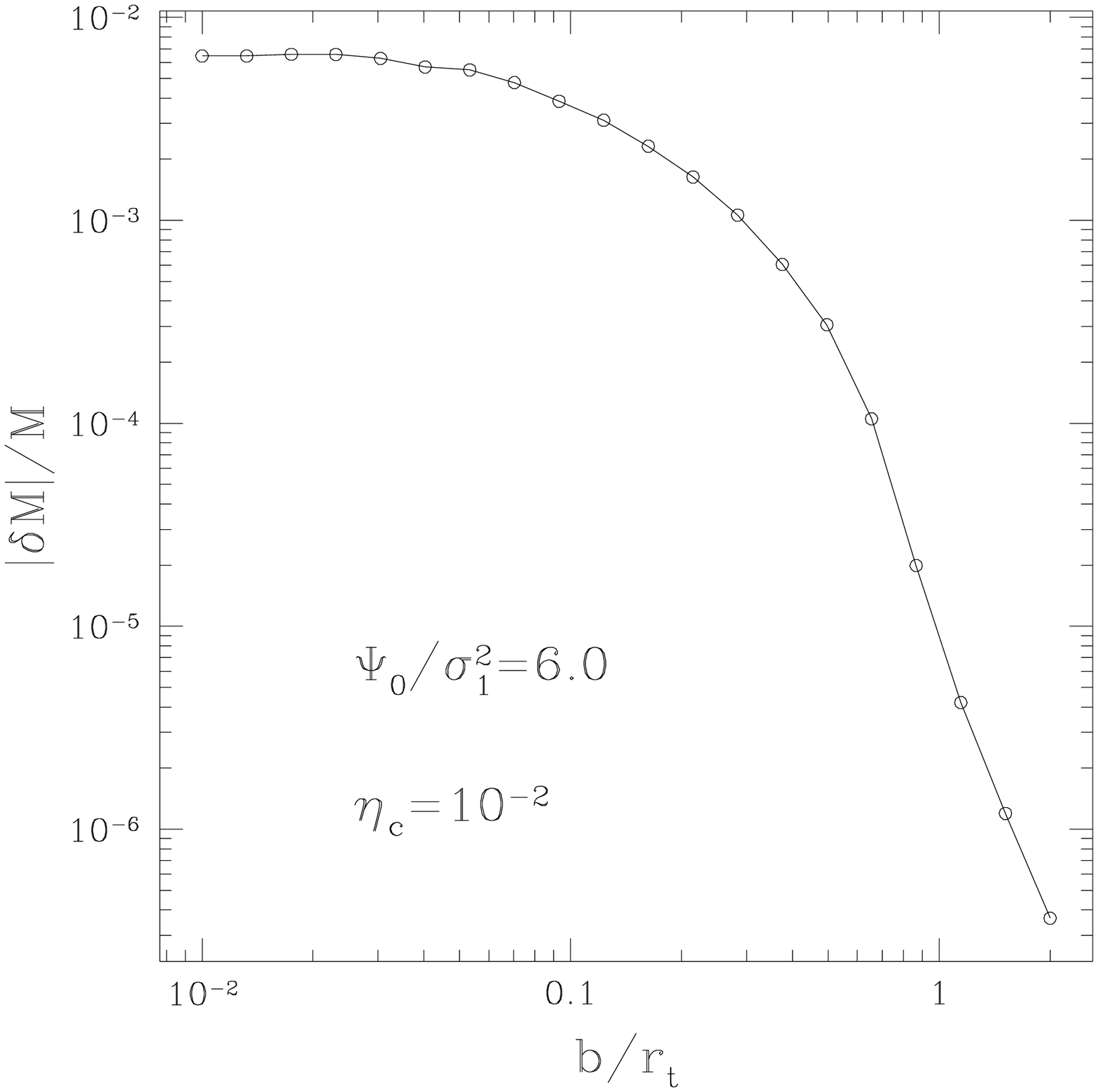}}
\end{picture}
 
\caption{ $| \langle \btu M(b,\eta_c) \rangle|/M$.
}
 
\label{dmofbplot}
\end{figure}
 
\setcounter{figure}{3}
\begin{figure}
 
\begin{picture}(468,468)(0,0)
\put(0,234){\epsfxsize=3.25in
\epsffile{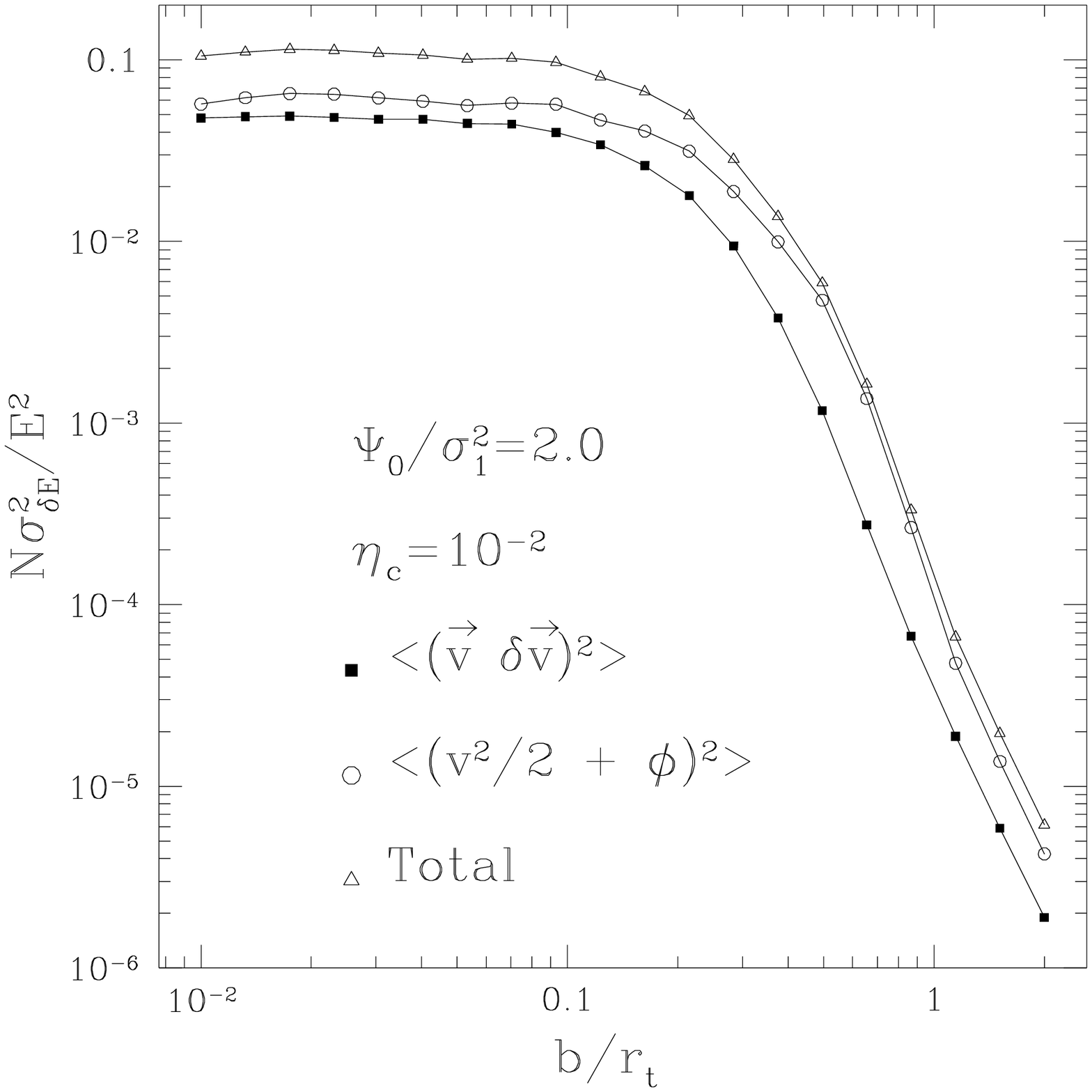}}
\put(234,234){\epsfxsize=3.25in
\epsffile{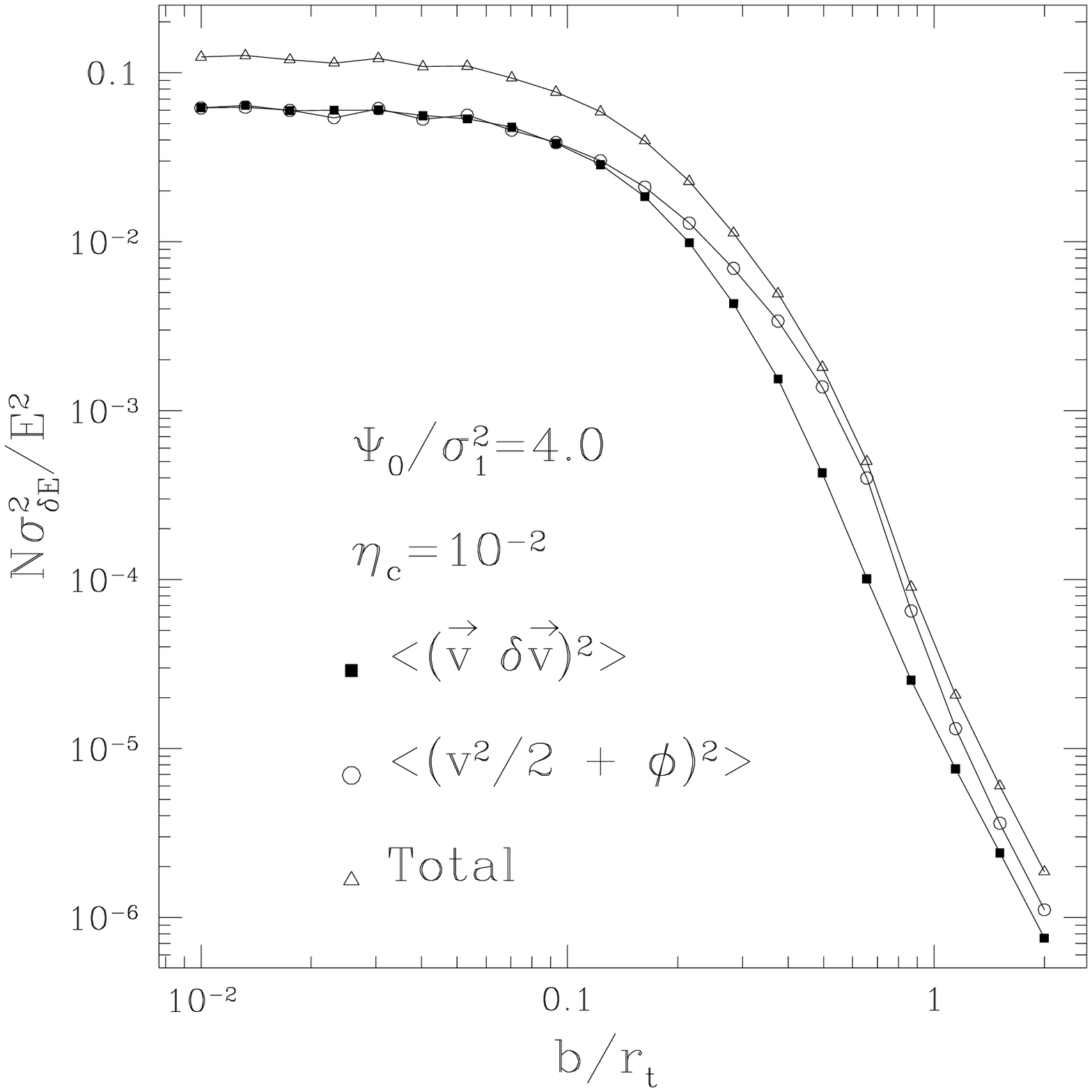}}
\put(117,0){\epsfxsize=3.25in
\epsffile{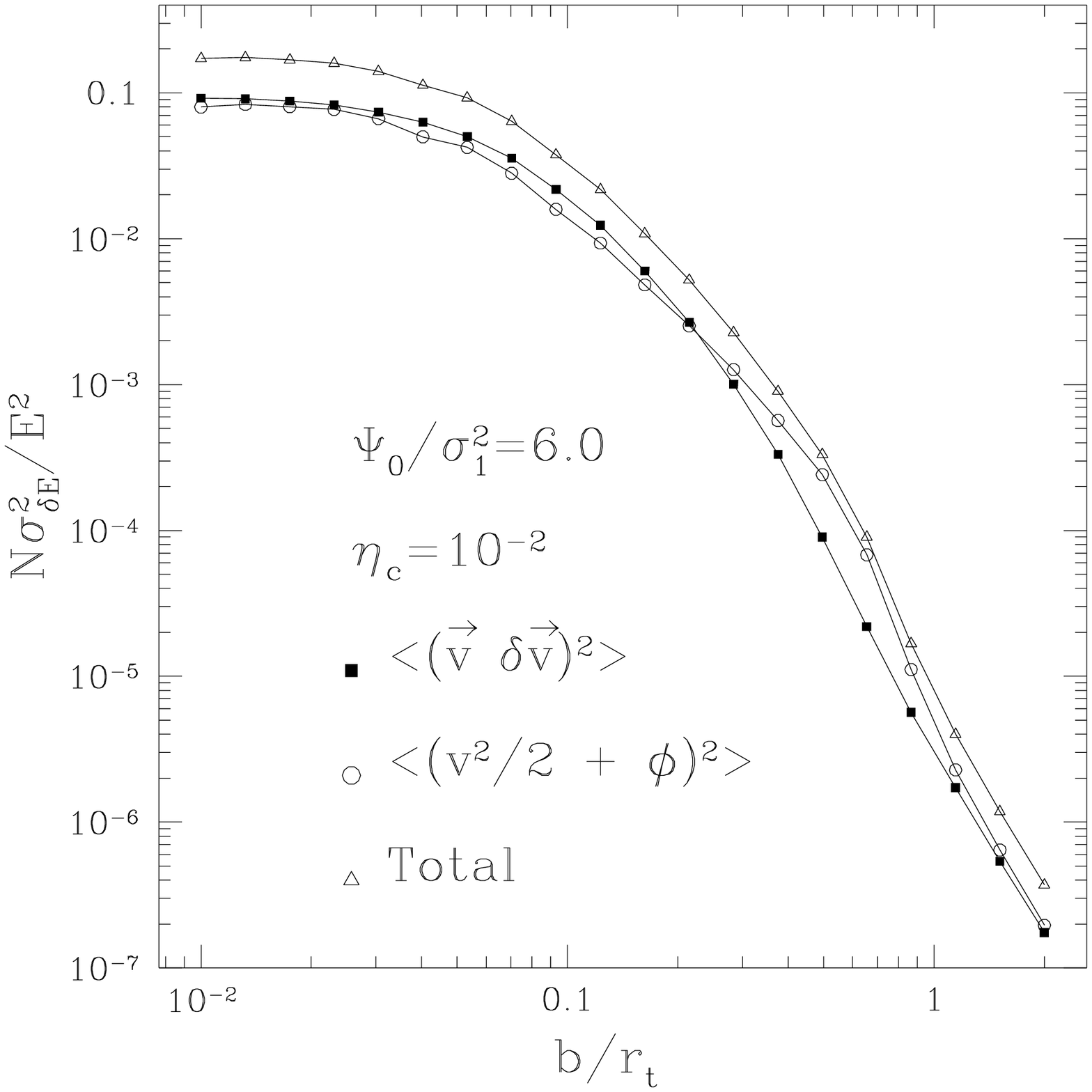}}
\end{picture}
 
\caption{ $N\sigma^2_{\btu E}(b,\eta_c)/E^2$.
}
 
\label{sigdeofbplot}
 
\end{figure}

The separate contributions to the energy input in eq. ($\ref{deofb}$)
have been plotted in Fig. $\ref{deofbplot}$. The contribution of 
mass loss is comparable to that due to the $\delta v^2/2$ heating
usually considered. Note too that the
$\langle \bv \cdot \dbv \rangle$ term is surprisingly
large for small $b$ encounters, contributing $\sim 5$ per cent to the 
energy input for impacts near the cluster center.
The variance in eq. ($\ref{vareofb}$) is also broken up into the 
heating part, $\bl \left( \bv \cdot \dbv 
\right)^2 \br_{<}$ and the mass loss part,
$\pej \bl \left( \frac{1}{2} v^2 + \phi \right)^2 \br_{>}$,
which are comparable in size.

\subsection{ The Tidal Limit \label{tidallimit} } 

The calculation of $\ol{\btu E}$, $\ol{\btu M}=\sqrt{N} 
\sigma_{\btu M}$, $\sigma^2_{\btu E}$, and $\sigma^2_{\btu E \btu M}$ 
is simplified considerably in the tidal limit, partly because mass
loss is restricted to particles with speeds very close to the
escape speed. Spitzer \shortcite{spit58} computed
$\ol{\btu E}$ without mass loss, however, mass loss contributes
significatly to $\ol{\btu E}$, sometimes exceeding the Spitzer term.
The expressions found below illustrate the
importance of mass loss explicitly and are also useful for understanding
disruption of clusters by high mass black holes.

Formally, we should expand in the two parameters $\eta_c$ and 
$\rt/b \ll 1$.
Here, we keep only the lowest powers of $\eta_c$ and $\beta=\rt/b$.
In this approximation, we find
\beqa
\frac{ \ol{ |\btu M| }(b,\eta_c,\tilde{\psi}_0) } { M }  & = & 
- C_M \frac{\eta_c^2}{\beta^4}
\eeqa
where
\beqa
C_M & =& \frac{8}{9 \sqrt{\pi} } \frac{ \mu_t } { \xi_t^4}
\frac{\tilde{\psi}_0^{7/2}}{f_cg_c}
\int_0^{\xi_t} \rmnd \xi \xi^4 \Theta(\xi)^{3/2},
\label{dmtidal}
\eeqa
and $\xi=r/a$, $\xi_t=\rt/a$, $a=(\Psi_0/4\pi G \rho_0)^{1/2}$, 
$\mu_t=M/M_0$, $M_0=4\pi \rho_0 a^3$, $f_c=\rho_0/\rho_1$,
$g_c=\sigma_0^2/\sigma_1^2$, and $\Theta(\xi)=\Psi(\xi)/\Psi_0$
[$\rho_1$ and $\sigma_1$ are defined in eq. ($\ref{kingpdf}$)]. 
Hence $\sigma^2_{\btu M}/M^2 = \pej/N = 
(C_{MM}/N)(\eta_c^2/\beta^4)$ where $C_{MM}=C_M$.
To ${\cal O}(\eta_c^2)$, 
\beqa
&& \frac{ \ol{ \btu E(b,\eta_c,\tilde{\psi}_0)}}{|E|}
 \simeq  \frac{M}{|E|} \left\{ \pej \frac{GM}{\rt}
+ \bl \frac{1}{2} \delta v^2 \br \right\} 
 \simeq  \pej \frac{GM^2}{\rt|E|} + 
\frac{4}{3} \frac{G^2\mbh^2M \bl r^2 \br}{|E|\vrel^2b^4} 
\equiv  C_E \frac{\eta_c^2}{\beta^4}
\label{detidal}
\eeqa
with
\beqa
C_E & =& C_M \frac{GM^2}{\rt|E|} + 
\frac{4}{3} \frac{G^2M^3 \bl r^2 \br}{|E|\sigma_0^2\rt^4}. 
\eeqa
In addition to the Spitzer energy loss $\langle \delta v^2/2 \rangle$, 
$C_E$ includes the energy carried away by the massloss,
$\btu M = - \pej M$,  from near the escape surface, where the
energy per mass is $-GM/\rt$. 
Note that the ejected mass can come from any spatial position
in the cluster as long as the velocity is close enough to the escape
velocity. Indeed, the mean radius from which mass is lost depends
weakly on cluster concentration, and is roughly
$2\rt/3$ for $0 \leq \psic \leq 8.5$.
The variance to the energy input is
\beqa
&& \frac{ \sigma^2_{\btu E}(b,\eta_c,\tilde{\psi}_0,N) } { E^2} =
\frac{1}{N} \pej \left( \frac{GM^2}{|E|\rt} \right)^2
+ \frac{8}{9}\frac{1}{N} \frac{G^2\mbh^2M^2}{E^2\vrel^2b^4}
\langle r^2v^2 \rangle
\equiv \frac{C_{EE}}{N}\frac{\eta_c^2}{\beta^4}
\eeqa
with
\beqa
C_{EE} & = & C_M \left( \frac{GM^2}{|E|\rt} \right)^2 + 
\frac{8}{9} \frac{G^2M^4}{E^2\sigma_0^2 \rt^4} 
\bl r^2v^2 \br,
\eeqa
and the mixed moment is
\beqa
&& \frac{ \sigma^2_{\btu E \btu M}(b,\eta_c,\tilde{\psi}_0,N) } {M |E|} =
\frac{M}{N|E|}
\left[
+ \pej \bl \frac{1}{2} v^2 + \phi \br_{>}
\right] 
\simeq  -\frac{M}{N|E|} \pej \frac{GM}{\rt} 
\equiv
- \frac{C_{EM}}{N}\frac{\eta_c^2}{\beta^4}
\eeqa
with $C_{EM}=C_M(GM^2)/(|E|\rt)$.

        Another extremely important quantity will be the change in
$\nu \equiv (E\rt)/(GM^2)$ in the tidal limit. Using the fact that
the cluster is tidally limited, $\nu$ can be expressed as
$\nu = (r_{t,0}/(GM_0^{1/3}))\times (E/M^{5/3})$ where the subscript
refers to some reference value. The change in
$\nu$ is then
\beqa
\btu \nu & =& \nu \left( \frac{\btu E}{E} - \frac{5}{3} \frac{\btu{M}}{M}
\right) = |\nu| 
\left( C_E - \frac{5}{3} C_M \right) \eta_c^2 \frac{\rt^4}{b^4}
\eeqa
which can be written in the usual form
\beqa
\ol{ \frac{\btu \nu (b,\eta_c,\tilde{\psi}_0)}{|\nu|} } & =& 
C_{\nu}\frac{\eta_c^2}{\beta^4}
\label{dnutidal}
\eeqa
with $C_{\nu} = C_E - 5C_M/3$.
The variance of the mean change in $\nu$ is given by
\beqa
\frac{ \sigma^2_{\btu \nu}(b,\eta_c,\tilde{\psi}_0,N) }{ \nu^2 }  & =& 
\frac{ \sigma^2_{\btu E}} { E^2}
+ \frac{5}{3} \frac{ \sigma^2_{\btu E \btu M} } {M |E|}
+\frac{25}{9} \frac{ \sigma^2_{\btu M} } { M^2} = 
\frac{C_{\nu \nu} }{N} \frac{\eta_c^2}{\beta^4} 
\eeqa
where $C_{\nu \nu} \equiv C_{EE} - 5C_{EM}/3 + 25C_{MM}/9 $. 

\setcounter{figure}{4}
\begin{figure}

\begin{picture}(324,324)(0,0)
\put(81,0){\epsfxsize=4.5in
\epsffile{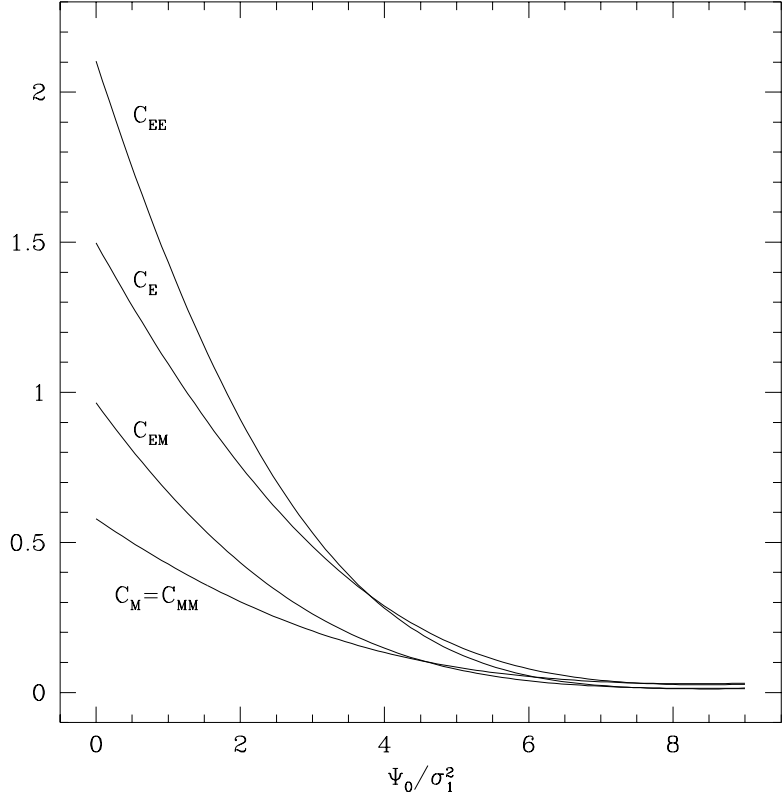}}
\end{picture}

\caption{ Dimensionless coefficients for fractional
energy input and mass loss in the tidal limit.}

\label{tlcoeff}
\end{figure}

\setcounter{figure}{5}
\begin{figure}
 
\begin{picture}(324,324)(0,0)
\put(81,0){\epsfxsize=4.5in
\epsffile{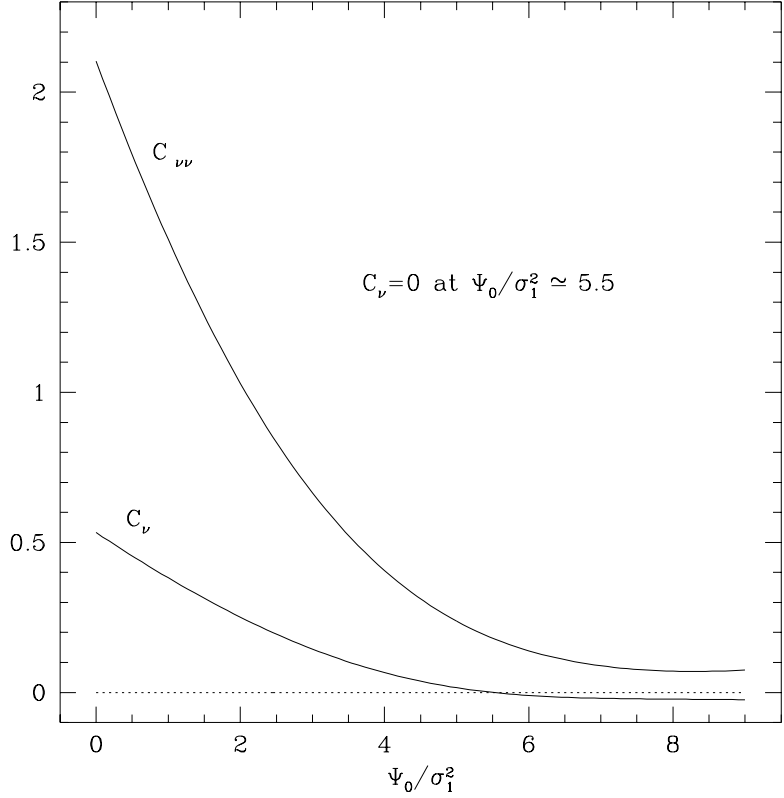}}
\end{picture}
 
\caption{ Dimensionless coefficients for fractional
changes in $\nu$ in the tidal limit.}
 
\label{nucoeff}
\end{figure}

	The seven coefficients for the tidal limit are plotted in 
Figs. $\ref{tlcoeff}$ and $\ref{nucoeff}$ as a 
function of $\tilde{\psi_0}$. All the curves show a
monotonic decrease as $\psi_0$ increases over the range $\tilde{\psi}_0
\in (0.0,8.5)$. Most noticeable is that {\it
$C_{\nu}$ becomes less than zero at roughly $\tilde{\psi}_0 \simeq 5.5$}. 
Since $\nu=\nu(\psic)$, whether 
an impact 
will make the cluster more or less concentrated is determined by
the value of $\tilde{\psi}_0$ for that cluster.
Clusters with 
$\tilde{\psi}_0 < 5.5$ are 
driven toward dissolution and clusters with $\tilde{\psi}_0 >
5.5$ are driven toward core collapse. This agrees well with
CKS who derived the same result in the context of perturbations
by giant molecular clouds.

For penetrating encounters, the situation is quite different. Over 
almost the entire range of $\tilde{\psi}_0$ and $\eta_{c0}$, the
diffusion coefficients yield positive changes
in $\nu$, implying dissolution.

\subsection{ Evaluation of the Diffusion Coefficients \label{diffcoeff} }

In $\S \ref{einputformulas}$, we found $\ol{\btu E}$, $\ol{\btu M}$, 
$\sigma^2_{\btu E}$, $\sigma^2_{\btu M}$, and 
$\sigma^2_{\btu E \btu M}$ for individual collisions as a function of
impact parameter $b$ and relative velocity $\vrel$. When $\mbh$ is
small enough that the changes in the cluster properties are always
small for single impacts,
then we may average over many encounters to find the changes in $E$ and
$M$ over a time period which includes many collisions, but for which 
the changes in cluster properties are still small. 

We weight the formulae for the various moments by the differential 
rate of collisions
\beqa
\rmnd \Gamma & = & \nbh 2\pi b \rmnd b f(\vrel) \vrel \rmnd \vrel
\eeqa
and integrate over $b$ and $\vrel$ 
where $f(\vrel)$ is given in eq. ($\ref{fofvrel}$).
The results may be written
\beqa
&& D_E  =  \int \rmnd \Gamma \ol{\btu E} = \Gamma_0 |E| d_E 
\nonumber \label{de} \\
&& D_M = \int \rmnd \Gamma \ol{\btu M }= - \Gamma_0 M d_M 
\nonumber  \label{dm} \\
&& D_{EE} = \int \rmnd \Gamma \sigma^2_{\btu E} = 
\frac{1}{N} \Gamma_0 E^2 d_{EE} 
\label{dee} \nonumber \\
&& D_{MM} = \int \rmnd \Gamma \sigma^2_{\btu M} = 
\frac{1}{N} \Gamma_0 M^2 d_M
\nonumber \label{dmm} \\
&& D_{EM} = \int \rmnd \Gamma \sigma^2_{\btu E \btu M } = 
- \frac{1}{N} \Gamma_0 |E|M d_{EM}
\label{dem}
\eeqa
where $\Gamma_0=\nbh\pi \rt^2 V_c$, and $d_E$, $d_M$, $d_{EE}$,
and $d_{EM}$ are dimensionless functions of $\psic$ and $\eta_{c0}=
(\mbh/M)(\sigma_0/V_c)$. 
Note that minus signs have been inserted in $D_M$ and $D_{EM}$ to 
make $d_M$ and $d_{EM}$ positive.

The diffusion coefficients $d_E$, $d_M$, $d_{EE}$, and $d_{EM}$
involve eight-dimensional integrals over phase space,
$b$, and $\vrel$. Integration over two velocity angles is elementary,
and the integral over velocity magnitude was done using Runge-Kutta. The
remaining five integrals (three spatial coordinates, $b$, and $\vrel$)
were done via Monte Carlo (to an accuracy of $5$ per cent). The value of 
$b_{\rmn max}$ was taken to be $10\rt$. 

\setcounter{table}{1}
\begin{table*}
\begin{minipage}{5in}
\label{kappacoeff} 
\caption{Fitting Formula Coefficients for $d_E$, $d_M$, $d_{EE}$, and $d_{EM}$.}
\begin{tabular}{rrrrrrrrr} \hline \hline 
$\tilde{\psi}_0$ & $\kappa_E$ & $\hat{\kappa}_E$ &
$\kappa_M$ & $\hat{\kappa}_M$ & $\kappa_{EE}$ &
$\hat{\kappa}_{EE}$ & $\kappa_{EM}$ & $\hat{\kappa}_{EM}$
\\ \hline 

   0.0& 12.8   & 70.5   & 4.14   & 22.8   & 20.8   & 114   & 7.29   & 40.1     \\
   0.5& 13.4   & 80.5   & 4.34   & 26.1   & 23.3   & 140   & 7.90   & 47.5     \\
   1.0& 11.0   & 73.2   & 3.38   & 22.5   & 18.3   & 122   & 5.89   & 39.2     \\
   1.5& 10.1   & 75.3   & 3.10   & 23.1   & 16.8   & 126   & 5.31   & 39.6     \\
   2.0& 7.05   & 60.1   & 2.06   & 17.6   & 10.0   & 85.3   & 3.07   & 26.2     \\
   2.5& 7.68   & 76.2   & 2.24   & 22.3   & 12.2   & 122   & 3.56   & 35.4     \\
   3.0& 6.42   & 75.8   & 1.85   & 21.9   & 9.64   & 114   & 2.72   & 32.2     \\
   3.5& 5.42   & 78.2   & 1.51   & 21.8   & 7.95   & 115   & 2.09   & 30.1     \\
   4.0& 4.85   & 88.0   & 1.39   & 25.2   & 7.33   & 133   & 1.93   & 35.0     \\
   4.5& 3.80   & 89.5   & 1.06   & 24.9   & 5.55   & 131   & 1.33   & 31.3     \\
   5.0& 2.85   & 90.2   &0.740   & 23.4   & 4.05   & 128   &0.985   & 31.2     \\
   5.5& 2.33   & 102   &0.702   & 30.8   & 3.36   & 147  &0.780   & 34.1     \\
   6.0& 1.88   & 117   &0.603   & 37.7   & 2.73   & 171   &0.631   & 39.4     \\
   6.5& 1.40   & 126   &0.499   & 45.0   & 2.03   & 183   &0.489   & 44.0     \\
   7.0& 1.06   & 137   &0.379   & 48.8   & 1.56   & 201   &0.329   & 42.4     \\
   7.5&0.836   & 141   &0.363   & 61.4   & 1.24   & 209   &0.308   & 52.0     \\
   8.0&0.677   & 134   &0.320   & 63.2   &0.971   & 192   &0.243   & 48.0     \\
   8.5&0.647   & 126   &0.362   & 70.7   &0.928   & 181   &0.277   & 54.0  
   \\ \hline
\end{tabular}
\footnotetext[0]{ Note: The dimensionless diffusion coefficients are
approximated by $d = \kappa \ln(1/\eta_{c0}) \eta_{c0}^2$
and $\hat{d} = \hat{\kappa} \ln(1/\hat{\eta}_{c0})
\hat{\eta}_{c0}^2$ where $\hat{d}=d(\rt^2/\rhm^2)$ and
$\hat{\eta}_{c0}=\eta_{c0}(GM/\rhm\sigma_0^2)^{1/2}$.}
\end{minipage} 
\end{table*}

For use in our Fokker-Planck calculations, the four diffusion 
coefficients were tabulated for almost all the values in the range
$\tilde{\psi}_0=0.0 \rightarrow 8.5$
and $\eta_{c0}=0.0001 \rightarrow 0.5$. 
Due to the extremely long integration times for small values of 
$\eta_{c0}$ and large $\tilde{\psi}_0$, a few
values were obtained
using analytic fitting formulae of the form
$d_E=\kappa_E \ln(1/\eta_{c0})
\eta_{c0}^2$, $d_M=\kappa_M \ln(1/\eta_{c0})\eta_{c0}^2$,
$d_{EE}=\kappa_{EE} \ln(1/\eta_{c0})\eta_{c0}^2$, and
$d_{EM}=\kappa_{EM} \ln(1/\eta_{c0})\eta_{c0}^2$, which represented
the data well; the coefficients for these fits are given in Table 2.
The fits were generally good to $\simless 5-30$ percent
and are useful for quick estimates, although
for more precise numerical work interpolation on tabulated
values was used. The values generated using the fitting formalae were
used only for the very smallest values of $\eta_{c0}$ for the cluster
PAL 5, where the fitting formulae are most accurate.
Fig. $\ref{diffcoeffplot}$ displays the various
coefficients for several values of 
$\eta_{c0}$ which are shown
increasing from bottom to top. The triangles represent data which was
generated by the Monte Carlo program itself, while the open circles
represent the data
estimated using the fitting formula.

\setcounter{figure}{6}
\begin{figure}

\begin{picture}(324,324)(0,0)
\put(81,0){\epsfxsize=4.5in
\epsffile{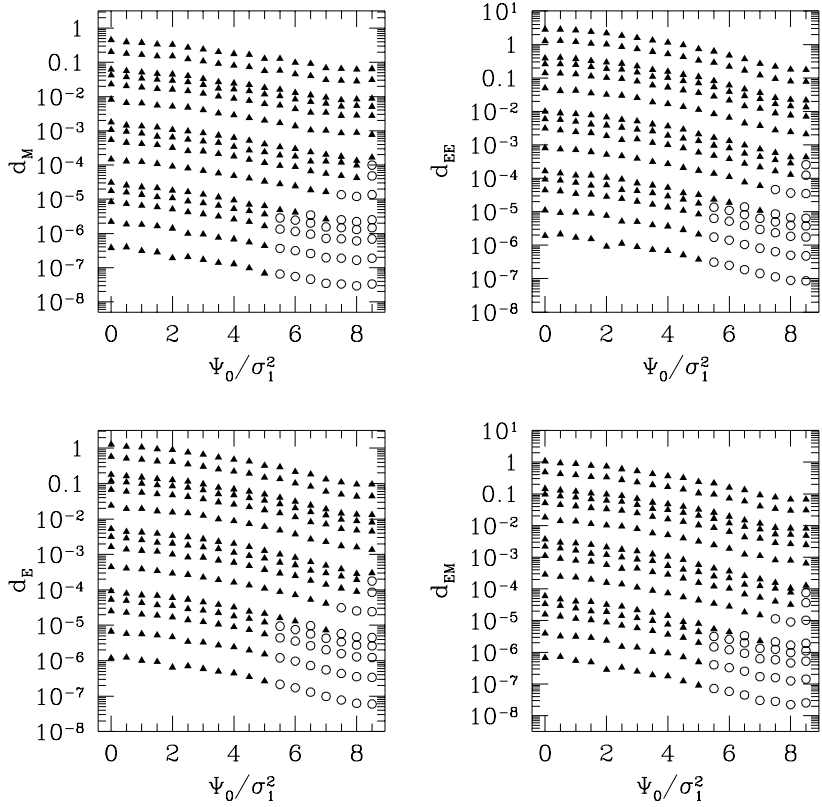}}
\end{picture}
 
\caption{ Dimensionless diffusion coefficients. The triangles represent
Monte Carlo results. The circles represent the incomplete data given 
by the fitting formulae. The values of $\eta_{c0}$ shown are $0.0001$,
$0.00025$, $0.0005$, $0.00075$, $0.001$, $0.0025$, $0.005$, $0.0075$,
$0.01$, $0.025$, $0.05$, $0.075$, $0.1$, $0.25$, $0.5$, increasing
from bottom to top.} 

\label{diffcoeffplot}
\end{figure}

It can be seen that
for a fixed value of $\eta_{c0}$, the diffusion
coefficients in Fig. $\ref{diffcoeffplot}$ vary by about an 
order of magnitude over the range
$\tilde{\psi}_0= 0.0 \rightarrow 8.5$. The coefficients for the
fitting formulae show a similar range of variation.
This spread can be reduced considerably if we recall
that many quantities vary little over the King sequence when
expressed in terms of the half-mass radius, $\rhm$. Choose 
a new unit of rate $\hat{\Gamma}_0=\nbh\pi \rhm^2 V_c$
and a new collision parameter $\hat{\eta}_{c0}=
\eta_{c0}(GM/\rhm\sigma_0^2)^{1/2}$. The constants $\hat{\kappa}$ in
the new dimensionless diffusion
coefficients $\hat{d}=\hat{\kappa}\ln(1/\hat{\eta}_{c0})
\hat{\eta}_{c0}^2=d(\rt^2/\rhm^2)$ 
then vary by about a factor of two over the
range $\tilde{\psi}_0= 0.0 \rightarrow 8.5$, as can be seen in Table 2.

\section{`Slow Heating' Lifetimes \label{slowheating} }

\subsection{ The Fixed Cluster Approximation \label{mean} }

When $\mbh<< \mhigh$, the properties of a given cluster only
change slightly over many encounters with black holes.
In this limit, cluster survival implies an upper bound to
$\fhalo\mbh$ (up to logarithmic corrections), 
or $\mbh$ assuming a certain value of $\fhalo$.
Very rough estimates of the critical black hole mass
for survival over time $T$ can be made if we neglect the evolution
of the cluster profile over time $T$, and require the mean energy 
or mass change not to exceed some threshold over that time span
(e.g. Moore 1993). The values found here are intended as benchmarks
for comparison to the more detailed results, obtained in the next
section, that include cluster evolution and statistical spreads in 
the energy and mass changes in time $T$. 

\setcounter{table}{2}
\begin{table*}
\begin{minipage}{5in}
\label{slowheatfixed}
\caption{Cluster Destruction for a Fixed Profile}
\begin{tabular}{lrrr} \hline \hline
Cluster Name& $\mbhcrit(E)/M_{\odot}$ &
$\mbhcrit(M)/M_{\odot}$ &
$\mbhcrit(\nu)/M_{\odot}$
\\ \hline
AM 4    &    1500&    7000&    710  \\
ARP 2   &    3100&    13,000&    4900  \\
NGC 5053&    4100&    18,000&    5200  \\
NGC 7492&    3400&    15,000&    6500  \\
PAL 4   &    170,000&  8,500,000&  350,000  \\
PAL 5   &    1100&    4300&    1100  \\
PAL 13  &    3100&    13,000&    6300  \\
PAL 14  &    14,000&  68,000&    15,000  \\
PAL 15  &    9100&    39,000&    6100  \\
\\ \hline
\end{tabular}
\footnotetext[0]{ Note: $\mbhcrit(E)/M_{\odot}$ is the critical black
hole mass for the criterion $\delta E/|E|=0.5$.
$\mbhcrit(E)/M_{\odot}$ is the critical black hole mass
for the criterion $|\delta M|/M=0.5$.
$\mbhcrit(\nu)/M_{\odot}$ is critical black hole
mass for $\nu$ to go from its initial value to $\nu=-0.6$.} 
\end{minipage}
\end{table*}

For a fixed cluster profile, 
$\rmnd E/\rmnd t$, $\rmnd M/\rmnd t$, and $\rmnd \nu /\rmnd t$
are independent of time, and $\langle \btu E(T) \rangle=
\Gamma_0 |E| d_E T$ and
$\langle \btu M(T) \rangle=-\Gamma_0 M d_M T$, with the coefficients $d_E$ and
$d_M$ defined and computed in $\S \ref{diffcoeff}$. 
Choosing a time
of $T=10^{10}$ years as the time over which the clusters have been
subjected to collisions, we defined three distinct values of 
$\mbhcrit$ by
$\langle \btu E(T,\mbhcrit) \rangle= |E|/2$, 
$\langle \btu M(T,\mbhcrit) \rangle= -M/2$, 
and $\langle \btu \nu(T,\mbhcrit) \rangle=
[D_E/|E|-(5/3)(D_M/M)]|\nu| T =  -0.6 - \nu$.
The results are presented in Table 3 for the sample of nine weakly 
bound clusters employed by
Moore (1993). Moore's results correspond most
closely to $\mbhcrit(E)$ as he used the criterion
$\langle \btu E(T=7 \times 10^{9} {\rm years},\mbhcrit) \rangle= |E|$. 
When comparing  to Moore's results one should multiply $\mbhcrit$
by a factor of $2/0.7$ to account for the larger value of $T$ and 
smaller $\btu E$ used here. Except for PAL 13, our
results agree with Moore's to within a factor of order a few, 
which is resonably close considering the improvements made
here (e.g. inclusion of mass loss, correct $\vrel$ distribution).

\subsection{ Gaussian Model with Evolution 
\label{gfp} } 

In $\S \ref{mean}$, two approximations were made. First, the cluster
was treated as having a fixed profile for which the diffusion 
coefficients did not change over time. The second approximation was that
the energy input and mass loss had sharply defined values over any time
interval. In this section, the evolution of the cluster is
followed over appropriately chosen intervals $\delta t$ and a 
Gaussian distribution of energy input and mass
loss is assumed. 

The characteristic
time for $N_{\rmn min}$ collisions to occur inside $b_{\rmn max}$ is 
\beqa
T_{\rmn coll} & =& \frac{ N_{\rmn min} }{\nbh\pi b_{\rmn max}^2 \xi_1 V_c} 
\propto \mbh R_g^2 N_{\rmn min}.
\eeqa
As most of the energy input is given by the penetrating encounters, 
a conservative estimate is to set $b_{\rmn max}=\rt$ in $T_{\rmn coll}$. 
If $\epsilon_{\rmn max}$ is the largest change allowed for the cluster in
a time period, then a time $T_{\rmn change}$ over which the cluster
evolves significantly is given by ($\S \ref{diffcoeff}$) 
\beqa
T_{\rmn change} & =&  \frac{ \epsilon_{\rmn max} } { \nbh \pi \rt^2 V_c d_E }
\propto \frac{ \epsilon_{\rmn max} R_g^2 }{ \mbh }.
\eeqa
The Gaussian approach is justified if there exists a $\delta t$ such
that $T_{\rmn coll}< \delta t < \delta T_{\rmn change}$, so the cluster properties 
change little over $\delta t$ but enough collisions occur
that the distribution of energy inputs and mass loss is approximately
Gaussian.
An estimate of the largest black hole mass, $M_{\rmn bh,fp}$, for which
the Gaussian approach is justified is
\beqa
M_{\rmn bh,fp} & \simless & M \frac{V_c}{\sigma_0} 
\left( \frac{\epsilon_{\rmn max}}{N_{\rmn min}} \right)^{1/2},
\eeqa
or $\eta_{c0} \simless (\epsilon_{\rmn max}/N_{\rmn min})^{1/2}$. The critical
mass can only be found in the present `diffusion' approximation when
$\epsilon_{\rmn max} \ll 1$ and $N_{\rmn min} \simgreat 1$ so that the cluster
must be destroyed for $\eta_{c0} \ll 1$.

Let the vector $ \btau = (\tau_1,\tau_2)=(\btu E,\btu M)$
where $\btu E$ and
$\btu M$ are the energy input and mass loss over the time interval 
$\delta t$; the expected values of $\btu E$ and $\btu M$ over the
time $\delta t$ are $\langle \tau_1 \rangle =  
\langle \btu E (\delta t) \rangle$ and
$\langle \tau_2 \rangle = \langle \btu M (\delta t) \rangle$. 
The normalized distribution of energy input and 
mass loss is then
\beqa
P(\btau) \rmnd^2\btau & = & \frac{ \sqrt{\det(\bXi)}}{ 2\pi } \rmnd^2 
\btau
\exp \left\{ - \frac{1}{2} 
(\btau- \langle \btau \rangle)^T \bXi (\btau-\langle \btau \rangle) \right\}, 
\eeqa
where $\bXi(\delta t)$ is the covariance matrix.
The matrix $\bXi^{-1}$ may be shown to be
\[
\bXi^{-1} = \left( \begin{array}{cc}
\langle (\tau_1 - \langle\tau_1 \rangle)^2 \rangle & 
\langle (\tau_1 - \langle \tau_1 \rangle)
(\tau_2 - \langle \tau_2 \rangle) \rangle  \\
 \langle (\tau_1 - \langle \tau_1 \rangle)
(\tau_2 - \langle \tau_2 \rangle) \rangle & 
\langle (\tau_2 - \langle \tau_2 \rangle)^2 \rangle
\end{array} \right);
\]
inverting implies
\[ 
\bXi = \frac{1}{\det(\bXi^{-1})} \left( \begin{array}{cc}
\langle (\tau_2 - \langle \tau_2 \rangle)^2 \rangle & 
- \langle (\tau_1 - \langle \tau_1 \rangle)
(\tau_2 - \langle \tau_2 \rangle) \rangle  \\
- \langle (\tau_1 - \langle \tau_1 \rangle)
(\tau_2 - \langle \tau_2 \rangle) \rangle & 
\langle (\tau_1 - \langle \tau_1 \rangle)^2 \rangle
\end{array} \right).
\]
These expressions reduce to the standard sum of Gaussian terms when
the correlation $\langle (\tau_1 - \langle \tau_1 \rangle)
(\tau_2 - \langle \tau_2 \rangle) \rangle$
is zero. The eigenvalues of $\bXi$ are
$\lambda_{\pm} = \frac{1}{2} \left[ \tr(\bXi) \pm
\sqrt{ (\tr(\bXi))^2 - 4\det(\bXi) } \right]$.

	In order to choose values for $\btu E$ and $\btu M$, 
we define a new set of variables $\bzeta=(\zeta_{+},\zeta_{-})$
along the eigenvectors of $\bXi$; the distribution of $\bzeta$ is
\beqa
P(\bzeta)\rmnd^2\bzeta & = & \frac{ \sqrt{ \lambda_{+} \lambda_{-}} }
{ 2\pi } \rmnd^2 \bzeta
\exp \left\{ - \lambda_{+} \zeta_{+}^2/2 - 
\lambda_{-} \zeta_{-}^2/2  \right\},
\eeqa
and  $\btau- \langle \btau \rangle = \bLambda \bzeta$, where the unitary
matrix
\[  
\bLambda  =  \left( \begin{array}{cc}
\Lambda_{1+}& \Lambda_{1-} \\
\Lambda_{2+}& \Lambda_{2-}
\end{array} \right)
=
\frac{1}{ \sqrt{ (\lambda_{-}-\Xi_{22})^2 + \Xi_{12}^2 } }
\left( \begin{array}{cc}
 \Xi_{12} & \lambda_{-} - \Xi_{22} \\
-( \lambda_{-} - \Xi_{22}) & \Xi_{12}
\end{array} \right).
\]
After choosing $\zeta_{\pm}$ from this Gaussian distribution, we multiply
by $\Lambda$ to get
\beqa
&&\btu E  = \langle \btu E \rangle + 
\Lambda_{1+} \zeta_{+} + \Lambda_{1-} \zeta_{-} \nonumber \\
&& \btu M  = \langle \btu M \rangle + \Lambda_{2+} \zeta_{+} + 
\Lambda_{2-} \zeta_{-}.
\eeqa

In the limit of infinite numbers of collisions $N_c$ in time $\delta t$
(the limit in which the Gaussian model is equivalent to a 2-D Fokker-
Planck equation)
\beqa
&& \langle \tau_1 \rangle=
 \Gamma_0 |E| d_E \delta t \nonumber \\
&& \langle \tau_2 \rangle = 
 -\Gamma_0 M d_M \delta t \nonumber \\
&& \Xi_{11}^{-1}(\delta t)  =
\frac{1}{N} \Gamma_0 E^2 d_{EE} \delta t \nonumber \\ &&
\Xi_{12}^{-1}(\delta t)  =
- \frac{1}{N} \Gamma_0 |E|M d_{EM} \delta t \nonumber \\ &&
\Xi_{22}^{-1}(\delta t)  = 
\frac{1}{N} \Gamma_0 M^2 d_{MM} \delta t.
\eeqa
For finite $N_c$, these expressions are accurate to ${\cal O}(N_c^{-1})$.

	Let us ignore the correlations of the change in mass and 
energy and focus solely on changes in cluster energy in order to 
get a qualitative feel for the `distribution of cluster states'.
The Fokker-Planck equation for the 
distribution of energies, $P(E,t)$,
of an ensemble of clusters is 
\beqa
\frac{ \partial P}{\partial t} & =&
- \frac{\partial}{\partial E} \left( D_E P \right)+ 
\frac{1}{2} \frac{\partial^2}{\partial E^2} \left( D_{EE} P \right).
\eeqa
If we ignore changes in the cluster properties and take $D_E$ and 
$D_{EE}$ to be constant, this equation has the solution
\beqa
P(E,t) & = & \frac{ 1 }
{ (2\pi D_{EE}t )^{1/2} }
\exp \left[ - \frac{ (E - E_0 - D_Et)^2 }{ 2 D_{EE}t }
\right]
\eeqa
where we have chosen the initial condition $E=E_0$ at $t=0$.
For small times $t < D_{EE}/D_E$, the width of the Gaussian is much
larger than the mean implying a very spread out set of cluster states,
and visa versa for times $t > D_{EE}/D_E$. As we have already shown
in $\S \ref{themany}$, for times $t \simeq T_{\rmn disrupt}$
the set of final states is 
sharply defined about the mean. Using the survival criterion
$\btu E < f|E_0|$, we find that the probability of survival after
time $t$ is given by
\beqa
P_s & = & \int_{-\infty}^{E_0+f|E_0|} \rmnd E P(E,t)
= \frac{1}{2} \left[
1 + \erf
\left( \frac{ f|E| - D_E t}{ \sqrt{2 D_{EE}t} } \right)
\right].
\eeqa
The same equation would have resulted if we had used the two-dimensional
Fokker-Planck equation for $E$ and $M$. In terms of the fitting
formulas of $\S \ref{diffcoeff}$, the characteristic mass range,
$\delta \mbhcrit$ over which $P_s$ changes from one to zero is
\beqa
\frac{\delta \mbhcrit }{\mbhcrit} & \simeq & 
\sqrt{ \frac{\kappa_{EE}}{N f \kappa_E} } \ll 1
\eeqa
so that in the $\mbh \ll \mhigh$ limit the value of $\mbhcrit$
found from just {\it one} history is likely to be close to the statistical 
mean.

\subsection{ Simulations and Results }

We chose the accuracy parameters from the previous section to be
 $\epsilon_{\rmn max} = 0.01$ and
$N_{\rmn min} = 10$. The time step was chosen so that the estimated change
in energy would be $0.01$. As the diffusion coefficients tended to 
increase as $\tilde{\psi}_0 \rightarrow 0$ and $M$ decreased, 
the time step was decreased if two events in a row occured with 
fractional changes in $\nu$, $M$, or $E$ greater than $0.01$. 
Also, the time step was increased if the estimated number of events
in $\delta t$ was less than $N_{\rmn min}$. If $\mbh>M_{\rmn bh,fp}$, as
described in the previous section, the simulation was stopped because
the approximations had broken down. 

For each of the nine weakly bound clusters from Moore \shortcite{moore93},
values of $\mbh$ were chosen and
the cluster was evolved
to one of the following outcomes:
(1) the Fokker-Planck assumption
broke down; (2) $T=10^{10}$ years was reached and the cluster survived;
(3) the quantity $\nu$ went out of the range $(-2.13,-0.6)$, signalling
either core collapse or `dissolution'. Survival probabilities and 
their uncertainties were derived from the number of surviving clusters
out of $1000$ simulations using Bayesian arguments. 
The results are shown in Fig. $\ref{gmc}$
for Moore's clusters as well as one
more `normal' cluster.
Note that all
curves asymptote to $\fdest=1.0$ as $\mbh \rightarrow \infty$
because single-event destruction has been ignored in these simulations.

\setcounter{figure}{7}
\begin{figure}

\begin{picture}(432,648)(0,0)

\put(0,432){\epsfxsize=3in
\epsffile{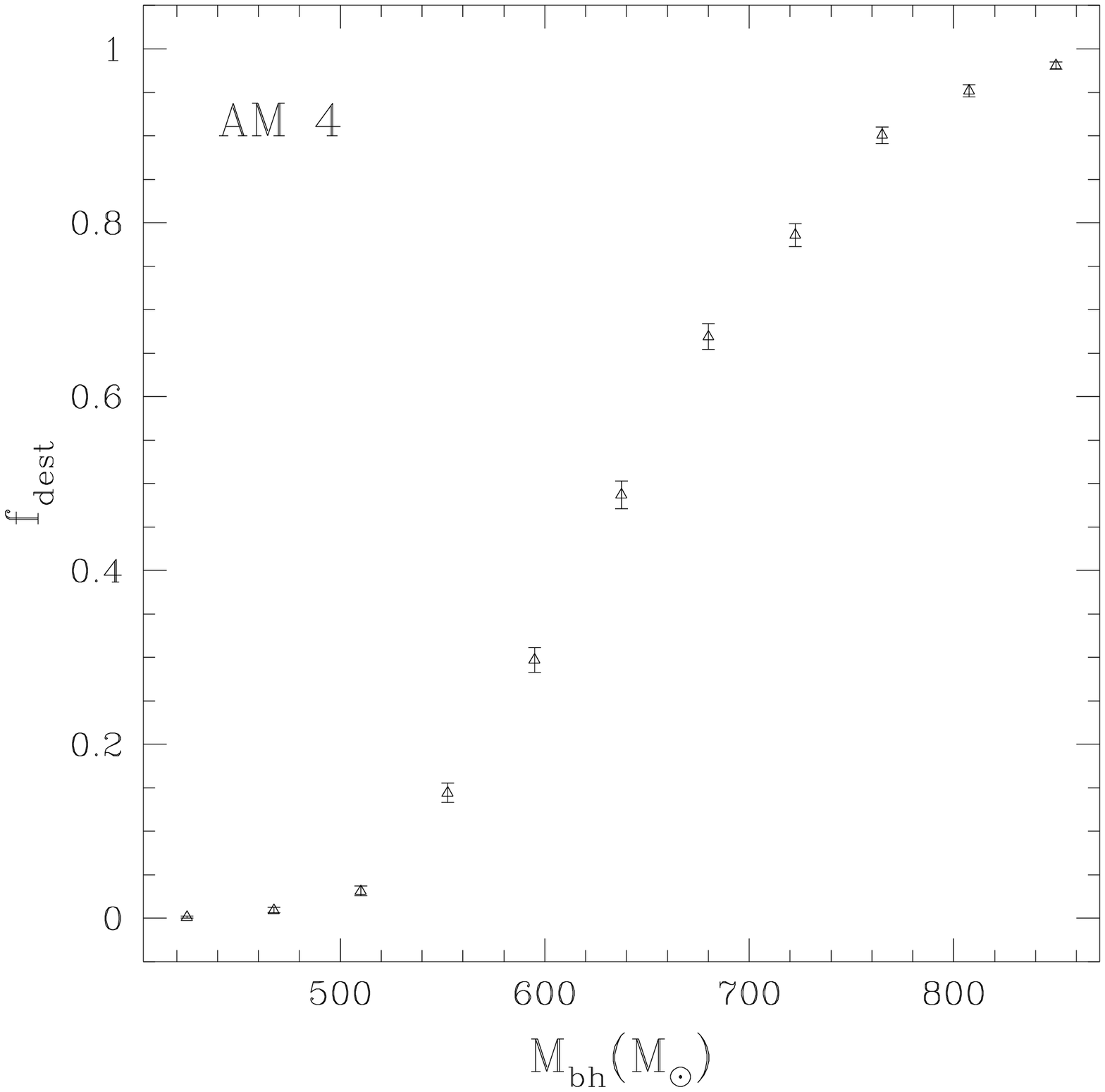}}

\put(216,432){\epsfxsize=3in
\epsffile{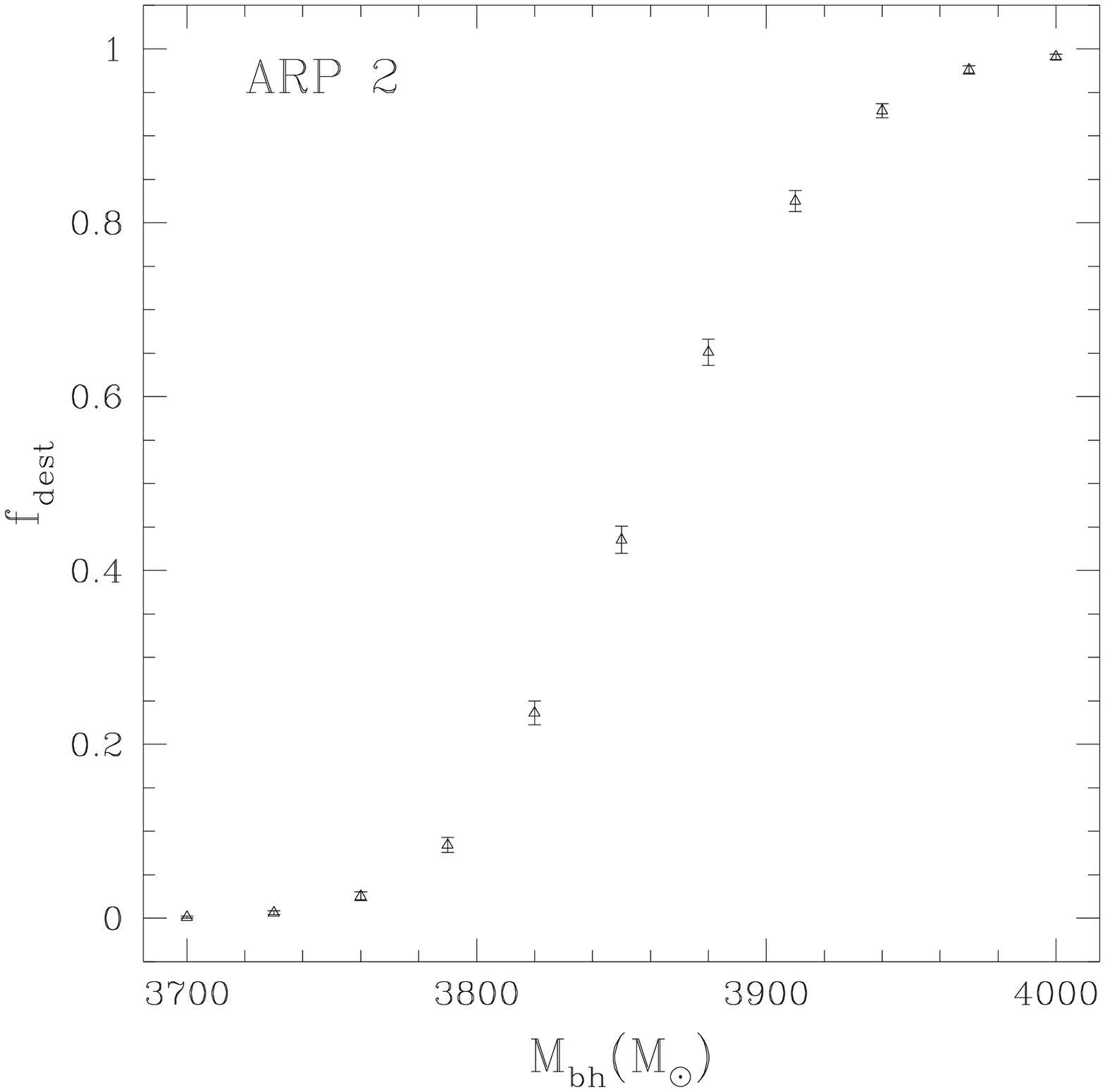}}

\put(0,216){\epsfxsize=3in
\epsffile{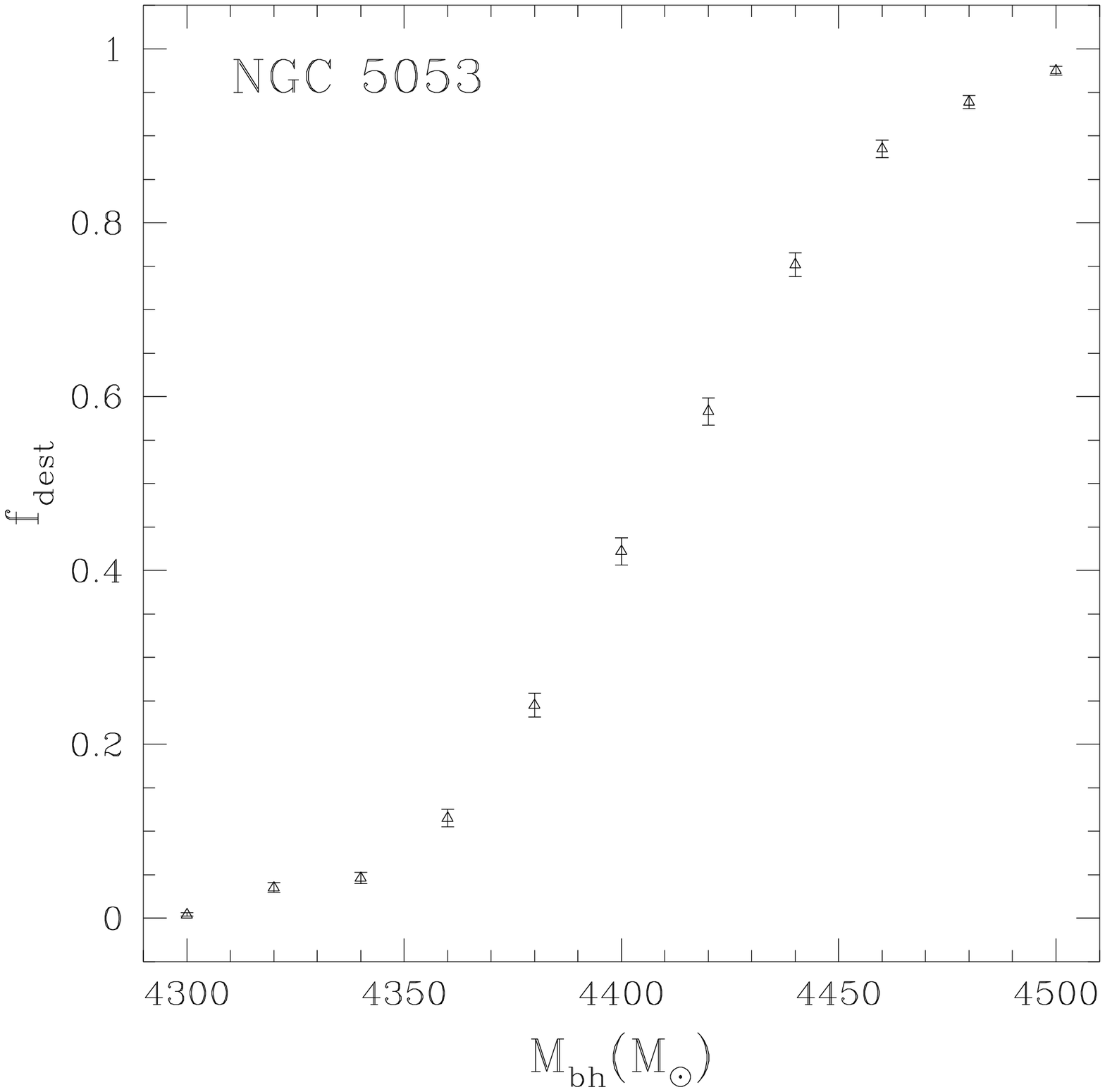}}

\put(216,216){\epsfxsize=3in
\epsffile{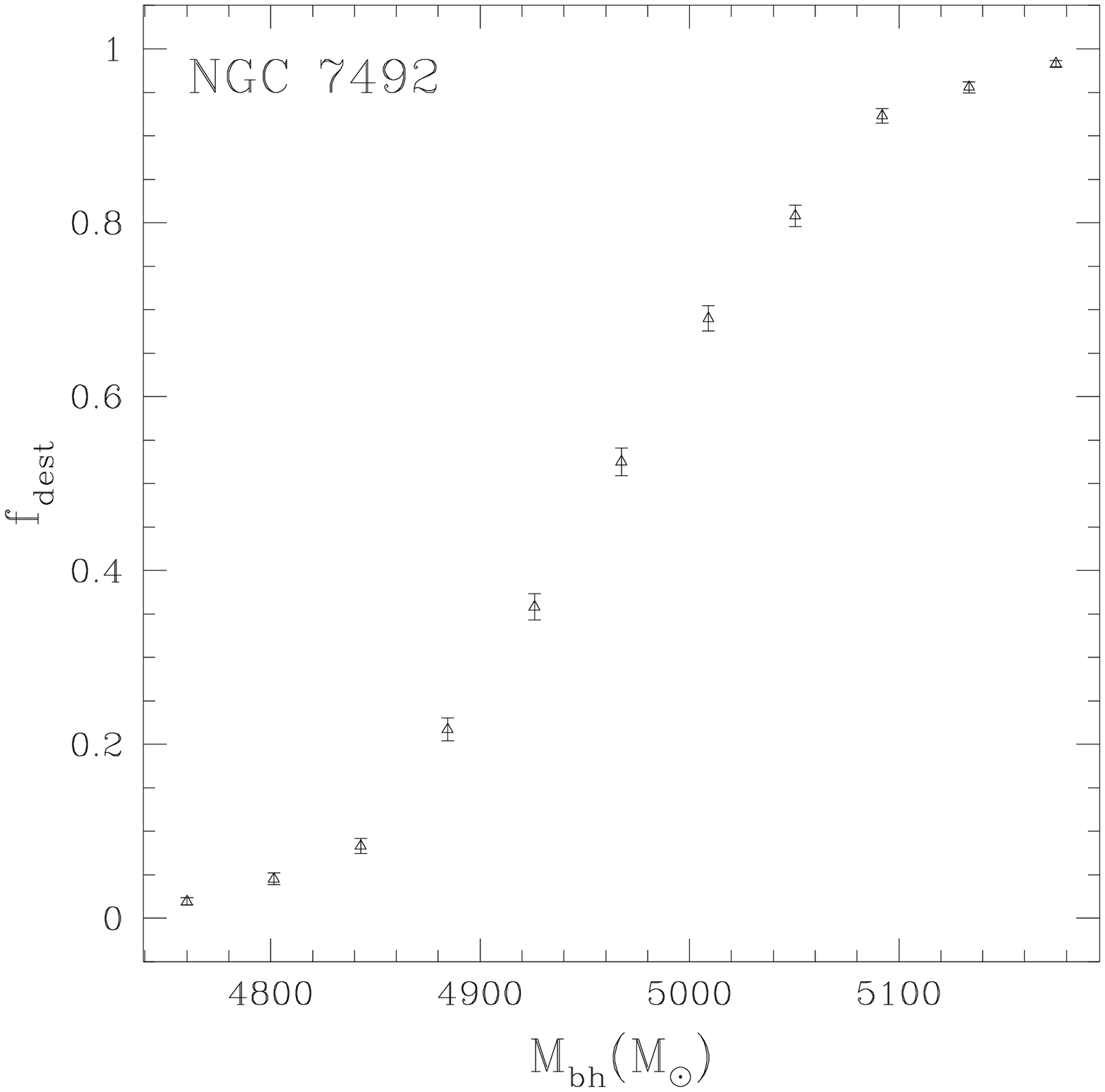}}

\put(0,0){\epsfxsize=3in
\epsffile{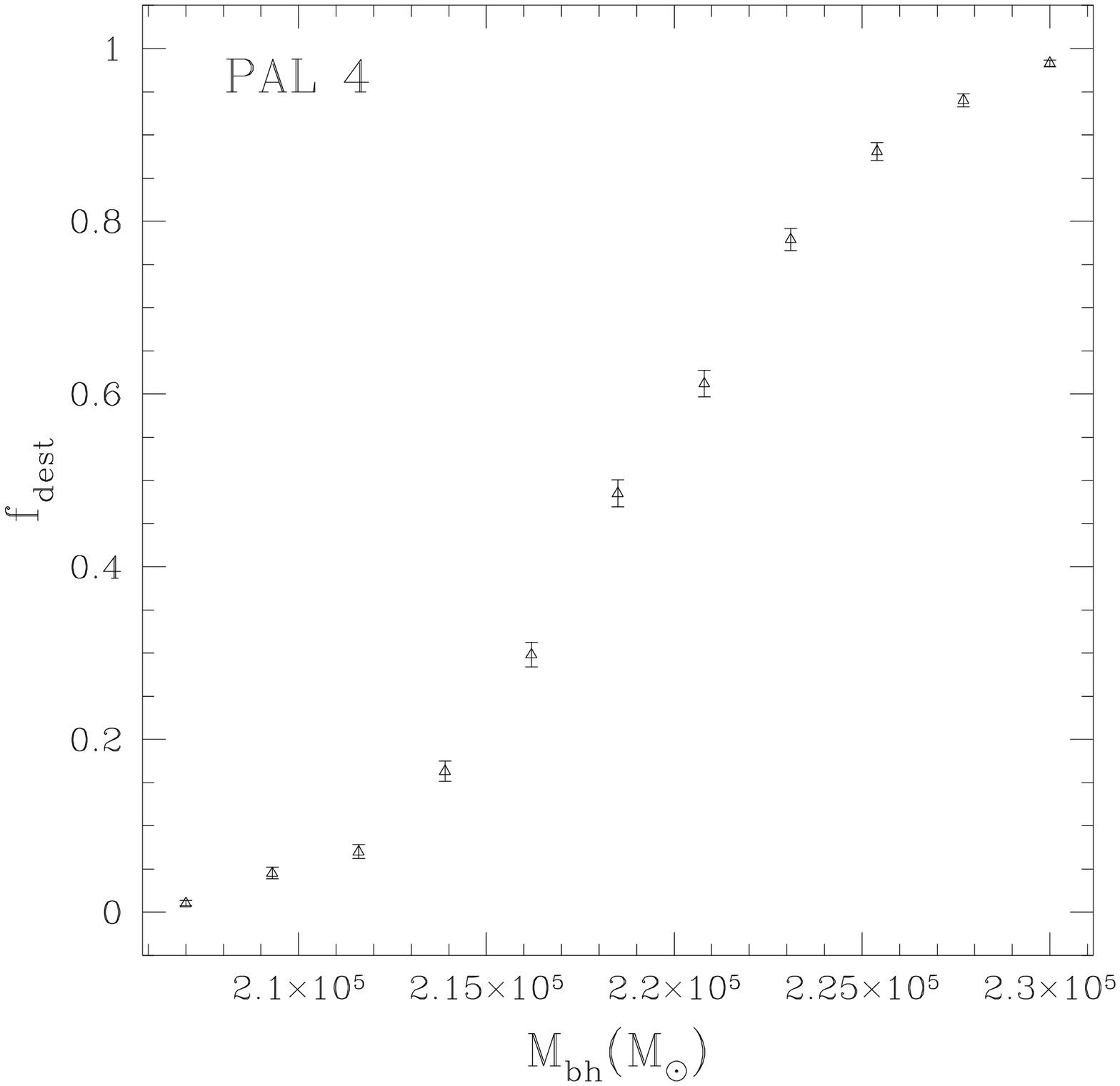}}
 
\put(216,0){\epsfxsize=3in
\epsffile{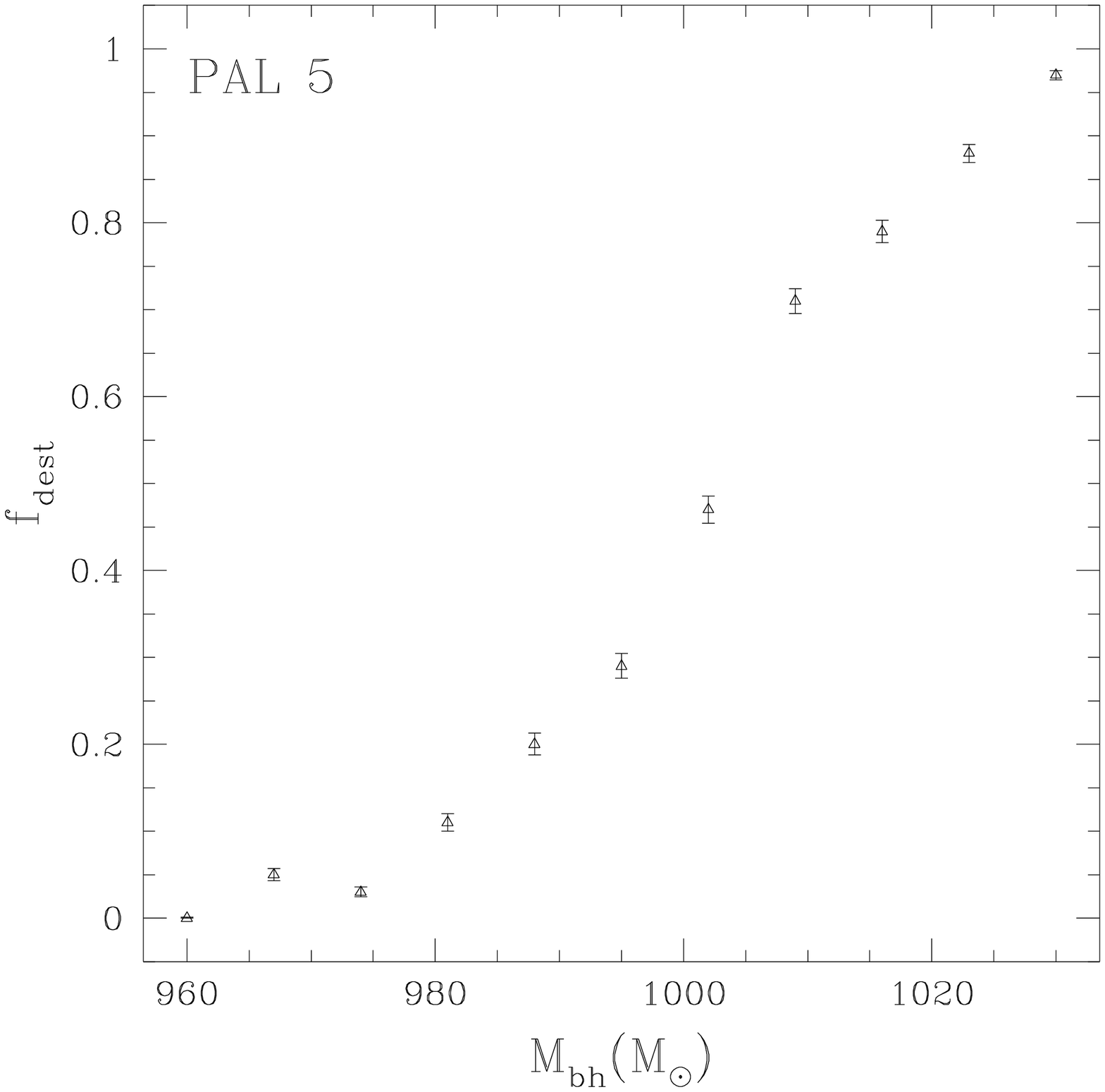}}

\end{picture}
\end{figure}
\begin{figure}
\begin{picture}(432,432)(0,0)

\put(0,216){\epsfxsize=3in
\epsffile{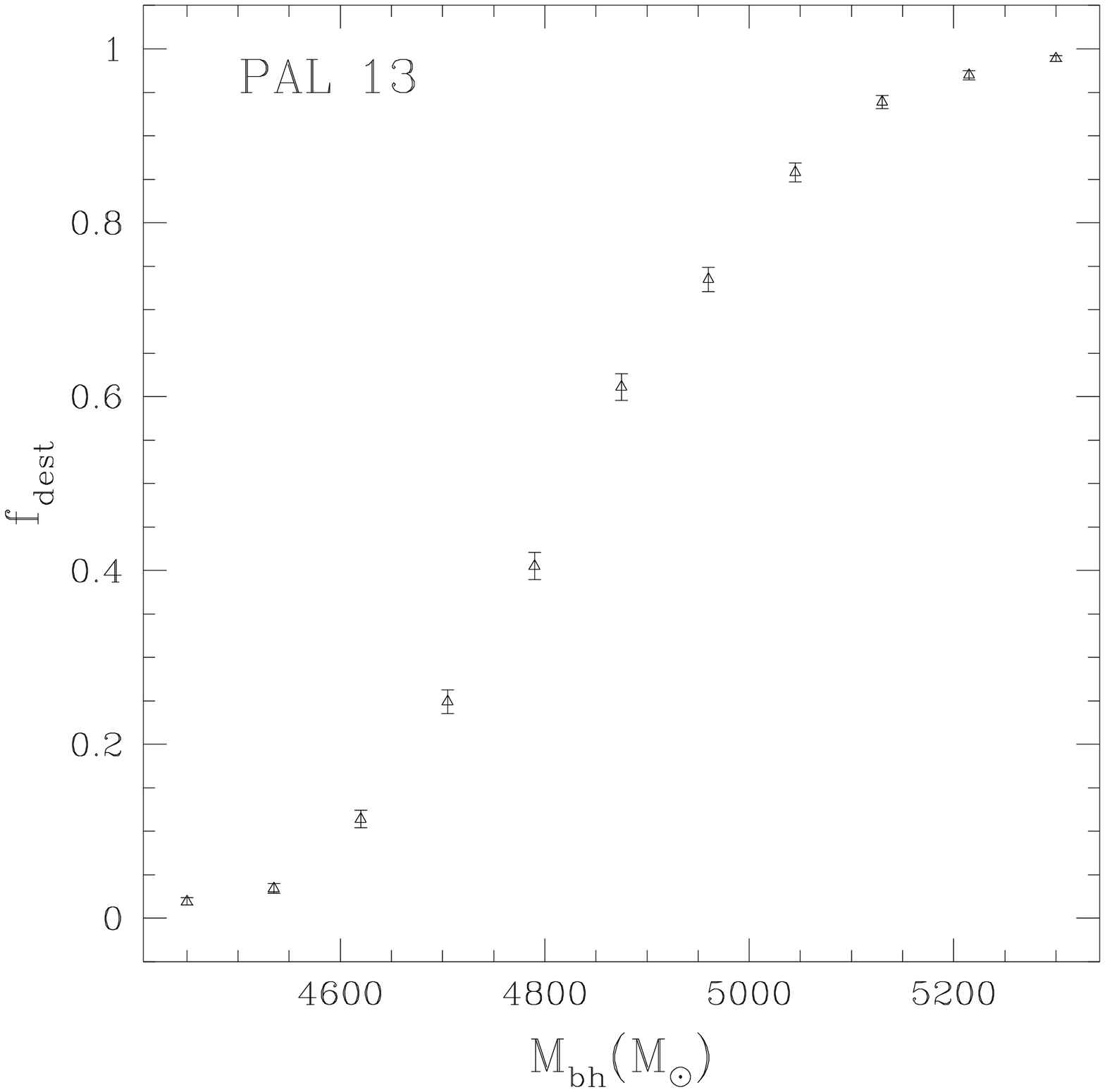}}
 
\put(216,216){\epsfxsize=3in
\epsffile{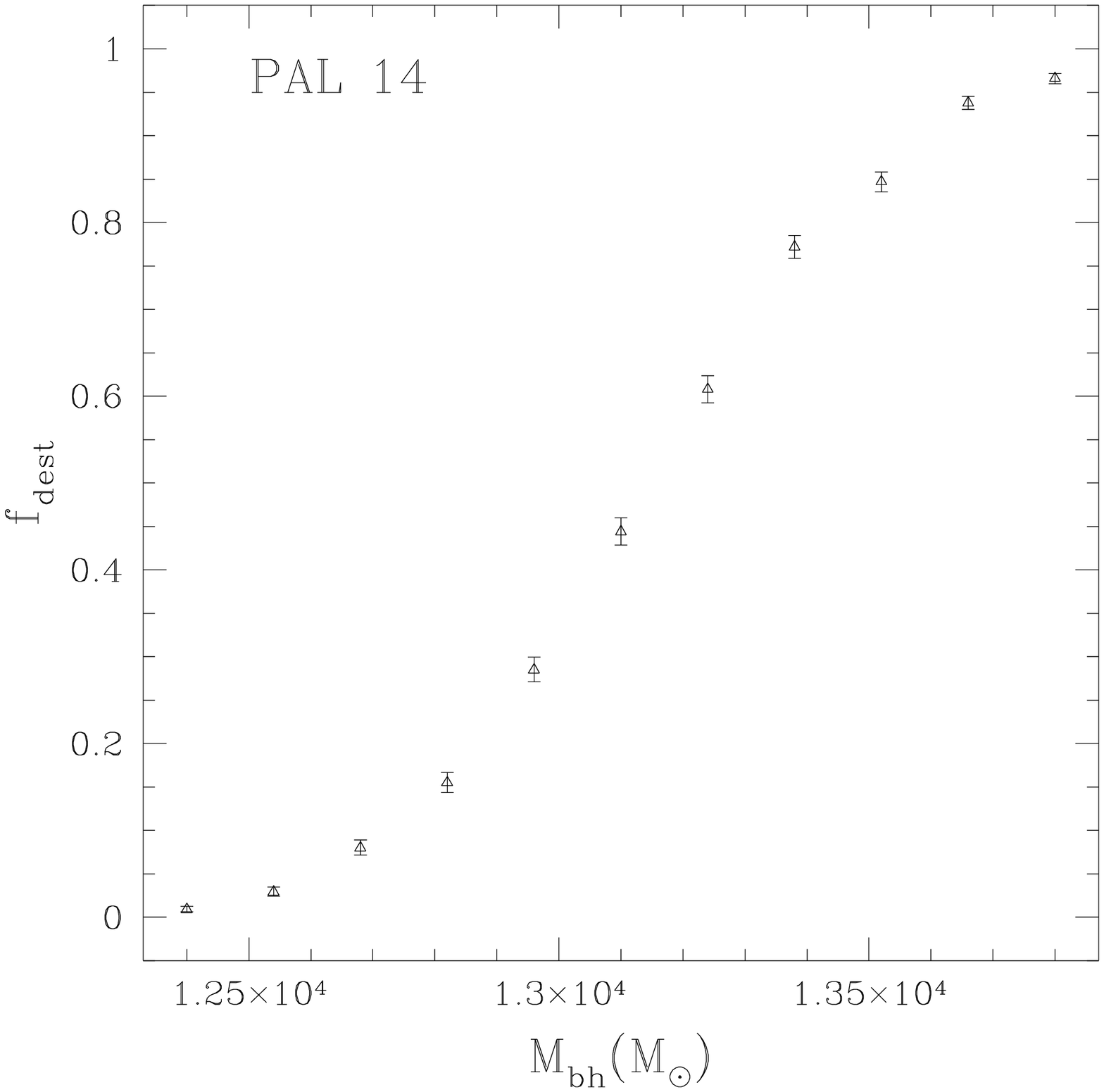}}

\put(0,0){\epsfxsize=3in
\epsffile{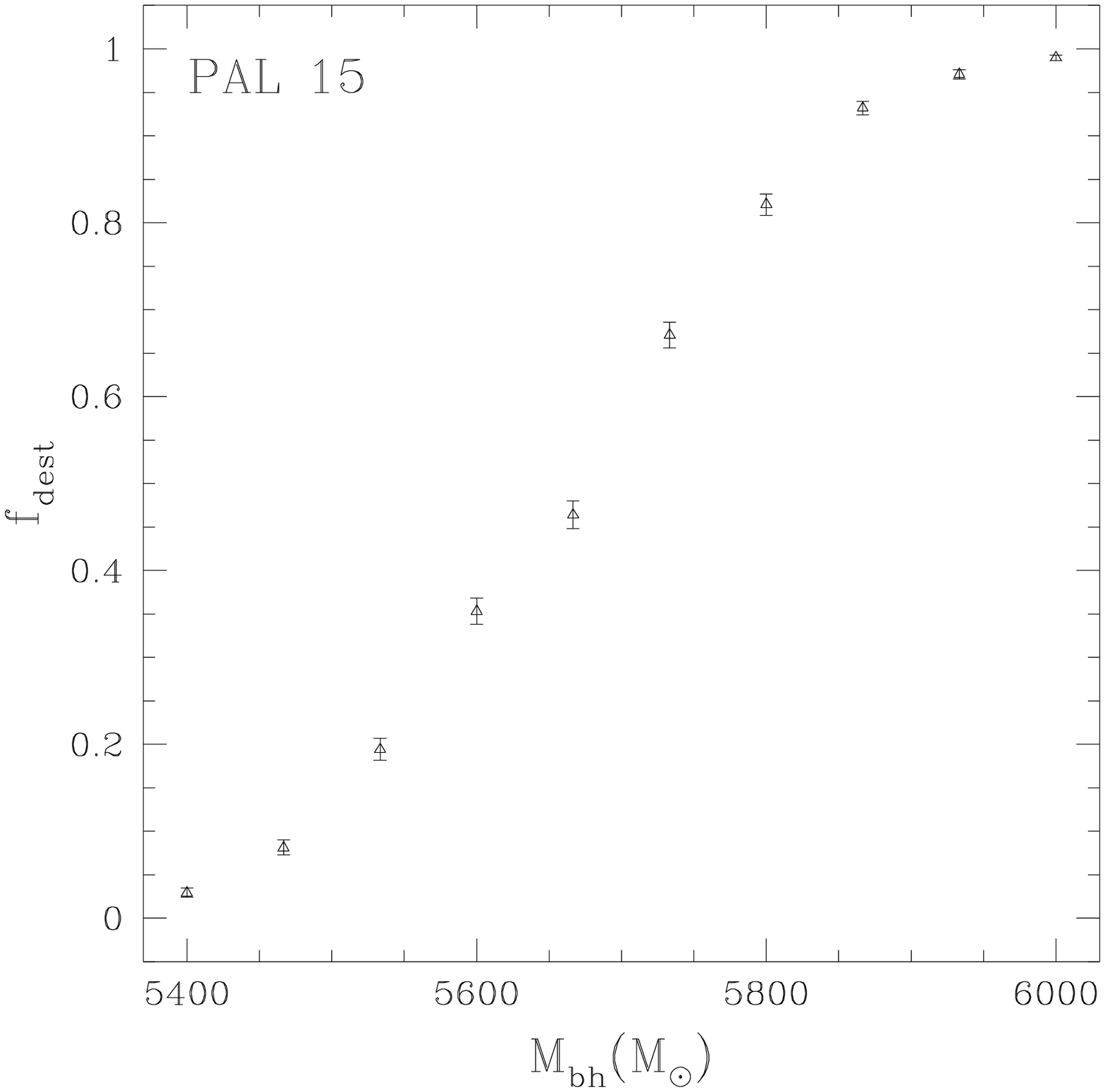}}
 
\put(216,0){\epsfxsize=3in
\epsffile{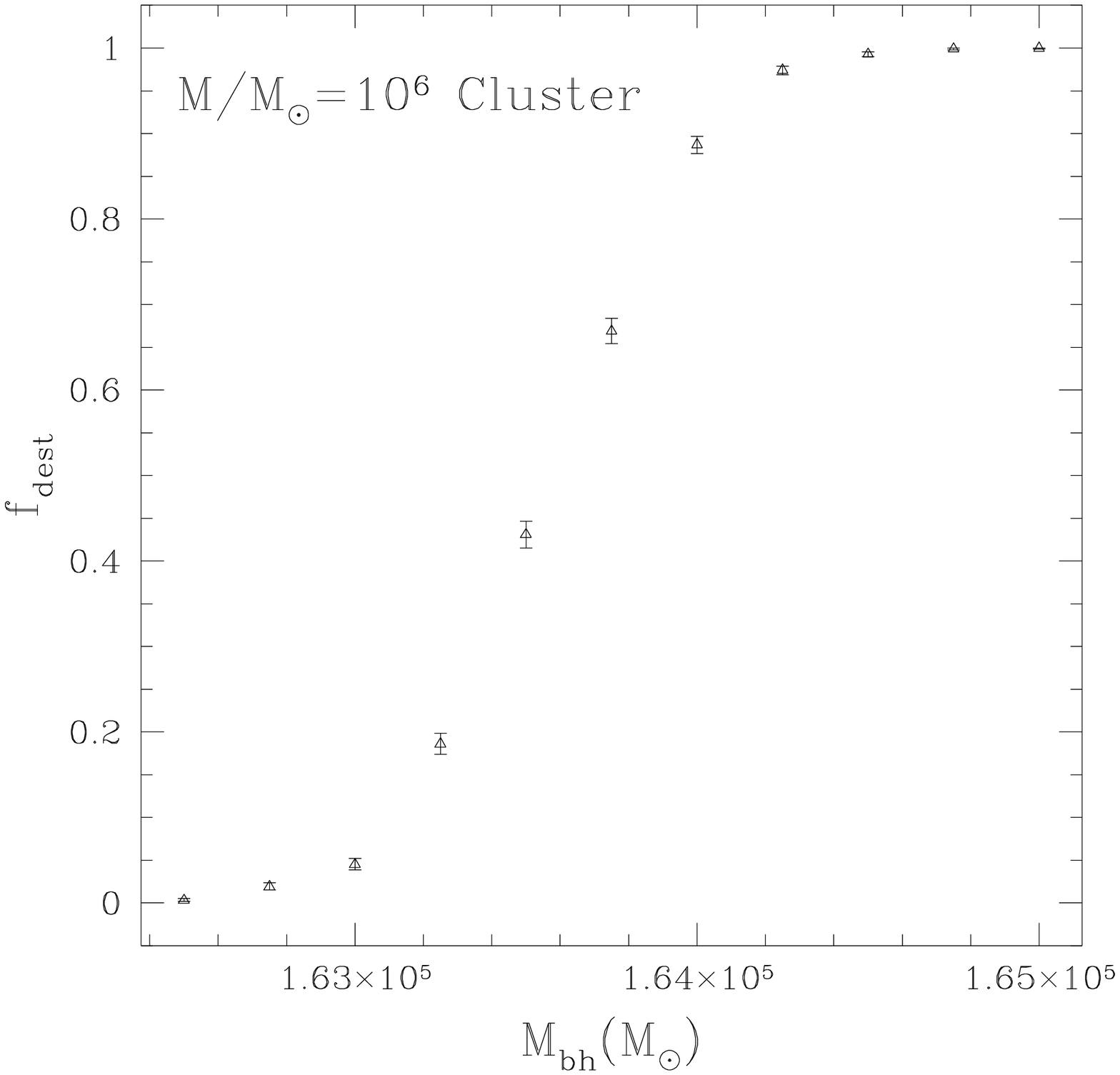}}

\end{picture}

\caption{ Fraction of clusters destroyed in $T=10^{10}$ years 
plotted against black hole mass. These curves were generated using the 
Gaussian Monte Carlo method.} 

\label{gmc}

\end{figure}

PAL 4 did {\it not} satisfy the Fokker-Planck assumption at the black
hole mass required to disrupt it in $T=10^{10}$ years. Since PAL 4
lies at such a great galactocentric distance, encounters are so
infrequent that disruption in $10^{10}$ years requires individually 
destructive collisions.
Hence a critical black hole mass $\mbhcrit$ cannot be determined
by the slow heating approximation for PAL 4; the more detailed 
collision-by-collision encounter history of $\S \ref{greensfunction}$ is
warranted for this case. 

As for the rest of Moore's clusters, we see that their values of 
$\mbhcrit$ agree well with the last column of Table 3; the
values obtained from the Gaussian Monte-Carlo simulation are
lower by less than a factor of two. As clusters are heated and 
lose mass, diffusion coefficients increase as $\psic \rightarrow 0$, 
and $\eta_{c0}$ decreases leading to smaller $\mbhcrit$ than for 
fixed cluster structure and mass. Also note that the fractional
width, $\delta \mbhcrit/\mbhcrit$, of the `transition region'
over which $\fdest=0.1 \rightarrow 0.9$
varies as $N^{-1/2}$, as predicted in $\S \ref{themany}$. 

We also simulated a large cluster with $M=10^6 M_{\odot}$, 
$N=1.4\times 10^6$, $\rt=50\ \pc$,
$\rcore=5\ \pc$, and $R_g=10\ \kpc$; hence
$\tilde{\psi}_0=4.85$ and the central one-dimensional velocity 
dispersion $\sigma_0=9.64\ \kms$. Fig. $\ref{gmc}$ shows $P_s$ as a 
function of $\mbh$ for this cluster.
The critical black hole mass
is quite a bit larger than that found with the small clusters from
Moore \shortcite{moore93} because 
$M$ is larger. Note that $\mbhcrit \sim
M/10$ for this cluster, which is on the verge of the limit in which
many black holes will be inside the cluster at any given time.

Klesson \& Burkert \shortcite{kb95} recently argued that Moore's 
determination of $\mbhcrit$ for his set of nine diffuse clusters 
was flawed because deviations from the mean energy input for each
encounter history would be quite large, implying a significant
scatter in the values of $\mbhcrit$. They contend that clusters are
destroyed as a consequence of a small number of encounters with large
energy inputs. Instead, we find that for eight of Moore's nine clusters,
all except PAL 4, disruption is due to numerous
encounters with $b \simless \rcore$ which cause small changes
individually. For these light clusters, 
the expected number of impacts, $N_{\rmn core}(\mbhcrit)$, inside 
$\rcore$ is large for $\mbh=\mbhcrit$.
We find that five of Moore's clusters have $N_{\rmn core}(\mbhcrit)
> 200$, two have $N_{\rmn core}(\mbhcrit) > 50$, 
one has $N_{\rmn core}(\mbhcrit) = 20$, and PAL 4 has
$N_{\rmn core}(\mbhcrit) \simless 1$. So (aside from PAL 4) the 
expected fluctuations in the $N_{\rmn core}(\mbhcrit)$ are $\simless
20$ per cent for eight of Moore's clusters, and $\simless 10$ per cent
for seven of them. Only for PAL 4, which sits at a large galactocentric
radius where few encounters occur, is the evolution near $\mbhcrit$
very stochastic. This case will be discussed in the next section.

\section{ The Large $\mbh$ Limit for the Survival Probability}
\label{largemp}

\subsection{Theory}

	The previous section treated the small $\mbh$ regime in which
many collisions slowly add energy and remove mass from a cluster. This
approximation will be valid for $\mbh<< \mhigh$, the mass at which
the cluster can be disrupted in a single collision. In the opposite limit
for which $\mbh \simgreat \mhigh$, 
any collision within a `destructive
impact parameter', $\bd$, will destroy the cluster. In addition, the
cumulative effect of many collisions outside $\bd$ can also destroy the
cluster. 

	Formulae for the changes in $E$, $M$, and $\nu$ for a single
collision in the tidal limit were derived in $\S \ref{tidallimit}$;
see eq. ($\ref{dmtidal}$), ($\ref{detidal}$), and ($\ref{dnutidal}$). 
The three formulae have
the same scaling with $\eta_c$ and $b$ and only differ by a
factor which depends on $\tilde{\psi}_0$. Hence, in the following 
derivations we only use the formula for energy input to 
derive results explicitly. To find the expressions for disruption by mass
loss or change in $\nu$ one must only substitute $C_E \rightarrow
C_M,C_{\nu}$. 

The energy input can be rewritten as
\beqa
\frac{\btu E}{|E|} & = & C_E \eta_c^2 \frac{\rt^4}{b^4} \equiv
f \left(\frac{\bd}{b} \right)^4
\eeqa
where the impact parameter for $\btu E/|E|=f$ is
\beqa
\bd & =& \rt \eta_c^{1/2} \left( \frac{C_E}{f} \right)^{1/4}.
\label{bd}
\eeqa
This energy input is sharply peaked about the mean
for $b \simeq \bd$ because the impact parameter at which 
$\btu E \simeq \sigma_{ \btu E }$ is larger than $\bd$
 by a factor of $N^{1/2}$.

A cluster has been destroyed if an impact occurs with $b<\bd$.
The critical impact parameter $\bd \geq \rt$ for $\mbh \geq \mhigh
=(M\vrel/\sigma_0)(f/C_E)^{1/2}$. Using $\langle \vrel \rangle
\simeq 1.47 V_c$ [see eq. ($\ref{vreltherm}$)], and 
$\sigma_0 \simeq 1-10 \kms$ implies $\mhigh \sim 10-100 M$.
 
	The mean number of destructive encounters, $N_d$, which occur 
in a time $T$ is given by the integrated rate of encounters inside 
$\bd(\vrel)$:
\beqa
&& N_d(T) = \pi \nbh T \int_0^{\infty} \rmnd \vrel f(\vrel) \vrel
\bd^2(\vrel)
= \left( \frac{C_E}{f} \right)^{1/2} 
\frac{ \rhobh }{M} \pi \rt^2 \sigma_0 T.
\label{nd}
\eeqa
Note that $N_d(T) \propto \fhalo$ but is independent
of $\mbh$. Consequently, when $\mbh>\mhigh$, limits can be
derived for $\fhalo$ but not $\mbh$ \cite{wiel88}.

For given $f$ and $T$, the Poisson probability of no individually
destructive events is given by
\beqa
P_{s} (\fhalo) & =& \exp[ -N_d(\fhalo) ]. 
\label{psold}
\eeqa 
The expression for $P_{s}$ becomes more complicated when we include
the cumulative effect of all the nondestructive encounters outside
$\bd$. As most collisions have $b>>\bd$, they do not significantly change
the cluster properties. These gentle collisions can be modelled using
diffusion coefficients.

The mean rate of change of $E$ due to impacts outside $\bd$ is given by
\beqa
\frac{\dot{E}}{|E|} & =& \nbh\int_0^{\infty} \rmnd \vrel f(\vrel)
\vrel \int_{\bd(\vrel)}^{\infty} 2\pi b \rmnd b 
\left[ f \frac{\bd^4(\vrel)}{b^4} \right] = f \Gamma_d
\eeqa
where $\Gamma_d=N_d(t)/t$.
Similarly, the time derivative of the variance is given by 
\beqa
\frac{ \dot{\sigma}^2_{E}}{E^2} & =& 
\nbh\int_0^{\infty} \rmnd \vrel f(\vrel) 
\vrel \int_{\bd(\vrel)}^{\infty} 2\pi b \rmnd b 
\left[ f^2 \frac{\bd^8(\vrel)}{b^8} \right] =
\frac{1}{3} f^2 \Gamma_d 
\eeqa
where we have ignored the small variance per collision; this variance
is due only to the variable number of encounters in a given time
period.
In addition, collisions inside $\bd$ act as a `sink' for clusters.
The appropriate equation for $P(E,t)$, the probability density at time
$t$ and energy $E$, is
\beqa
\frac{ \partial P}{\partial t} & =& - \Gamma_d P - 
\dot{E}\frac{ \partial P}{\partial E} + \frac{1}{2} \dot{\sigma}^2_{E}
\frac{ \partial^2 P}{\partial E^2}
\eeqa
where again the diffusion coefficients are treated as constants
for simplicity. The solution with $E=E_0$ at $t=0$ is
\beqa
P(E,t) & = & \frac{ \exp[- \Gamma_d t] }
{ (2\pi \dot{\sigma}_E^2t)^{1/2} }
\exp \left[ - \frac{ (E - E_0 - \dot{E}t)^2 }{ 2 \dot{\sigma}_E^2t } 
\right].
\eeqa
This expression reduces to the $\mbh<\mhigh$
solution when $\bd \rightarrow 0$.
The probability of survival is then the cumulative probability that
the change in $E$ has been less than $f|E|$, or
\beqa
P_{s} & =& \frac{1}{2} e^{- \Gamma_d T} \left[ 1 + \erf 
\left( \frac{ f|E| - \dot{E}T}{ \sqrt{2 \dot{\sigma}^2_{E} T}} \right)
 \right] 
= \frac{1}{2} e^{- N_d} \left[ 1 + \erf
\left( \frac{1-N_d}{ \sqrt{2 N_d / 3} } \right)
 \right].
\label{psnew}
\eeqa
For small $N_d$, $P_{s} \simeq \exp(-N_d)$, as in eq. ($\ref{psold}$), 
but for large $N_d$,
$P_{s}  \simeq  (6\pi N_d)^{-1/2} \exp(-5N_d/2)$, 
which decreases much faster. 
Analagous results can be derived for changes
in $\nu$ and $M$.

This decrease in the survival probability below the simple exponential
decline due to the effects of distant encounters has
been noted by previous investigators. In Wielen \shortcite{wiel88}, 
Fig. 5 presented
a numerical determination of the survival probability in the large
$\mbh$ limit for clusters bombarded by both black holes and giant
molecular clouds. Using $N_d(T) \simeq 0.5 \times T/T_{1/2}$, where
$T_{1/2}$ is the time over which half the clusters have dissolved, 
our eq. ($\ref{psnew}$) is a rather good fit to the figure.

The results of this section do not agree with Klesson \& Burkert 
\shortcite{kb95} since the
curves of $P_s$ against $\mbh$ in their Figs. 8, 9, and 10 show
no systematic evidence of an asymptotic $P_{s}$ independent of 
$\mbh$. This is surprising as they have included
exponential adiabatic damping [Spitzer \shortcite{spit58}, 
but see also Murali \& Arras \shortcite{muraliarras98} and
Weinberg (1994) for a more recent version 
of these effects] which would 
decrease the energy input per encounter and hence {\it increase} 
$P_{s}$. The lack of encounters with small energy input in 
Klesson \& Burkert's \shortcite{kb95} simulations could 
result if their method of choosing the 
collision parameters oversampled small $b$ and $\vrel$ compared
to our method.

\setcounter{table}{3}
\begin{table*}
\begin{minipage}{5in}
\label{largembhps}
\caption{Limits on $\fhalo$ in the Large $\mbh$ Limit}
\begin{tabular}{llrrrr} \hline \hline
Cluster Name & Criterion$^{(a)}$ & $\mhigh/M_{\odot}$ &
$\fhalo(P_s=0.9)$ & $\fhalo(P_s=0.5)$ &
$\fhalo(P_s=0.1)$
\\ \hline
AM 4     & $\nu$&   $3.8\times 10^5$& 0.028&  0.14&  0.33  \\
ARP 2    & $\nu$&   $1.6\times 10^7$& 0.014&  0.07&  0.16  \\
NGC 5053 & $\nu$&   $1.8\times 10^7$& 0.011&  0.06&  0.13  \\
NGC 7492 & $\nu$&   $1.5\times 10^7$& 0.029&  0.15&  0.34  \\
PAL 4     & $\nu$&   $1.4\times 10^7$& 0.174&  0.90&  $>$1.0  \\
PAL 5    & $\nu$&   $8.6\times 10^6$& 0.005&  0.02&  0.05  \\
PAL 13   & $\nu$&   $6.4\times 10^6$& 0.053&  0.27&  0.63  \\
PAL 14   & $\nu$&   $8.7\times 10^6$& 0.035&  0.18&  0.41  \\
PAL 15   & $\nu$&   $4.4\times 10^6$& 0.025&  0.13&  0.30  \\
AM 4     & 20\%M& $4.1\times 10^5$& 0.031&  0.20&  0.67  \\
ARP 2    & 20\%M& $7.6\times 10^6$& 0.006&  0.04&  0.14  \\
NGC 5053 & 20\%M& $1.0\times 10^7$& 0.006&  0.04&  0.13  \\
NGC 7492 & 20\%M& $4.9\times 10^6$& 0.009&  0.06&  0.20  \\
PAL 4    & 20\%M& $8.9\times 10^6$& 0.109&  0.72&  $>$1.0  \\
PAL 5    & 20\%M& $5.8\times 10^6$& 0.003&  0.02&  0.07  \\
PAL 13   & 20\%M& $2.1\times 10^6$& 0.018&  0.12&  0.39  \\
PAL 14   & 20\%M& $6.1\times 10^6$& 0.025&  0.16&  0.54  \\
PAL 15   & 20\%M& $3.9\times 10^6$& 0.023&  0.15&  0.50  \\
\hline
\end{tabular}
\footnotetext[0]{ (a) The criterion `$\nu$' means that neither single
destructive events nor many nondestructive events caused $\nu \rightarrow
-0.6$. The criterion `$20\%M$' means that no single events of
$20$ per cent mass loss occured.}
\end{minipage}
\end{table*}

Table 4 contains the allowed values of $\fhalo$ derived from the
fixed cluster approximation of this section 
for the nine clusters found in Moore \shortcite{moore93}. 
Two different destruction
criteria were used: (1) $\nu$ out of bounds in $T=10^{10}$ years
due to the combined influence of single destructive collisions and
multiple nondestrucive
collisions [eq. ($\ref{psnew}$)], and (2) a single episode of $20$ per cent mass
loss in $T=10^{10}$ years [eq. ($\ref{psold}$)].
Also given are
the values of $\mhigh$, the black hole mass above which the cluster
can be destroyed in a single collision, appropriate to each criterion.
Three values were used for the probability of survival:
$P_s=0.1,0.5, $ and $0.9$.

For $50$ per cent survival probability using the criterion $\nu \rightarrow
-0.6$, the limiting values of $\fhalo$ range from $0.02$ (PAL 5) to
$0.90$ (PAL 4). It is unlikely that $\fhalo>0.3$ since $P_s<0.1$
for most of the clusters in that case.
These limits depend sensitively
on galactocentric radius and cluster size, as
is evident in the scatter in the values of $\fhalo$. 

The results of this section will be tested in $\S \ref{greensfunction}$
using the
results of the full Monte Carlo simulations. One wrinkle which appears
in the results is that the different criteria for destruction can
compete with each other so that the results of this section are not
always good approximations to $P_{s}(N_d)$. Only in the cases where one
method of destruction completely dominates over all the others do 
the results agree accurately. 

\section{ Monte Carlo Simulations of Individual Collisions}
\label{greensfunction}

\subsection{ The Set-up}

        Previously, $\mbhcrit$ and $\fhalo$ were calculated
for the two
limiting cases of slow heating (small $\mbh$) and single-event 
destruction (large $\mbh$). The major
simplification for both methods was the neglect of detailed, 
collision-by-collision evolution of the cluster. Both calculations
relied on approximations. The Gaussian
Monte-Carlo method relied on restricting collisions to very small
energy and mass changes individually, and involved averaging over many
collisions to streamline the computations. In this section we
present a Monte-Carlo simulation of individual black hole encounters.
We make no approximations for the energy input besides the impulse
and straight line approximations; consistent with the former, we 
neglect displacement of cluster stars during the encounter. This 
treatment is simpler than an N-body simulation because the evolution
of the cluster is mapped by the sequence of King models. As before, the 
King sequence limits the evolution we allow.

The calculations amount to simulating the `Green's function' of the
cluster. A cluster in a given initial configuration is subjected
to collisions with the halo black hole population. After each 
collision the King model for the cluster is altered, using its
post-collision $E$ and $M$, and assuming $\rt \propto M^{1/3}$. 
An evolutionary history of the cluster is mapped out over $10^{10}$
years, and a final cluster state is found. As this process depends
on many random variables, a range of final states is possible
for each initial state, and many realizations are needed to find
the Green's function.

\subsection{The Probability Distributions for $t$, $b$, and $\vrel$
\label{foftbv} }

In $\S \ref{bhmodel}$, we derived the distribution of relative
cluster-black hole speeds assuming an isothermal black hole halo.
For the Monte Carlo simulations, we need to know the probability
that the {\it next} collision suffered by a cluster occurs a time
interval $t$ to $t+\rmnd t$ after a given encounter, and involves an
impact parameter in $(b,b+\rmnd b)$ and a relative speed in $(\vrel,\vrel
+\rmnd \vrel)$. 
Since the number of collisions inside $b$ is $\propto b^2$, 
we chose a maximum impact parameter $b_{\rmn max}$, and define
$\sigma_{\rmn max} = \pi b_{\rmn max}^2$. Let us first consider collisions
with a single relative velocity, $\vrel$. 
The expected number of collisions in a time $t$ is
\begin{eqnarray}
\ol{N} & = & \int_{0}^{t}\nbh\sigma_{\rmn max} \vrel \rmnd \tau,
\end{eqnarray}
so the Poisson probability that 
there are no collisions for the time interval
$(0,t)$ and then one collision in $(t,t+\rmnd t)$ is
\begin{equation}
\rmnd P_1(t) = \exp \left[-\int_{0}^{t}\nbh\sigma_{\rmn max} \vrel 
\rmnd \tau \right]
\cdot \nbh\sigma_{\rmn max} \vrel \rmnd t.
\end{equation}
Multiplying by the probability $2\pi b db/\sigma_{\rmn max}$ that the
collisions is in $(b,b+db)$ gives
\begin{eqnarray}
&& \rmnd P_1(t,b)  
 =  \left\{ \exp \left[- \int_{0}^{t} \nbh \sigma_{\rmn max}
\vrel \rmnd \tau \right]
\times \nbh \sigma_{\rmn max} \vrel \rmnd t \right\}
\times \left\{ \frac{2\pi b \rmnd b}{\sigma_{\rmn max}}  \right\}.
\end{eqnarray}
The first bracketed factor can be used to choose $t$,
while the second
bracketed factor can be used to choose $b$.

Next, suppose there is a set of black hole populations, each with a
single relative velocity $\vreli$; let $f_i$ be the fraction of all
black holes with $\vreli$. Then the probability that the next 
collision has relative speed $\vreli$ is
\begin{eqnarray}
\rmnd P_1(t,i,b) & = &
\left\{ \exp \left[-\int_{0}^{t}\nbh\sigma_{\rmn max}
\langle \vrel \rangle d\tau \right]
\nbh \sigma_{\rmn max} \langle \vrel \rangle \rmnd t \right\} \times
\left\{ \frac{f_i\vreli}{\langle \vrel \rangle} \right\} \times
\left\{ \frac{2\pi b \rmnd b}{\sigma_{\rmn max}} \right\} 
\eeqa
where $\langle \vrel \rangle  =  \sum_i f_i \vreli$.
Take the continuum limit by substituting
$f_i \rightarrow f(\vrel) d\vrel$ to get
\begin{eqnarray}
\rmnd P_1(t,b,\vrel) & = & \left\{
\exp \left[-\int_{0}^{t} \nbh\sigma_{\rmn max} \langle \vrel \rangle 
\rmnd \tau
\right]
\nbh\sigma_{\rmn max} \langle \vrel \rangle \rmnd t \right\} 
\times \nonumber \\
& & \left\{ \frac{f(\vrel)\vrel \rmnd \vrel} {\langle \vrel \rangle}
\right\}
\times
\left\{ \frac{2\pi b \rmnd b}{\sigma_{\rmn max}} \right\}
\label{bigprob}
\end{eqnarray}
The values
of $t,\vrel$, and $b$ may be chosen from the distributions in the 
first, second, and third bracketed factors, respectively. 

Note that our derivation does not require
$\ol{\Gamma} \equiv \nbh \sigma_{\rmn max} \langle \vrel \rangle$
to be time independent, so
eq. ($\ref{bigprob}$) holds for any orbit, not just 
circular ones.
Also, the velocity distribution found in
the middle brackets is {\it not} $f(\vrel)$, but is the
distribution of speeds measured for the first collision
after some chosen reference time.
For example, the mean of this distribution is
\begin{eqnarray}
\ol{\vrel} & = & V_c \left[
\frac{ 3/2 + x} { \left( x + \frac{1}{2x}\right) \mbox{erf}(x) +
\frac{1}{\sqrt{\pi}}\exp\left(-x^2\right)   }
\right]
\label{vrelscatt}
\end{eqnarray}
where $x=\vcl/V_c$.
As $x \rightarrow 0$, $\ol{\vrel}/V_c \rightarrow 3\sqrt{\pi}/4$, and
as $x \rightarrow \infty$, $\ol{\vrel}/V_c \rightarrow 1$;
for circular orbits eq. ($\ref{vreltherm}$) and
eq. ($\ref{vrelscatt}$) imply
$\ol{\vrel}(x=1) / \langle \vrel(x=1) \rangle \simeq 1.16 $ , or,
$\ol{\vrel}(x=1)\simeq 1.70\times V_c$.

        For globular clusters on circular orbits,
the rate of encounters, 
$\Gamma = \nbh\sigma_{\rmn max} \langle  \vrel \rangle $
inside $\sigma_{\rmn max}$ is constant.
The normalized cumulative distribution for the collision time is then
$P_{\rmn cdf,t}(t)  = 1 - e^{- \Gamma t }$.

\subsection{ Evolving the Cluster   }

Distinct histories are computed for individual clusters at fixed
$R_g$ with identical initial values of $M$, $\rt$, $\rcore$, and $N$,
the number of cluster stars. The mass per star is held fixed at
$m=(M/N)|_{t=0}$ even as $M$ (and $N$) evolve; multiple species with
different particle masses and spatial distributions were not considered.
The maximum impact parameter $b_{\rmn max}$ is also held fixed.

At any time $t \geq 0$, the cluster has known values of $M$, $\rt$, and
$E$ which allow determination of $\psic$, and hence the smooth King
model phase space distribution. From this distribution, we choose random
positions and velocities for the $N$ stars in the cluster at the time
of its {\it next} encounter with a black hole, $t_{\rmn next}>t$ (to
keep fluctuations to a minimum, we actually choose velocities first in
symmetric pairs $\pm \bv$ and then choose positions from their
distribution given the velocity). The values of $t_{\rmn next}$, $b$, and
$\vrel$ are chosen according to eq. ($\ref{bigprob}$), and velocity
kicks $\btu \bv_i$ are computed for each cluster star using eq. 
$\ref{delvimp}$ (by definition $\bb=b \be_x$ and $\bvrel
=\vrel \be_z$).

Given the $\btu \bv_i$, the center of mass velocity kick is
$\ol{\btu \bv}=(1/N)\sum_{i=1}^N \btu \bv_i$; 
even stars which are ejected are included in the
sum. Star $i$ is 
ejected from the cluster if $ \left( \bv_i+ \btu \bv_i -
\ol{\btu \bv} \right)^2 > 2\psi (r_i)$, where $\psi (r_i)$ is
the pre-collision potential at $r_i$. If $N'$ stars remain in the 
cluster then $M'=M(N'/N)$ is the new mass of the cluster. The new
energy of the cluster is the total energy input to the remaining
stars ($i=1,2,...,N'$) plus the contribution from the ejected stars
($j=1,2,...,N-N'$);
\begin{equation}
\btu E=\sum_{i=1}^{N'} m \left(\bv_i \cdot \dbv_i
+ \frac{1}{2} \left| \dbv_i \right|^2
\right) + \sum_{j=1}^{N-N'} m \left(\frac{GM}{\rt} +\psi_j
- \frac{1}{2} v_j^2 \right).
\end{equation}
The new parameters $\rt'$ and $\psic'$
are then found by the steps outlined in
$\S \ref{clustermodel}$. In this way, 
we have a new nondimensional model of the
cluster after the collision.

These steps are repeated until either $t=10^{10}$years or the
cluster `dies'. We keep track of six different conditions for
destruction. For two of these, the simulation is stopped at
once. The first, `$\nu$ out of bounds', 
triggers if $\nu$ goes out of the 
range $(-2.13,-0.6)$ (since all clusters simulated have 
$\psic \simless 5.5$), signalling that no unique 
member of the King sequence can be 
found to represent the cluster. In our simulations, clusters always
went out of bounds at $\nu \rightarrow -0.6$, so below we refer to 
this condition as `$\nu \rightarrow -0.6$'. 
The second criterion for an immediate
halt, called `$M_{20}$', is realized if a {\it single} event of 
$|\btu M| \geq 0.20 M$ occurs which would greatly distort the
cluster and invalidate our approximation of small $\btu M/M$. There
are four other criteria which are recorded {\it the\  first\  time\  
they\ occur}, but do not stop the simulation. These are:
`$M_{10}$', a single occurence of $10$ per cent mass loss;
`$\frac{1}{2}M$', decrease of the mass of the cluster
to half its original value; `$\frac{1}{2}E$', 
change of the energy of the cluster up or down by half of its original 
value; and `$\frac{1}{2}E/M$', change of the quantity $E/M$ up or down
by half of its original value. Survival of the cluster will be 
called criterion `$s$'. Note that the criteria
$M_{10}$, $\frac{1}{2}E$, $\frac{1}{2}M$, and $\frac{1}{2}E/M$
can happen repeatedly before $\nu \rightarrow -0.6$, $M_{20}$, or $s$,
but are only recorded the first time they occur. The two conditions
$\frac{1}{2}E$ and $\nu \rightarrow -0.6$ correspond most closely to 
the criteria used by previous investigators, but since our simulations
include mass loss the correspondence is not exact. It is possible
for a cluster to be destroyed in less than $10^{10}$ years by $M_{20}$
or $\nu \rightarrow -0.6$ even though $\frac{1}{2}M$, $\frac{1}{2}E$,
$M_{10}$, or $\frac{1}{2}E/M$ might not have had a chance to occur
yet. Moreover, $M_{20}$ may occur before $\nu \rightarrow -0.6$. When
analyzing the results, the competition among the various criteria
must be kept in mind. 

This simulation ignores a number of possibly important effects. The most
restrictive approximation is the use of the King sequence to model the
evolution of the cluster. Relatively small changes in mass and energy
may lead to $\psic \rightarrow 0$, so a normal cluster
could have a lifetime larger than is found here. For example, a 
$M \sim 10^6 \msun$ cluster
could lose $30$ per cent of its binding energy and mass, 
but no longer be fit well by a King model; yet you would still have a 
$M \sim 7 \times 10^5 \msun $ cluster. Our treatment assumes that
clusters become unstable to rapid dissolution when $\psic \rightarrow 0$.

To narrow the focus to the effects of halo black holes, we have
neglected the disk and bulge components of the galaxy;
destructive effects such as disk shocking and
collisions with molecular clouds
have also been suppressed entirely. 
Internal evolution of the cluster and non-spherical
galactic tidal fields are not treated. Clusters are kept at single
$R_g$, so time dependence of the density of halo black holes and 
relative velocites along cluster orbits is neglected.
Lastly, we will make fractional errors of order $\btu M/M$ in our method.
 
Nevertheless, our model advances previous attempts to constrain
properties of a hypothetical population of halo black holes via
their effects on clusters. We include mass loss which, as was shown
in $\S \ref{tidallimit}$, contributes significantly to cluster
heating, sometimes slightly more than `Spitzer' heating. Moreover, 
since the mean mass loss is comparable to the mean energy input,
the evolution of $\nu \propto E/M^{5/3}$ is driven by both 
$\btu E$ and $\btu M$. Here no simplifications of the energy input
are employed except the impulse approximation and the straight line
orbit approximation. Hence, the Monte Carlo method for finding 
$\btu E(b)$ includes the important `shocking' effect of the 
$\bv \cdot \dbv$. Lastly, the correct expression for the
rate of collisions in eq. ($\ref{fofvrel}$) is used. We
also note that many criteria for the cluster to be disrupted have been
used in the past; the most popular is $\btu E(t)/|E|=1$. This choice
is usually implemented with no regard to mass loss and evolution. Here,
we test a variety of criteria for cluster destruction in order to find
the most restrictive.

        One last technical detail which must be discussed is the value of
$b_{\rmn max}$, the maximum impact parameter for our 
scattering experiment. Since the number of collisions that must
be simulated $\propto b_{\rmn max}^2$, it is essential to choose as small
a $b_{\rmn max}$ as possible without losing accuracy. First
examine the small $\mbh$ case in which the cluster cannot be destroyed in
a single pass. In this slow heating limit, 
$ \btu E(b) \propto b^{-4}$ outside the core
and $\int_0^{b_{\rmn max}} 2\pi b \rmnd b
\btu E(b)$ converges as $b_{\rmn max}^{-2}$. 
Nearly all of the heating results from
penetrating encounters and $b_{\rmn max}\simeq \rt$ will be a good 
approximation. However, in the large $\mbh$ case there is a critical 
impact parameter $\bd^2\propto \mbh$ 
inside of which the cluster is destroyed, and we must choose
$b_{\rmn max}> \bd \propto \mbh^{1/2}$. An estimate for $M_{high}$,
the value of $\mbh$ above which $b_{\rmn max}$ must increase 
$\propto \mbh$, is discussed in $\S \ref{largemp}$.

\subsection{ Description of the Simulations }

	The time necessary to realize a cluster with $N=1000$ stars
and compute the energy input was $\sim 1 \rm sec$ on a Sun workstation. 
The computation time needed to map out the history for one cluster is
$\ol{N_{ev}} \times 1 \rm sec 
\sim (80 hrs) (M/\mbh)(10pc/\rt)$. This severely limited
practical choices for $N$ and $m$. In order for the simulations to be 
realistic, we chose to simulate the full range of $\mbh$ for 
AM 4 only, using $m=0.7
M_{\odot}$ and $N=1000$ initially.

To study much more massive clusters, the time limitation would force
the number of stars used to be a small fraction of the physical value.
As a consequence, the variance in energy input and mass loss would
be unrealistically large in the simulations, and hence the 
Green's function would be spread over too broad a range of
final states.
Too small a number of stars will decrease
the number of stars in the portion of phase space from which particles
are ejected resulting in an incorrect evaluation of mass loss,
especially if the total mass lost in the simulation is small. Uneven
sampling of phase space, arranged with finer sampling near the 
escape surface, could alleviate this problem.

For a given $\mbh$, $b_{\rmn max}$, and $\fhalo$, a number $N_t$ trials 
were performed. The number of clusters destroyed, $N_{\rmn dest}$,
by each of the six criteria was recorded and the 
fraction $\fdest=N_{\rmn dest}/N_t$ of clusters destroyed computed.
In addition to $\fdest$ for each criterion, the sum of $\fdest(\nu
\rightarrow -0.6)+\fdest(M_{20})$ is computed because
the sum of the fraction destroyed by these two conditions is not
subject to competition effects which appear for the six criterion 
separately. 

At least two runs with different $b_{\rmn max}$ were done for each
cluster at a certain $\mbh$. The runs with the larger $b_{\rmn max}$
are presented here. The variation in $\fdest$ for the two
runs was in all cases within the error bars shown in the figures.  
The value of $b_{\rmn max}$ used for the results presented here
was $b_{\rmn max}=2\rt$ for 
$\mbh \simless 100 M$ and $b_{\rmn max}=2\rt(\mbh/100\msun)^{1/2}$ for
$\mbh > 100 M$. These values of $b_{\rmn max}$ 
are much larger than is needed, as $\mhigh \simgreat 4 \times 10^5
M_{\odot}= 570M$ (see Table 3), 
and the destructive radius
is not outside $\rt$ until $\mbh > \mhigh$. 

The number, $N_t$, of trials ranged between $N_t=40$ for 
small $\mbh$ and $N_t=1000$ for large $\mbh$. The number of 
trials was restricted by the run time, which was greatly increased 
for small $\mbh$ because of the large number of collisions
$\propto \fhalo/\mbh$. Bayesian methods were used to compute $P_s$
and its uncertainty. The error bars in the figures span the range of
$P_s$ around the peak of its posterior containing $68$ per cent probability.

\subsection { Results and Discussion }

The first set of runs was for $\fhalo=1$. The fractions of trial
clusters destroyed by our various criteria are
plotted in Fig. $\ref{fmcam4}$.

\setcounter{figure}{8}
\begin{figure}
 
\begin{picture}(432,648)(0,0)

\put(0,432){\epsfxsize=3in
\epsffile{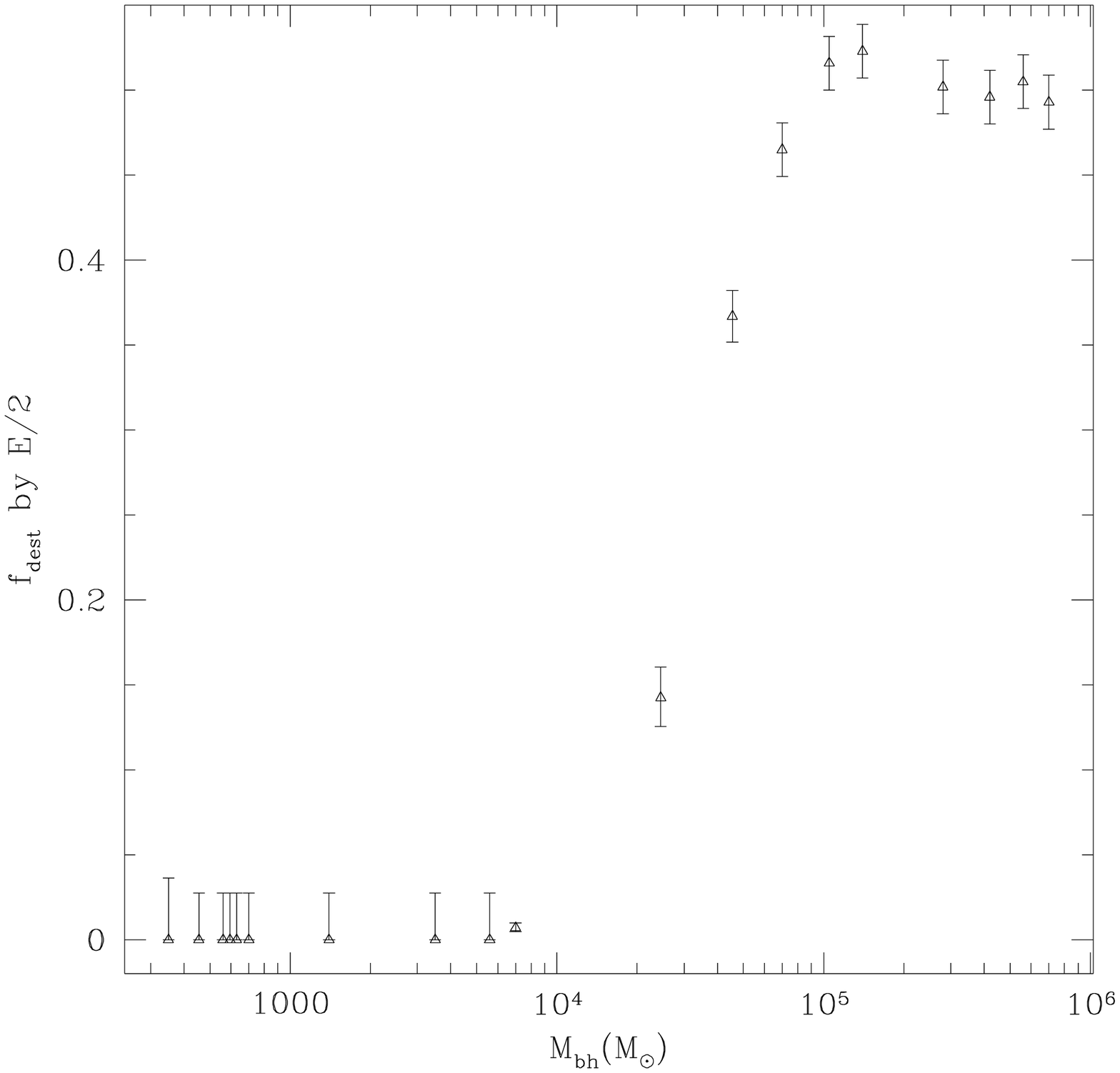}}
 
\put(216,432){\epsfxsize=3in
\epsffile{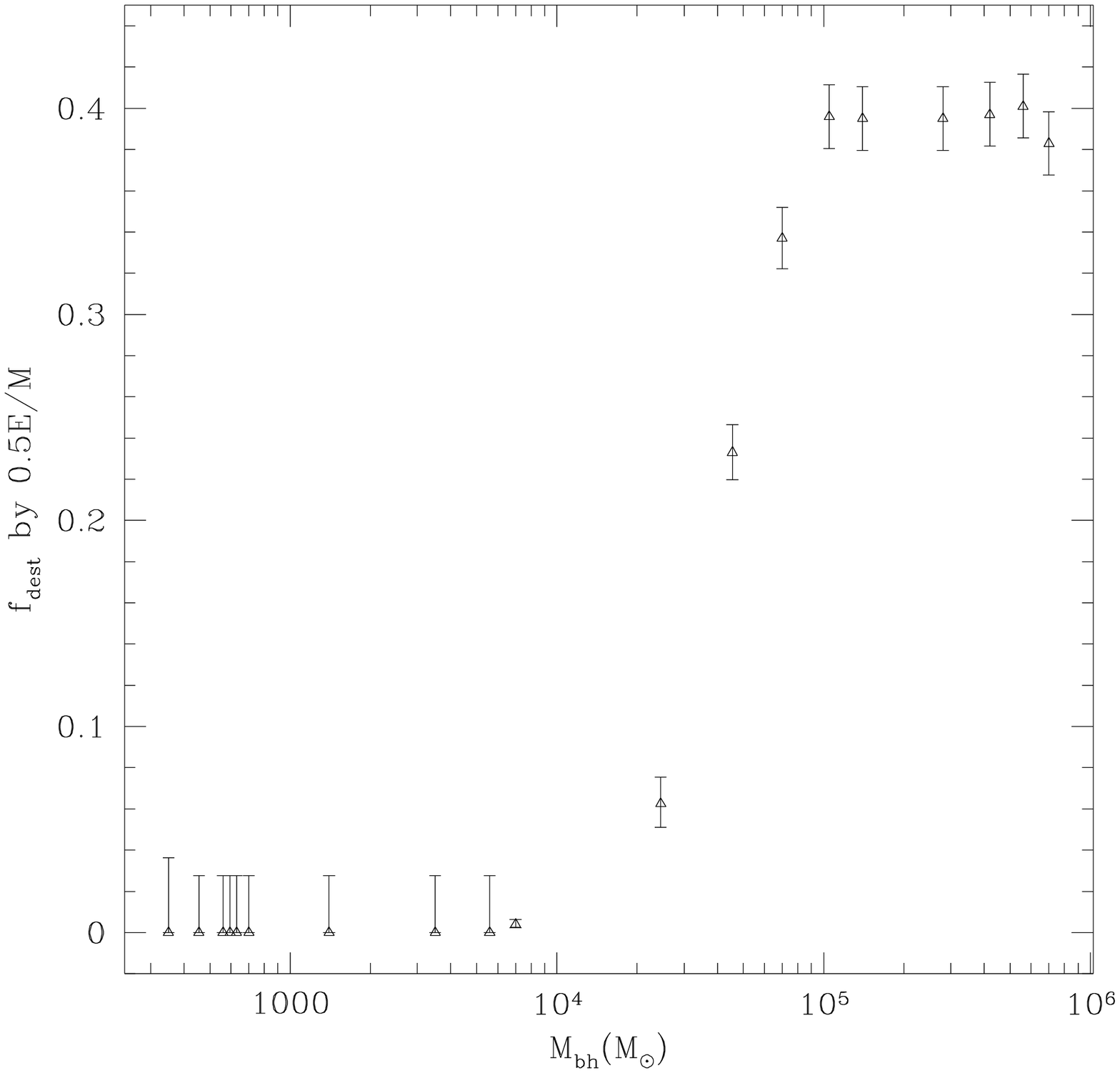}}
 
\put(0,216){\epsfxsize=3in
\epsffile{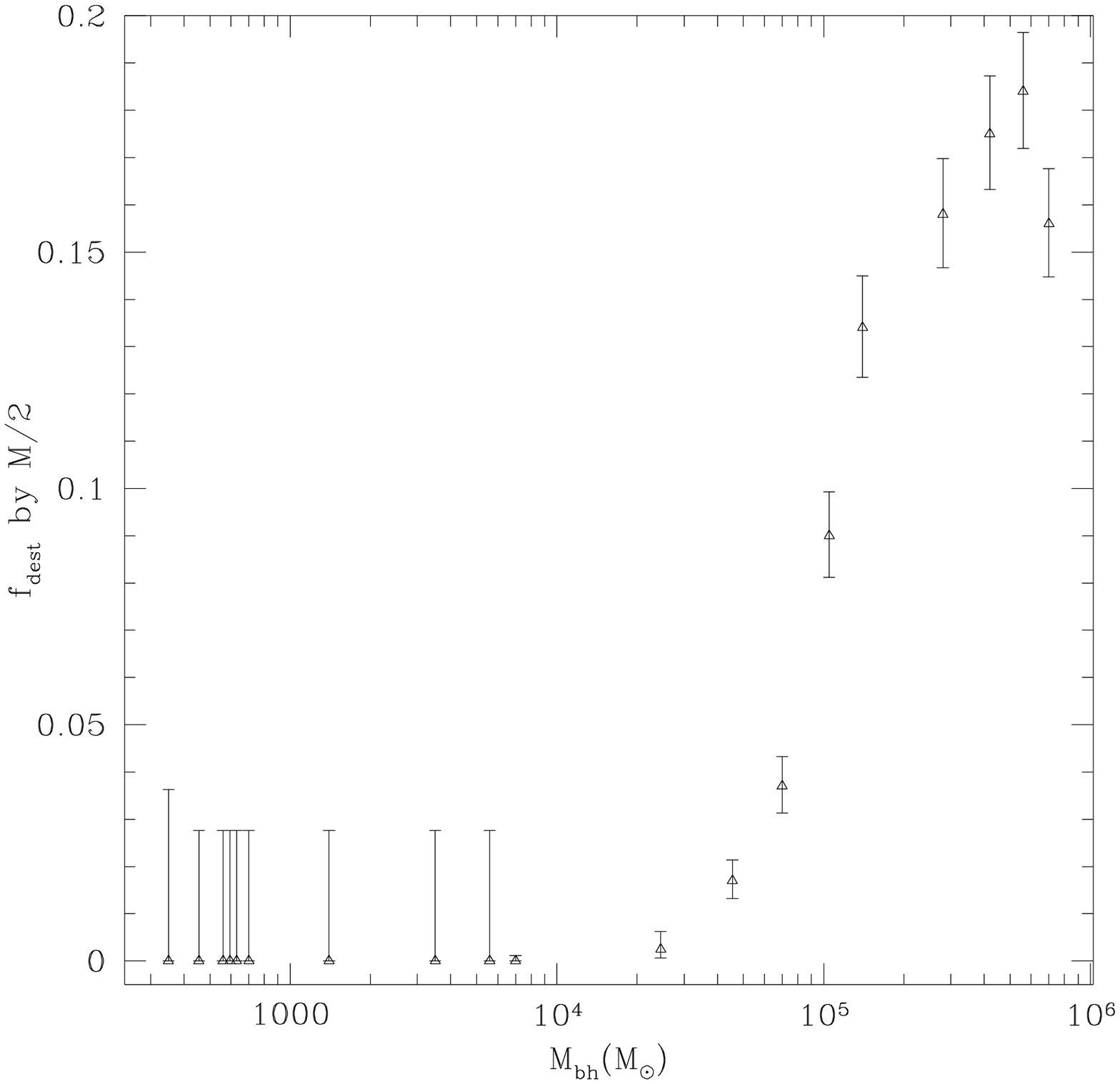}}
 
\put(216,216){\epsfxsize=3in
\epsffile{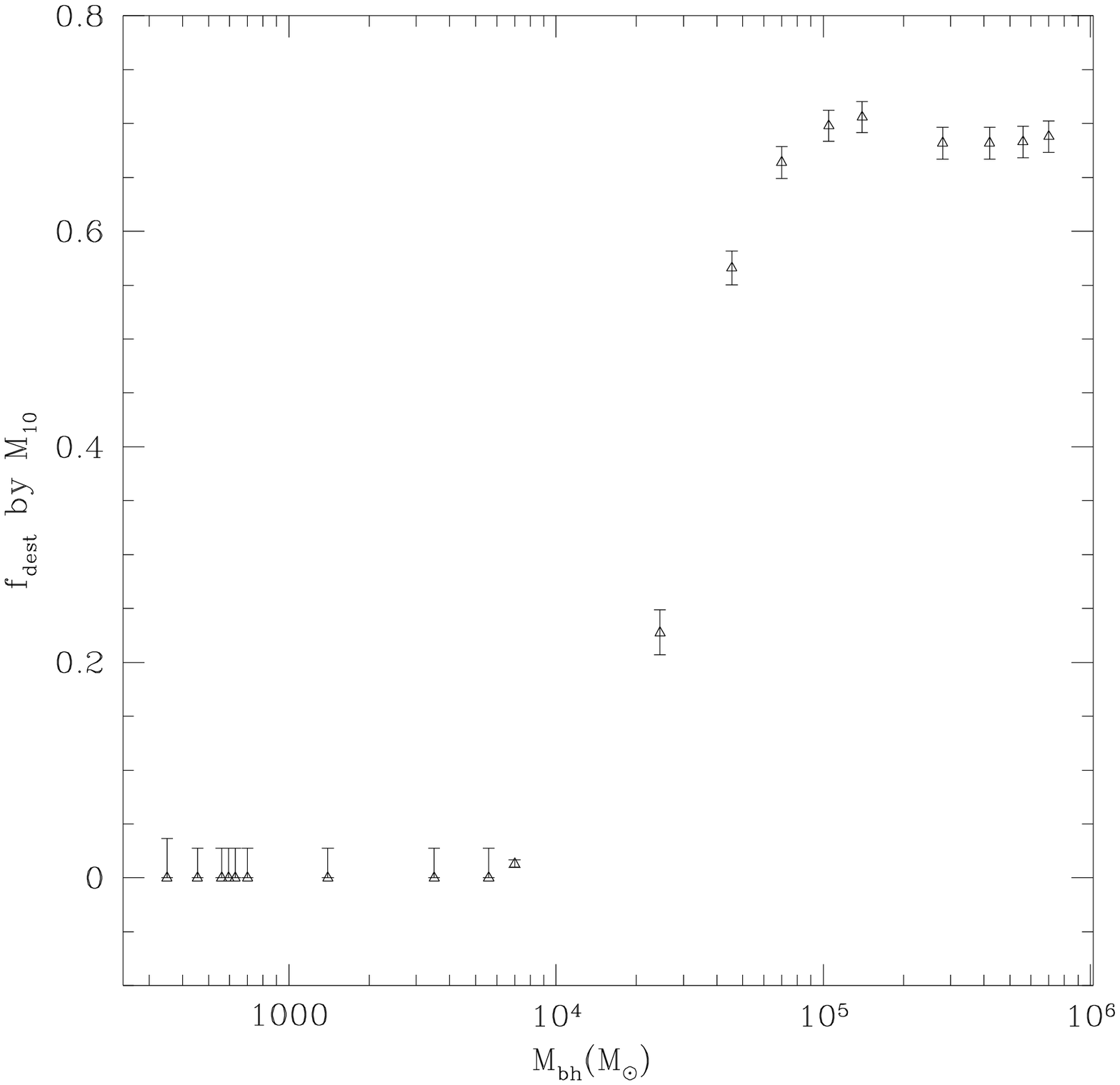}}
 
\put(0,0){\epsfxsize=3in
\epsffile{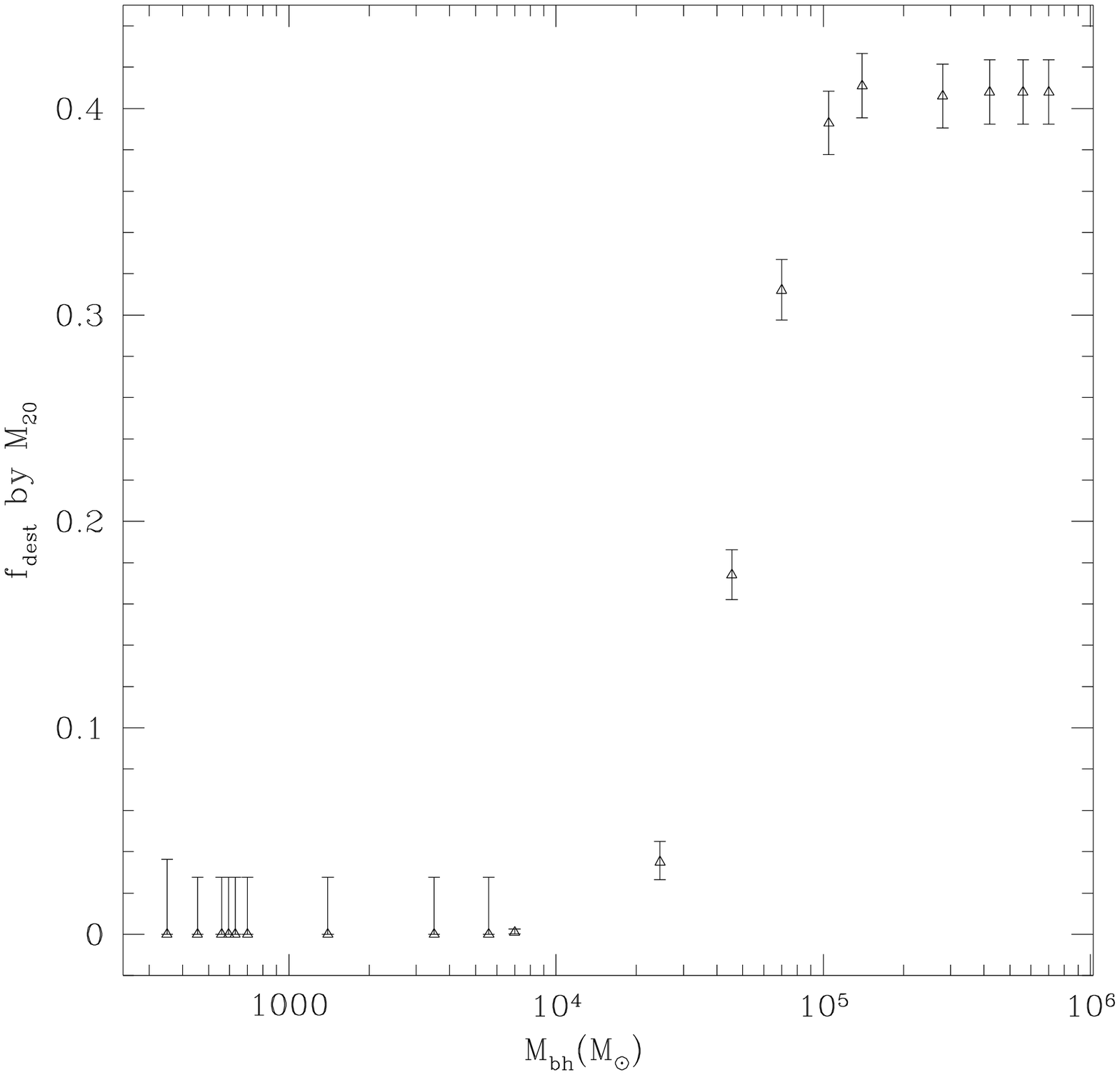}}
 
\put(216,0){\epsfxsize=3in
\epsffile{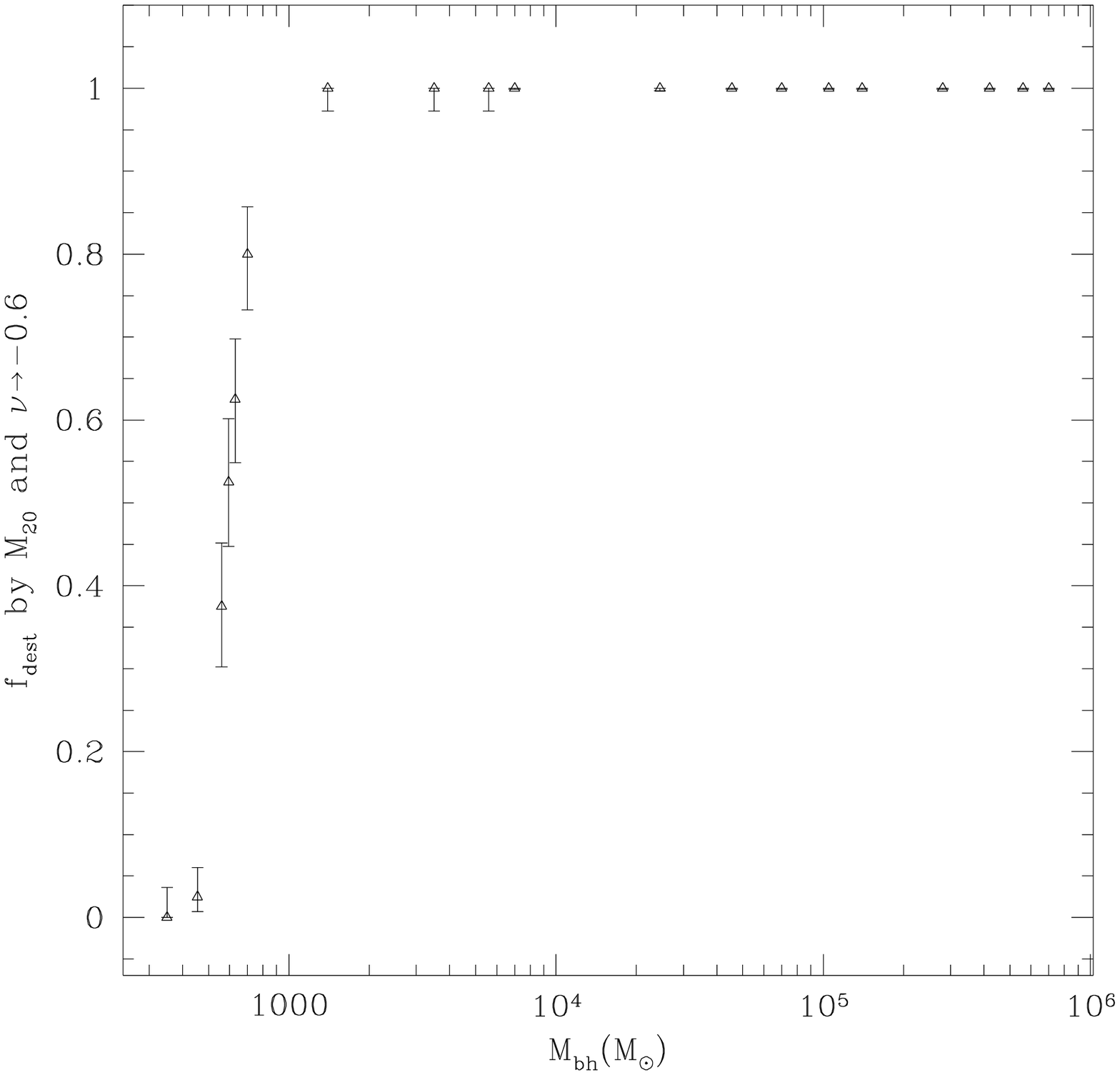}}
 
\end{picture}
\end{figure}
\begin{figure}
\begin{picture}(216,216)(0,0)
 
\put(108,0){\epsfxsize=3in
\epsffile{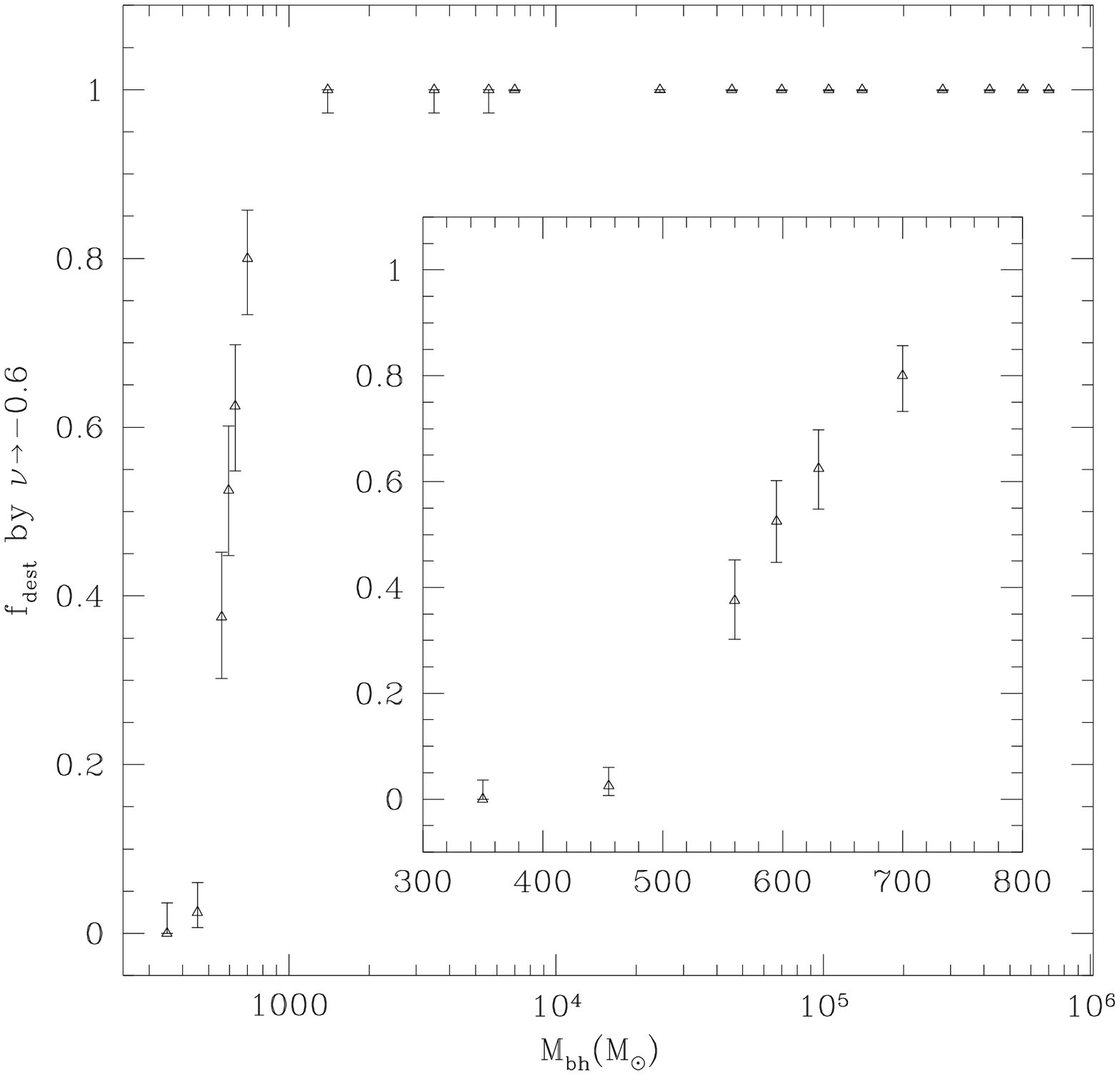}}
 
\end{picture}
 
\caption{ Fraction of initial AM 4 clusters destroyed according to
the various criteria for destruction as a function of $\mbh$.
These curves were made using the full Monte Carlo method.}

\label{fmcam4}
 
\end{figure}

The dominant destruction mechanism by far is $\nu \rightarrow -0.6$. 
For this mechanism, the black hole mass at which $50$ per cent 
of the clusters were destroyed is
\beqa
\mbhcrit(50\%) & \simeq & 600 M_{\odot} = 0.86 M 
\eeqa
The other destruction mechanisms only come into play once
$\mbh \simgreat 5 \times 10^{4} M_{\odot}$.
At this point, $\nu \rightarrow -0.6$ still destroys
$100$ per cent of the clusters (as predicted in Table 4 for 
$\fhalo=1.0$); even the approximately $40$ per cent that die via $M_{20}$
do so in the same encounter that pushes $\nu \rightarrow -0.6$.
If $\nu \rightarrow -0.6$ were not stopping the  simulations simulations
the other destruction criteria would be more important, and $\fdest$
would not necessarily asymptote to small values as in Fig. 
$\ref{fmcam4}$.

We can compare the results in Fig. $\ref{fmcam4}$ for the criterion
$\nu \rightarrow -0.6$ with the Gaussian
Monte-Carlo simulations in Fig. $\ref{gmc}$. There the critical
black hole mass is $\mbhcrit(50\%) \simeq 650 M_{\odot}$, which
is different from the full Monte-Carlo simulation by $\simless 10$ per cent.
Considering that the error bars in $\fdest$ are of order $10-15$ per cent,
the two numbers agree. In addition, the mass range for which
$0.9 \geq \fdest \geq 0.1 $ is $\delta \mbhcrit \sim 250 M_{\odot}$
for both the Gaussian Monte-Carlo
and full Monte-Carlo cases. The agreement
of these two methods shows that the assumptions of (1) a Gaussian 
distribution of energy input and mass loss, and (2) smooth Fokker-Planck
evolution, are accurate in the small $\mbh$ regime. In addition, 
the ansatz for the variance of the number of stars ejected in 
$\S \ref{newcks}$ gave similar results are the full Monte-Carlo
treatment, as evidenced by the comparable widths $\delta \mbhcrit$.

\setcounter{figure}{9}
\begin{figure}
\begin{picture}(216,216)(0,0) 
  
\put(108,0){\epsfxsize=3in 
\epsffile{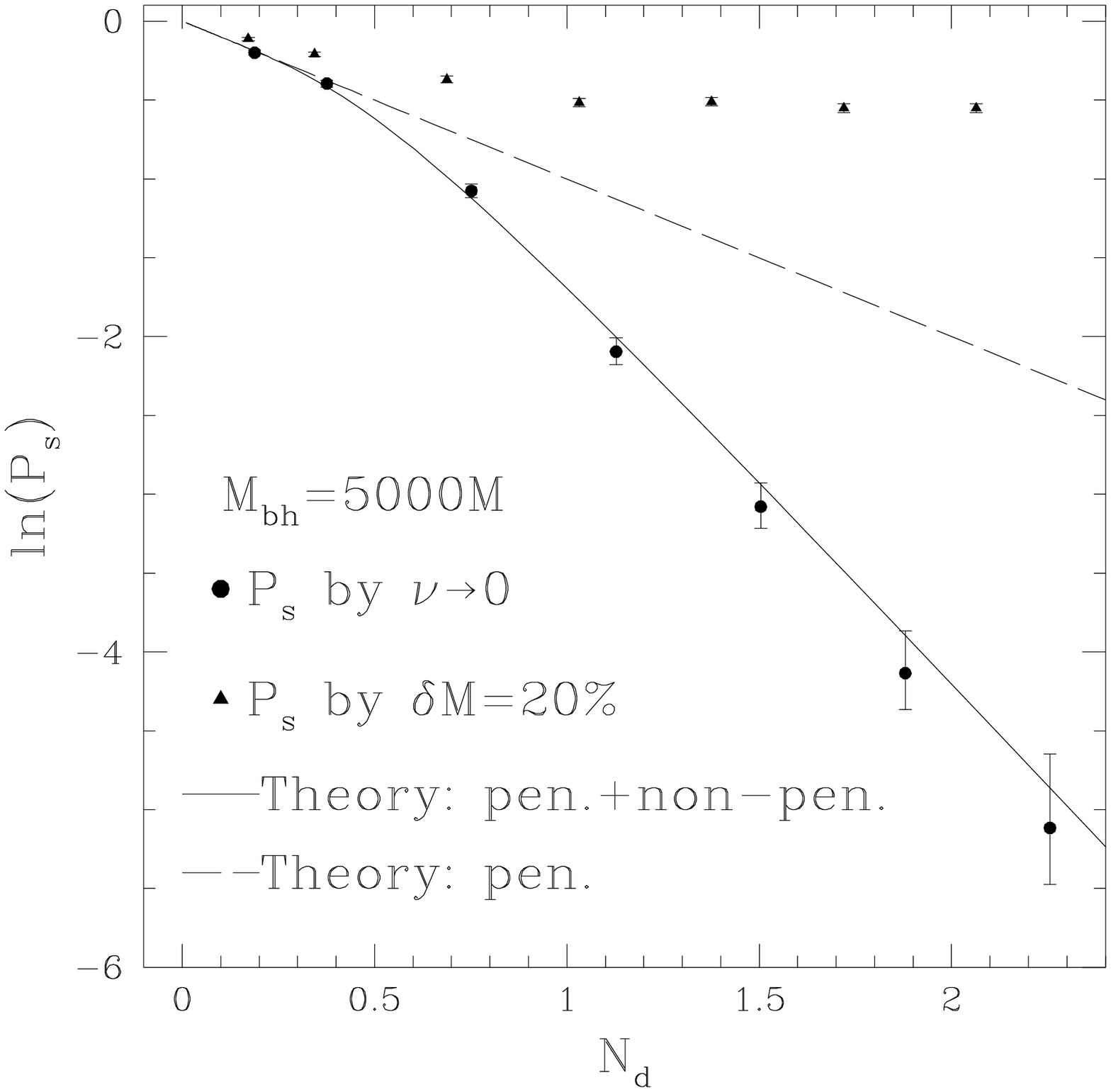}}  
\end{picture} 
  
\caption{ Comparison of full Monte Carlo simulations against theory for
AM 4 with $\mbh=5000 M$.
The circles are the points of $P_s$ vs. $N_d$ for the destruction
criterion $\nu \rightarrow -0.6$. The triangles are
for $P_s$ vs. $N_d$ with the destruction criterion of
$20$ per cent mass loss in a single encounter.}
 
\label{fmcam4largembh5000} 
  
\end{figure} 

\setcounter{figure}{10}
\begin{figure}
\begin{picture}(216,216)(0,0) 
  
\put(108,0){\epsfxsize=3in 
\epsffile{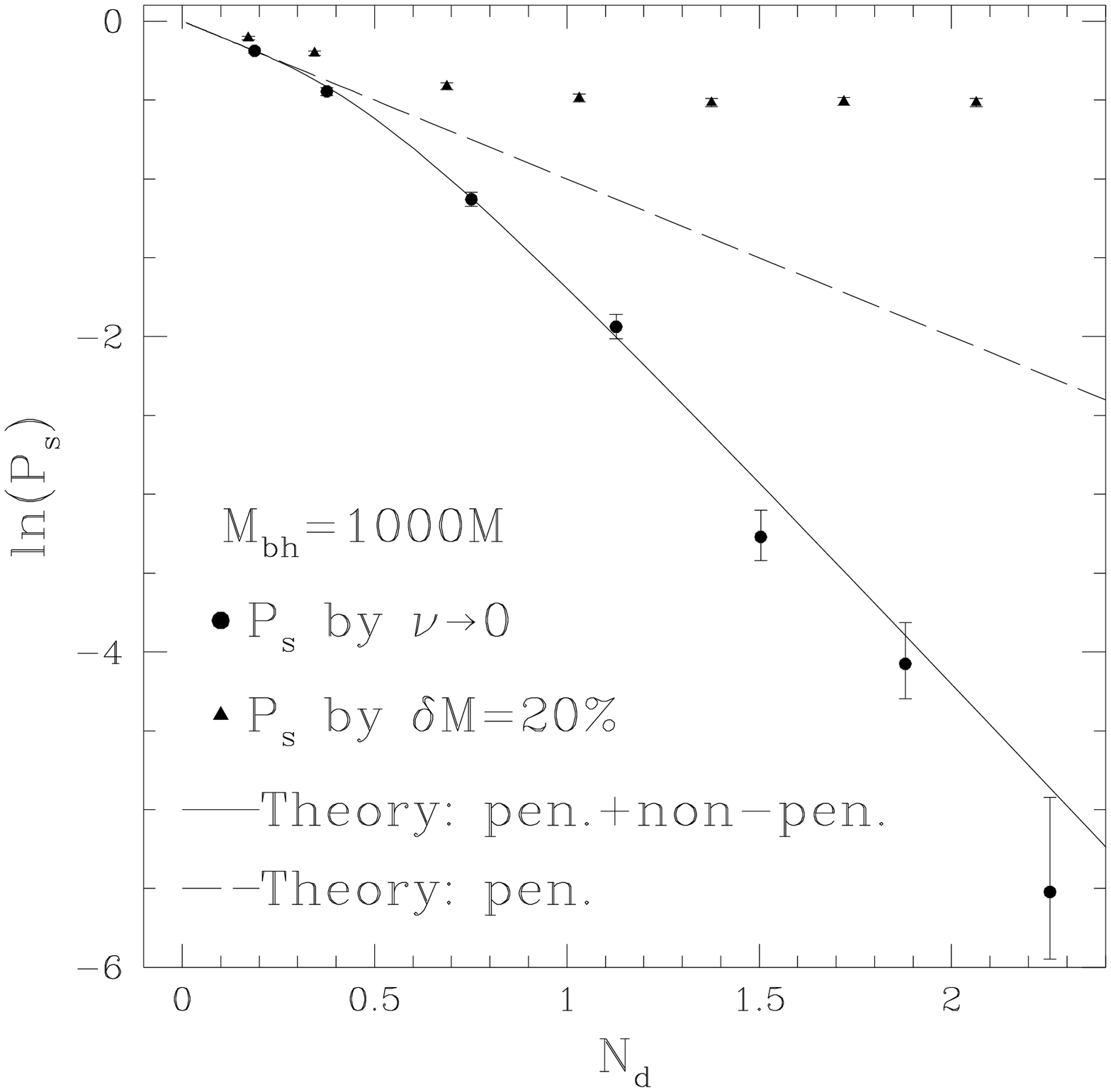}}  
\end{picture} 
  
\caption{ Comparison of full Monte Carlo simulations against theory for
AM 4 with $\mbh=1000 M$.
The circles are the points of $P_s$ vs. $N_d$ for the destruction
criterion $\nu \rightarrow -0.6$. The triangles are
for $P_s$ vs. $N_d$ with the destruction criterion of
$20$ per cent mass loss in a single encounter.} 
 
\label{fmcam4largembh1000} 
  
\end{figure} 

\setcounter{figure}{11} 
\begin{figure}
\begin{picture}(216,216)(0,0)
 
\put(108,0){\epsfxsize=3in
\epsffile{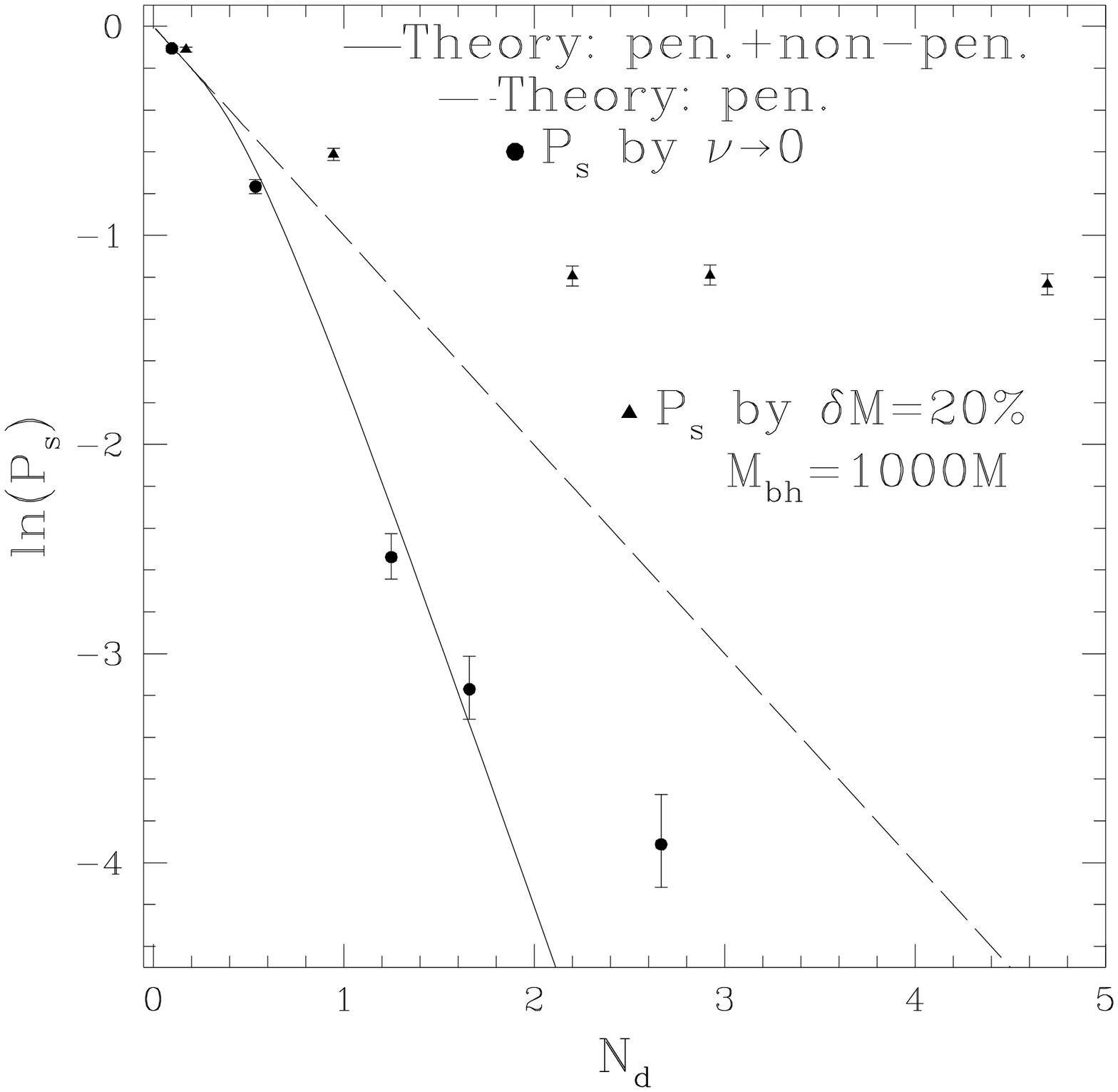}}
\end{picture} 
 
\caption{ Comparison of full Monte Carlo simulations against theory for
NGC 5053 with $\mbh=1000 M$.
The circles are the points of $P_s$ vs. $N_d$ for the destruction
criterion $\nu \rightarrow -0.6$. The triangles are
for $P_s$ vs. $N_d$ with the destruction criterion of
$20$ per cent mass loss in a single encounter.}
 
\label{fmcngc5053largembh1000} 
 
\end{figure} 

The second set of runs explores the large $\mbh$ limit of 
$\S \ref{largemp}$. Two runs for AM 4 at $\mbh=1000M=7 \times 10^5
M_{\odot}$ and $\mbh=5000M=3.5 \times 10^6M_{\odot}$ were performed
as discussed in the previous section. The results are shown in Fig. 
$\ref{fmcam4largembh5000}$ and $\ref{fmcam4largembh1000}$. In addition, 
we also tried NGC 5053 since it has
a fairly large mass, $M=37700 M_{\odot}$, and yet it restricts
$\fhalo$ to be small. Because 
the run time would have been prohibitively long using the correct
number of stars in this cluster, we chose $N=1000$ for NGC 5053. As
discussed previously, this should not change the values of $P_s$ much 
since we will be in the limit in which the collisions are quite
strong, and the variances are expected to be small compared to the 
means. A single black hole mass of 
$\mbh=1000M=3.77\times 10^7 M_{\odot}$
was used to get the results shown in Fig. $\ref{fmcngc5053largembh1000}$.

The values of $\fhalo$ used in Fig. $\ref{fmcam4largembh5000}$
and $\ref{fmcam4largembh1000}$ are (from left to right), $\fhalo=
0.05$, $0.1$, $0.2$, $0.3$, $0.4$, $0.5$, $0.6$. 
The values of $N_d$ computed with eq.($\ref{nd}$) for 
each cluster will depend on the criterion for destruction through
the values of $f$ and $C$ chosen. This is why the two curves for the
two criteria $\nu \rightarrow -0.6$ and $\btu M/M=20$ per cent do not have the
same values of $N_d$ for each data point of a given $\fhalo$.
We have converted the values of $\fhalo$ to the values of $N_d$ shown
in the graphs using eq. ($\ref{bd}$) and ($\ref{nd}$). The values of 
$\fhalo$ used for NGC 5053 in Fig. $\ref{fmcngc5053largembh1000}$
are (from left to right), $\fhalo=0.01,0.055,0.128,0.17,0.273$.  

As the results for AM 4 and NGC 5053 were a bit different, let us
discuss AM 4 first.
Notice that the full Monte-Carlo $P_s$ curves in Figs.
$\ref{fmcam4largembh5000}$ and $\ref{fmcam4largembh1000}$ agree quite
well with each other. This was predicted in $\S \ref{largemp}$ from
the fact that $N_d$, the number of destructive encounters, is 
independent of $\mbh$. Next notice that the curves for the 
$\nu \rightarrow -0.6$ destruction criterion agree quite well with 
eq. ($\ref{psnew}$) which was used to draw the solid lines in the figures.
The predicted $P_s$ for $\btu M/M=20$ per cent did
{\it not}, however, agree very well with eq. ($\ref{psold}$). This can
be attributed to the fact that $\nu \rightarrow -0.6$ generally
occurs well before any collision with $\btu M/M=20$ per cent. In fact, the
good agreement of the numerical results with eq. ($\ref{psnew}$)
for $\nu \rightarrow -0.6$ is partly a consequence of the relative
rarity of collisions with $|\btu M|/M=0.2$ for AM 4.

For NGC 5053, the fraction of clusters destroyed by $\btu M/M=20$ per cent
is much larger, and hence neither curve agrees well with 
either eq. ($\ref{psnew}$) or ($\ref{psold}$).
For $N_d \simless 1.5$, since few
clusters were destroyed by $\btu M/M=20$ per cent the agreement for 
$\nu \rightarrow -0.6$ was good, but competion between the two
destruction criteria is apparent for $N_d \simgreat 1.5$.
However, note that the survival probability against 
$\nu \rightarrow -0.6$ is smaller than for $|\btu M|/M=0.2$. Since
the latter occurs {\it only} in a single catastrophic encounter, 
whereas the former also involves the cumulative effect of numerous
gentle collisions, it is incorrect to ascribe cluster destruction at
large $\mbh$ entirely to the effect of the single most destructive
collision. 

\section { Conclusions and Prospects for Future Work \label{conclusions} }

Our results confirm the basic picture of cluster evolution proposed by
Wielen \shortcite{wiel88} 
in which there are two distinct regimes
based on the size of the black hole mass; for $\mbh$ too small to 
destroy the cluster in a single collision, the evolution can be modelled
by a smooth average over many collisions while for $\mbh$ large 
enough to destroy the cluster in a single collision, 
the survival probability over $10^{10}$
years depends both on the heating from non-destructive 
encounters and the stochastic effect of the individually destructive
collisions. 

Several technical improvements have been made over previous 
investigations for the $\mbh \ll \mhigh$ limit. 
The cluster's structure was allowed to change by 
evolving it along a King sequence; comparison of the values of 
$\mbhcrit$ in Fig. 8 and the last column of Table 3 
show that evolution accelerates the dissolution
process for Moore's weakly bound clusters, leading to a stricter limit.
The inclusion of mass loss gave energy changes comparable to the
usual $\delta v^2/2$ heating. Finite $N$ effects have been included, 
and shown to be relatively unimportant in determining $\mbhcrit$ 
for the $\mbhcrit \ll \mhigh$ case since 
`the range of final states' is quite small when the cluster is evolved
over times long enough to disrupt it. The Fokker-Planck
model ($\S \ref{gfp}$) for the evolution of an ensemble of initial
clusters agreed closely with the full Monte-Carlo simulations
($\S \ref{greensfunction}$) showing that the the `slow heating'
approximation is indeed an accurate representation of the evolution.
Indeed, the simple model of $\S \ref{gfp}$ with constant diffusion
coefficients given by the convenient formulas in Table 2 gives 
quick, analytical results accurate to within a factor of a few (Table 3).

The final fate of the clusters studied in this paper was always 
dissolution. This occured for three reasons. First, we ignored
internal relaxation which tends to drive a cluster to core 
collapse. The role of internal relaxation is currently being studied
(Murali et al. 1998).
Second, when the cluster is heated slowly by many penetrating
encounters, the final fate is dissolution independent of cluster
concentration. Third, in the large $M_{bh}$ limit in which all non-
destructive encounters are tidal, clusters with $\psic \simless 5.5$
dissolve and those with $\psic \simgreat 5.5$ core collapse. All the
clusters examined here had $\psic \simless 5.5$.

Our results for the slow-heating, $\mbh \ll \mhigh$ limit are given
in Table 3, Fig. $\ref{gmc}$, and Fig. $\ref{fmcam4}$. The strictest
limit on $\mbh$ comes from the full Monte-Carlo calculation for
AM 4 with $\mbhcrit \simeq 600 M_{\odot}$. For the $\nu$ out of bounds 
criterion, several of Moore's clusters are disrupted at $\mbh \sim
1000 M_{\odot}$ and seven of nine are disrupted for $\mbh \sim
6000M_{\odot}$. For the $\btu E/|E|=0.5$ criterion, six clusters die
at $\mbh \sim 5000$ and two die for $\mbh \sim 1500$. To summarize,
in this regime it is extremely unlikely that all the clusters could
survive unscathed for $10^{10}$ years if $M_{bh}$ is greater than a 
few thousand solar masses.

For the $\mbh \gg \mhigh$ limit, the probability of survival $P_s$
does not tend to zero as $\mbh \rightarrow \infty$, but instead 
asymptotes at a nonzero value (which does not have to be small). A 
simple Fokker-Planck model to determine $P_s$ 
has been developed ($\S \ref{largemp}$) in which
both the close, destructive collisions and the distant, non-destructive
collisions are included. The distribution of energies for an initial
ensemble of clusters is shown to depend only on the expected number of 
destructive encounters for the cluster as a function of time, and the
resultant $P_s$ gives good agreement with the
full Monte-Carlo simulations (when the competition of the various survival 
criteria is small). Inclusion of the distant,
non-destructive encounters leads to smaller values of $P_s$ which 
in turn gives much tighter limits 
on the allowed fraction of the halo in black holes.  

The results for the $\mbh \gg \mhigh$ limit are presented in Table 4
and Fig. $\ref{fmcam4}$, $\ref{fmcam4largembh5000}$, 
$\ref{fmcam4largembh1000}$, and $\ref{fmcngc5053largembh1000}$.
For a particular cluster and a given value of $P_s$, one can place
limits on $\fhalo$. For $P_s=0.5$, Table 4 gives values of $\fhalo$
for Moore's clusters ranging from $0.02$ to 0.9. In this ``tidal" regime,
it is unlikely that
$\fhalo > 0.3$ since then $P_s<0.1$ for most of Moore's clusters. 

The existence for $10^{10}$ years 
of the set of nine tenuous globular clusters studied by
Moore \shortcite{moore93} places severe restrictions the 
allowed mass and fraction of the halo mass
in massive black holes. 
Left unanswered in our
paper is the question of whether Moore's clusters were {\it initially}
similar in size to what we see today, of if they have been `whittled
down' to their present small stature. Indeed, even in isolation,
the least massive of these clusters, AM 4, PAL 13, and perhaps NGC 7492
might be expected to have evaporated in $\sim 10^9$ years,
suggesting that they are relatively young 
(Murali 1998; Murali et al. 1998; Spitzer 1987).
To answer this question fully, 
one must evolve a representative population of {\it initial} clusters
and attempt to reproduce the observed population today. The work for
this project has already begun \cite{murali98}.
Significantly tighter limits may result from these
new investigations by the introduction of other sources of cluster
evolution, such as internal relaxation, evaporation, and disk shocking.

\section*{Acknowledgments}
We would like to thank David Chernoff and Chigurupati Murali for 
many informative discussions, 
Mark Scheel for invaluable computer assistance, Jamie Lombardi
for helpful comments, and Peter Brooks for the title. This research 
was supported in part by NSF grants no. AST 93-15375 and
AST 95-30397.

\appendix
\section{ The Center of Mass Velocity Kick \label{cmkick} }

For a continuous cluster with spherical mass density $\rho(r)$, the 
velocity kick to the cluster center of mass takes on a very simple form.
For $\bb=b \be_x$, $\bvrel=\vrel \be_z$, and
projected cluster position $\bR=R\cos(\phi) \be_x + 
R\sin(\phi) \be_y$, eq. ($\ref{delvimp}$) gives the mean velocity kick
\beqa
&& \left \langle \btu \bv \right \rangle  =  
\frac{2G\mbh}{\vrel} \frac{1}{M} \int \rmnd^3x \rho(r)
\frac{ (b-R\cos(\phi))\be_x -R\sin(\phi) \be_y }
{b^2+R^2-2bR\cos(\phi)}
\nonumber \\ && =
\frac{2G\mbh}{\vrel} \frac{1}{M} \int_0^{\rt} \rmnd R R \Sigma(R)
\int_0^{2\pi} d\phi
\frac{ (b-R\cos(\phi))\be_x -R\sin(\phi) \be_y }
{b^2+R^2-2bR\cos(\phi)}
\nonumber \\ && =
\frac{2G\mbh}{\vrel} \frac{1}{M} \int_0^{\rt} \rmnd R 2\pi R \Sigma(R)
\frac{ \theta(b-R)}{b} \be_x
\nonumber \\ && = 
\frac{2G\mbh}{b\vrel} \frac{\mcyl(b)}{M} \be_x
\label{cmkick}
\eeqa
where $\Sigma(R)=2\int_0^{(\rt^2-R^2)^{1/2}} \rmnd z \rho(\sqrt{R^2+z^2})$
and $\mcyl(b)=\int_0^{b} \rmnd R 2\pi R \Sigma(R)$
is the mass enclosed within a cylindrical distance $b$ from the center
of the cluster.
The $b$ dependent function $\mcyl(b)/bM$ is zero at the cluster
center, reaches a maximum inside the cluster and is equal to $1/b$ 
outside the cluster since $\mcyl(b \geq \rt)=M$, recovering the usual
velocity kick for a structureless particle.

\label{lastpage}

\end{document}